%% file: cft_DC_group_theory.tex
\documentclass{article}
\usepackage{jheppub}
\pdfoutput=1
\usepackage{amsmath,amssymb,amsthm}
\usepackage{bm}
\usepackage{pdflscape}
\usepackage[usenames,dvipsnames]{xcolor}
\usepackage{graphicx}
\usepackage[para]{threeparttable}
\usepackage{longtable}
\usepackage{booktabs}
\usepackage{float}
\usepackage{array}
\usepackage{arydshln}
\usepackage{multirow}
\usepackage{rotfloat}
\usepackage{caption}
\usepackage{subcaption}
\usepackage[normalem]{ulem}
\usepackage{hyperref}
\usepackage{cleveref}
\usepackage{tikz}
\usepackage{tikz-cd}
\usepackage{comment}

\hypersetup{
	pdftitle={Group Theory and the CFT Distance Conjecture: N=2 Tensionless Strings Have No (Co)Weight},
	pdfauthor={F. Baume, F. Mantegazza},
	pdflang={en-GB}
}

\newtheorem{theorem}{Theorem}

\newtheorem{classification}{Classification}
\theoremstyle{plain}

\input{tikz_data.tex}

\title{Group Theory and the CFT Distance Conjecture: \\ {\large $\mathcal{N}=2$ Tensionless Strings Have No (Co)Weight}}

\author[a]{Florent Baume,}
\author[a,b,c]{Fabio Mantegazza}

\affiliation[a]{II. Institut f\"ur Theoretische Physik, Universit\"at Hamburg, Notkestr. 9, 22607 Hamburg, Germany}
\affiliation[b]{Deutsches Elektronen-Synchrotron DESY, Notkestr. 85, 22607 Hamburg, Germany}
\affiliation[c]{Zentrum f\"ur Mathematische Physik, Universit\"at Hamburg, Bundesstr. 55, 20146 Hamburg, Germany}

\emailAdd{florent.baume@desy.de}
\emailAdd{fabio.mantegazza@desy.de}

\abstract{We perform a systematic survey of the Hagedorn behaviour at infinite-distance points in the conformal manifold of four-dimensional large-$N$ $\mathcal{N}=2$ Superconformal Field Theories admitting a Lagrangian description. Many properties of these theories can be understood in terms of the Lie algebra encoding the shape of their quiver. We find that in the overall-free limit, the Hagedorn temperature is determined by the largest eigenvalue of an affine or finite Cartan adjacency matrix. This defines two types of universality classes of theories sharing the same high-energy exponential growth of states characteristic of string-like spectra. The first and largest is the affine case, corresponding to orbifold and orientifold projections of $\mathcal{N}=4$ super-Yang--Mills, and all share the same temperature. The others fall into universality classes following an ADE classification with a temperature set by the dual Coxeter number, and can be obtained by deforming the affine case. Our results apply to all large-$N$ $\mathcal{N}=2$ quivers with any classical gauge symmetry, including those with matter charged beyond bifundamental representations. We further discuss the string-theoretic construction of these theories and some of the holographic implications, as well as how our methods extend to broad families of theories with less supersymmetry. We also consider limits where only part of the theory becomes free, and find lower and upper bounds on the exponential rate predicted by the CFT Distance Conjecture. Both these bounds and the Hagedorn temperature are set by the same eigenvalue, and when the quiver has a single gauge node the lower bound is saturated, giving a natural explanation for the three universality classes recently found in the literature.}

\begin{document}
\begin{flushright}
		{\tt ZMP-HH/26-12}\\
		{\tt DESY-26-077}\\

\end{flushright}
\maketitle

\newpage

\section{Introduction}

Conformal Field Theories (CFTs) have long played a special r\^ole in the study
of Quantum Field Theory. As endpoints of Renormalisation-Group flows
\cite{Polchinski:1987dy, Luty:2012ww}, they provide a natural way to organise
theories in terms of their low- and high-energy regimes. They need not,
however, be isolated points in the space of theories, but can in principle
occur in continuous families related by exactly-marginal deformations, forming
a space called the conformal manifold. Yet, the presence of operators associated
with an exactly-marginal coupling is difficult to establish. For
instance, quantum corrections in Lagrangian theories will generically lead to
an anomalous dimension for classically-marginal operators. In fact, the only
known cases of conformal manifolds involve supersymmetry in spacetime dimensions
larger than two, as non-renormalisation theorems may prohibit such corrections.
The possible candidates for deformations preserving superconformal invariance
then depend on both the number of supercharges and spacetime dimensions 
\cite{Minwalla:1997ka, Cordova:2016xhm}. For instance, in five and six
dimensions, the superconformal algebra prohibits the existence of marginal
operators altogether, and these Superconformal Field Theories (SCFTs) cannot be part of a
 conformal manifold \cite{Louis:2015mka, Buican:2016hpb,
Cordova:2016xhm}.

In lower dimensions on the other hand, such operators are known to exist. In particular,
four-dimensional SCFTs admit two classes of marginal operators which are
related to superpotential-type deformations, or to the tuning of possible gauge
couplings \cite{Leigh:1995ep, Green:2010da, Cordova:2016xhm}. The most
celebrated example of an SCFT admitting a conformal manifold is $\mathcal{N}=4$
super-Yang--Mills, where the continuous marginal parameter is the complexified
gauge coupling.

Moreover, the conformal manifold encodes more than the possible deformations of
a given CFT, and its structure can be used to obtain information about at least
part of the conformal data. For instance, central charges and certain conformal
anomalies are invariant under a change in marginal couplings, and more generally
certain correlators are covariantly constant, or must satisfy certain
consistency conditions imposed by conformal invariance, see e.g.
\cite{Osborn:1991gm, Baume:2014rla, Gomis:2015yaa, Schwimmer:2019efk,
Niarchos:2020nxk, Niarchos:2021iax, Andriolo:2022lcb} and references therein.
Understanding the structure of the conformal manifold then enables one to
compute these quantities at a given point of the conformal manifold---such as
one associated with a weakly-coupled regime where perturbative methods are
available---and obtain their values at any other point, at least in principle.
Another prime example is the superconformal index \cite{Kinney:2005ej}, which
does not change under exactly-marginal deformations and can therefore be obtained
in the free-field regime, when it exists \cite{Gadde:2011uv}. 

Furthermore, via holography the conformal manifold maps to the moduli space of
massless fields in the AdS bulk gravity dual, and its structure is therefore of
particular interest to probe gravitational features. The renewed interest in
the study of the constraints imposed by quantum gravity on low-energy effective
field theories has in the past decade led to a better understanding of these
moduli spaces, particularly on a flat-space background. For instance, the
Swampland Distance Conjecture \cite{Ooguri:2006in} states that at infinite
distance in moduli space one should find an infinite tower of states becoming
massless exponentially fast with the distance. While it has not yet been
proven, it has been tested in a plethora of examples and to an increasing
degree of mathematical rigour.

Concurrent advances in the Swampland Programme and the Conformal Bootstrap have
further motivated studies of the consequences of the Swampland Distance
Conjecture on conformal manifolds, in a line of study initiated in references \cite{Baume:2020dqd,
Perlmutter:2020buo}. This has led to a conformal version of the Distance
Conjecture stating that infinite-distance points in the conformal manifold
correspond to decoupling limits of the underlying CFT, and vice-versa.
Loci on the boundary of that space are therefore associated with a subsector of the
theory becoming free, where conformal symmetry enhances to a larger
higher-spin algebra generated by an infinite number of higher-spin conserved
currents \cite{Maldacena:2011jn, Boulanger:2013zza, Alba:2013yda,
Hartman:2015lfa, Alba:2015upa, Li:2015itl}. Using properties of these
currents, it has been shown that such decoupling points are always at infinite
distance for any $d>2$ without assuming supersymmetry \cite{Baume:2023msm}.
However, it is not yet known if the converse also holds, or whether there can
be exotic infinite-distance points that do not admit higher-spin conserved
currents.

Using the holographic dictionary, these operators 
map to an infinite number of massless higher-spin fields in the bulk,
prohibiting any effective field theory description. Away from these special loci
however, the higher-spin operators develop an anomalous dimension and are no
longer conserved. The CFT Distance Conjecture also posits constraints on the
behaviour of their anomalous dimension $M_\text{HS}L\sim\gamma_\text{HS}^{1/2}$, which is expected to vanish
exponentially fast with the distance:
\begin{equation}\label{CDC-decay}
		\gamma_\text{HS} \sim e^{-\kappa\,\text{dist}}\,.
\end{equation}
For four-dimensional theories, it has been proposed that the exponential rate is bounded from below, $\kappa\geq
\frac{1}{\sqrt{2}}$, in a normalisation corresponding to Planck units in the bulk dual. It is saturated by
$\mathcal{N}=4$ super-Yang--Mills, although it is also known to take large values 
\cite{Baume:2020dqd, Perlmutter:2020buo, Baume:2023msm,
Calderon-Infante:2026rkj, Calderon-Infante:2024oed}.\footnote{In the literature
related to the Swampland Distance Conjecture, it is common to write the
behaviour of the tower of massless states as $m\sim \text{exp}(-\alpha\,
\text{dist})$. The set of all possible values $\alpha_i$ for the different
types of infinite-distance limits is then often referred to as ``$\alpha$-vectors''. As
we will be interested in the relation between the CFT Distance Conjecture and
group theory in this work, we reserve the symbol $\alpha^i$ for the simple roots of
a Lie algebra, and denote the exponential rates by $\kappa_i$.}

In flat space a refinement of the Swampland Distance Conjecture has been
proposed, giving additional information about the nature of the infinite number
of states becoming massless \cite{Lee:2019wij}. It posits that this tower must
be either associated with Kaluza--Klein states and therefore correspond to a
decompactification limit to a higher-dimensional description, or the
excitations of a fundamental string becoming tensionless. A possible diagnostic
to distinguish these two towers is their growth of states: the former grows
polynomially with the mode number, while the latter grows exponentially. 

A natural question is then whether this refinement remains valid for AdS
backgrounds, and the possible interpretation of the tower of states. Of course,
one expects the behaviour of the tower to be drastically different than in flat
space: in units of the AdS radius, the masses of bulk fields are proportional
to the deviation of the conformal dimension from its unitarity bound
\cite{Metsaev:2003cu}, $mL\propto \Delta- \Delta_\text{bound}$. This means the
bulk counterpart of an operator of the form $\text{Tr}\varphi^n$ has mass
$m_nL\sim (n-1)$ in the free-field limit. In that regime, the tower will
therefore generically not collapse to zero mass, but rather be regularly
spaced, and only states associated with higher-spin conserved currents will
form a massless tower. While it can be shown that this tower has polynomial
growth, see e.g. references \cite{Calderon-Infante:2026rkj,Mantegazza:2026spd},
it cannot be associated to a Kaluza--Klein tower as it involves fields with
larger and larger spins. 

On the other hand, by studying multiplet recombination rules at threshold, it
has recently been argued that for certain theories, the higher-spin tower is
always accompanied by another tower of states starting at the AdS scale. These
new states come from long multiplets becoming BPS at the free point. They form a
second tower that does show an exponential growth of states, albeit not
massless ones \cite{Mantegazza:2026spd}. The interpretation is then that they
correspond to modes of a tensionless string propagating in AdS. The
prototypical example is type-IIB string theory on an $\text{AdS}_5\times S^5$
background, which is dual to $\mathfrak{su}(N)$ $\mathcal{N}=4$
super-Yang--Mills. In that case, the marginal parameter is set by the string
coupling in the dual picture, $g_\text{YM}^2 = 4\pi g_s$, and the tension in
units of the AdS radius is given by $(L^2/\alpha')^2 \sim g_s\,N$. 

We stress that the word ``tensionless'' can be somewhat of a misnomer. In the
supergravity regime $g_s\,N\gg1$, the spectrum of the string can be obtained by
quantisation of a sigma model on $\text{AdS}_5\times S^5$
\cite{Metsaev:1998it}, but the worldsheet description reproducing the spectrum
of the free $\mathcal{N}=4$ SCFT is very different \cite{Gaberdiel:2021jrv,
Gaberdiel:2021qbb}. In the absence of a controlled interpolating worldsheet
theory, it is arduous to define a notion of tension, and we will use that term
to define the bulk string with massless higher-spin excitations characteristic
of tensionless critical strings in flat space, and whose CFT dual has a free
subsector.

The precise nature of these strings in generic decoupling limits remains
elusive. To obtain an estimate of the growth of states and characterise them, a
possibility is to compute the thermal partition function of the
four-dimensional conformal field theory, defined as:
\begin{equation}\label{density-thermal}
		Z[x] = \text{Tr}_{\mathcal{H}_{S^{3}}}\, x^{H} = \int_{0}^\infty\,dE\,x^E\,\rho(E)\,,\qquad x = e^{-\frac{1}{T}}\,,
\end{equation}
with $T$ the temperature setting the length of the thermal circle $S^1_T$, $H$
the Hamiltonian of the theory and $\mathcal{H}_{S^{3}}$ its Hilbert space
quantised on $S^{3}$.

If at high energy the density of states has an exponential behaviour
$\rho(E)\sim e^{E/T_\text{H}}$, the partition function exhibits a Hagedorn
behaviour \cite{Hagedorn:1965st} characteristic of string-like spectra
\cite{Fubini:1969qb, Huang:1970iq, Atick:1988si}. In four dimensions, this was
first observed for $\mathcal{N}=4$ super-Yang--Mills in the free-field limit
\cite{Sundborg:1999ue, Aharony:2003sx}. More recently, it was found that all
Lagrangian SCFTs with a single simple gauge group admitting a large-$N$ limit
\cite{Calderon-Infante:2024oed}, as well as classes of $\mathcal{N}=2$ unitary
quivers \cite{Calderon-Infante:2026rkj}, also exhibit a Hagedorn temperature
$T_\text{H}$ in that limit. Intriguingly, in the one-node case it was shown
that there are three universality classes sharing the same Hagedorn
temperature and set solely by the ratio $a/c$ of the central charges. The
same ratio also defines the value of the exponential rate $\kappa=\kappa(a/c)$
in equation \eqref{CDC-decay}.

\subsection{Outline and Summary of Results}

The goal of this work is to extend this line of research, and perform a
systematic study of the Hagedorn behaviour in the overall free-field limit for
all $\mathcal{N}=2$ four-dimensional SCFTs admitting both a Lagrangian
description and a large-$N$ limit for all gauge symmetries. We focus on these
theories due to their attractive features. First, through
superconformal-representation arguments the only possible
$(\mathcal{N}=2)$-preserving marginal deformations correspond to changes in
Yang--Mills couplings and therefore constrain the allowed conformal manifolds
\cite{Leigh:1995ep, Green:2010da, Cordova:2016xhm}. Furthermore, as we review
in detail in Section \ref{sec:classification-quivers}, large-$N$ Lagrangian
SCFTs with eight supercharges are described in terms of quiver gauge theories
whose shape resembles that of Dynkin diagrams. This deep connection with group
theory will enable us to drastically simplify certain computations by calling
on properties of Lie algebras and their root systems.

In Section \ref{sec:classification-quivers} we review the classification of
large-$N$ $\mathcal{N}=2$ Lagrangian SCFTs, giving a brief introduction to
quiver theory. We take a slightly non-standard bottom-up approach, which
highlights how the adjacency matrices of quivers involving only bifundamental
hypermultiplets are related to Cartan matrices. We also discuss how, for
quivers involving only classical algebras $\mathfrak{su}(k_i),
\mathfrak{so}(k_i), \mathfrak{usp}(k_i)$, the vanishing of $\beta$-functions
relates gauge and flavour symmetries through the shape of the quiver.

The reader familiar with quiver theories can safely skip Section
\ref{sec:classification-quivers} and jump to Classification
\ref{classification-scft}, which summarises it and sets our notation. It states
that up to a few outliers involving representations beyond the bifundamental,
these SCFTs can be described by a triplet 
\begin{equation}
		\text{$\mathcal{N}=2$ quivers with bifundamental representations:}\qquad (\mathfrak{b}, S^i, f^i)\,. 
\end{equation}
The Lie algebra $\mathfrak{b}$ sets the shape (the ``base'') of the quiver,
$S^i\in\{0,1,-1\}$ gives the type of symmetry on each node, and $f^i$ is the
dimension of the fundamental representation of each flavour node. The rest of
the gauge data is then obtained by demanding we have an SCFT.

This triplet simplifies the computation of the thermal partition function and
the associated Hagedorn temperature, which we perform for all quivers in
Section \ref{sec:thermal}. In the large-$N$ limit, it can be written as a
Gaussian integral \cite{Aharony:2003sx, Imamura:2016abe,
Calderon-Infante:2024oed} and---no matter how the limit is performed---the
Hagedorn temperature is solely set by the shape of the quiver. This is perhaps
expected, as the large-$N$ limit washes away a large portion of the gauge
data.

More precisely, the Hagedorn temperature depends solely on the largest
eigenvalue $\lambda_\text{max}$ of the matrix $A = 2\mathbf{1}-C$ where $C$ is
the Cartan matrix of $\mathfrak{b}$. It is then obtained by finding the
smallest value of the temperature $T$ satisfying the following simple algebraic
relation:
\begin{equation}\label{hagedorn-constraint-intro}
		1- z_\text{V}(x) = \lambda_\text{max}\, z_\text{H}(x)\,,\qquad
		\begin{cases}
				\lambda_\text{max} = 2\,,\qquad &\mathfrak{b}\text{ affine}\,;\\
				\lambda_\text{max} = 2\,\text{cos}(\frac{\pi}{h^{\vee}_{\mathcal{U}(\mathfrak{b})}})\,,\qquad &\mathfrak{b}\text{ finite}\,,
		\end{cases}
\end{equation}
where $z_\text{H}$ and $z_\text{V}$ are fixed functions of the temperature
$x=e^{-1/T}$ associated with the contribution of free hyper- and vector
multiplets. 

We therefore find that there are two broad classes of theories, depending on
whether the algebra $\mathfrak{b}$ is of finite or affine type:
$\lambda_\text{max}$ is either universal for affine quivers, or set only by the
dual Coxeter number $h^{\vee}_{\mathcal{U}(\mathfrak{b})}$ of an ADE algebra
obtained by ``unfolding'' the quiver, a standard operation in group theory. 

We further find that the few outliers not falling in the classification above, e.g.
those involving (anti-)symmetric representations of a gauge algebra, have
nonetheless a Hagedorn temperature set by equation
\eqref{hagedorn-constraint-intro} and can be similarly explained in terms of
group theory. In Section \ref{sec:N=1}, we further discuss how our results can
also be applied to a broad class of $\mathcal{N}\leq1$ SCFTs. 

The shape of the quiver therefore defines universality classes of SCFTs sharing
the same string-like spectrum in the large-$N$ limit. In Section
\ref{sec:orientifolds} we discuss the brane realisations of such theories. When
$\mathfrak{b}$ is affine, these SCFTs are obtained as orbifolds and/or orientifolds of
$\mathcal{N}=4$ super-Yang--Mills, and the associated bulk tensionless string is
that of type-IIB string theory propagating on an $\text{AdS}_5\times S^5$
background. The remaining SCFTs are associated with finite Lie algebras, and
can be obtained from an affine quiver by decoupling the gauge algebra
associated with its affine node. 

The resulting quivers fail to satisfy the necessary conditions to admit a
weakly-coupled Einstein-gravity bulk dual, and an interpretation of the bulk
tensionless string is therefore difficult. In Section \ref{sec:holography} we
discuss how the holographic conditions can also be related to
$\lambda_\text{max}$, and how $\mathcal{N}=2$ large-$N$ quiver SCFTs can be
obtained via compactification of six-dimensional theories on a torus. These are
known to have an F-theory geometric engineering, and can be realised as
deformations of Little String Theories (LSTs). While LSTs are neither conformal
nor local, they are non-critical six-dimensional strings. We discuss how
$\mathcal{N}=2$ quivers are related to LSTs, how those associated with a
finite algebra can be obtained through a certain double-scaling limit, and how
this limit is related to ``little-string holography''.

Finally, in Section \ref{sec:bound-rate} we study the exponential rate $\kappa$
defined in equation \eqref{CDC-decay} and how it can be written in terms of
group-theoretical quantities related to the shape of the quiver. Our main
finding is that if the node of the quiver with gauge symmetry $\mathfrak{g}_i$
is decoupled, it is bounded from below by
\begin{equation}\label{intro-bound}
		\kappa_i = \kappa_\text{min}\sqrt{\frac{\text{dim}(\mathfrak{g}_\text{tot})}{\text{dim}(\mathfrak{g}_i)}}\,,\qquad
		\kappa_\text{min} \geq \sqrt{\frac{1}{2} + \frac{1}{12}(2-\lambda_\text{max})}\,, 
\end{equation}
where $\lambda_\text{max}$ is the same as above. From equation
\eqref{hagedorn-constraint-intro}, we recover the weaker bound $\kappa_i\geq
\kappa_\text{min}\geq \frac{1}{\sqrt{2}}$. The bound on $\kappa_\text{min}$ is saturated for affine
quivers, and a similar upper bound can be found for finite quivers. If the
total dimension of the gauge group $\text{dim}(\mathfrak{g}_\text{tot})$ is much
		larger than the one decoupling, which may easily occur when the
		quiver has a very large number of gauge nodes, the rate becomes large.

Our results give a simple explanation for the presence of the three
universality classes found in the ``mini-landscape'' of one-node large-$N$
SCFTs discovered in reference \cite{Calderon-Infante:2024oed}: we show in
Section \ref{sec:mini-landscape} that all such theories correspond either
to orbifolds and/or orientifolds of $\mathcal{N}=4$ super-Yang--Mills and therefore share the same
Hagedorn temperature, or are obtained by decoupling the affine node of quivers
associated with $\mathbb{Z}_2$ or $\mathbb{Z}_3$ orbifolds and related to an
algebra $\mathfrak{b}$ of finite type. The three universality classes are then
labelled by $\lambda_\text{max}\in\{2,0,1\}$, respectively. 

Furthermore, the bound in equation \eqref{intro-bound} becomes an equality for
one-node quivers. Since $\mathfrak{g}_\text{tot}=\mathfrak{g}_1$ in that case, it
follows that both the Hagedorn temperature \emph{and} the exponential rate are
set by the same $\lambda_\text{max}$ in those cases. We however stress that
while $\kappa_i$ and the Hagedorn temperature are ultimately both determined by
Lie-algebraic properties of $\mathfrak{b}$, only the latter depends only on
$\lambda_\text{max}$, and in general $\kappa_i$ depends on the full spectrum of the
adjacency matrix, rather than only its largest eigenvalue.

We give our conclusions and possible future directions in Section
\ref{sec:conclusions}. Finally, we have collected various definitions related
to Lie algebras and their symmetric polynomials in Appendix
\ref{app:lie-algebras}. In particular, the interested reader can find
additional details on the computation of the large-$N$ thermal partition
function in Appendix \ref{app:Haar}.

\paragraph{Note Added:} While this work was being finalised, we became aware of
a work by Calder\'on Infante, Uranga, and Valenzuela \cite{Calderon-Infante:2026zmd},
which also explores the nature of the universality classes defined by the
Hagedorn temperature, in their case through brane models. We coordinated
submission to appear on the same day on the arXiv.

\section{\texorpdfstring{Large-$N$ $\mathcal{N}=2$}{Large-N N=2} Superconformal Gauge Theories}\label{sec:classification-quivers}

In this section, we review the classification of four-dimensional
$\mathcal{N}=2$ SCFTs admitting both a quiver description and a large-$N$ limit
for all gauge algebras. We also give a concise introduction to quiver theories
intended for readers interested in the CFT Distance Conjecture who may not be
familiar with them, as well as to set our notation. It will also illustrate
how the methods we will utilise in later sections can be recast in terms of
Lie-algebraic properties encoded in the shape of the quiver. The reader
familiar with quivers can skip ahead to Classification
\ref{classification-scft}, which summarises the classification and sets our
conventions.

The study of Lagrangian $\mathcal{N}=2$ SCFTs traces back to the second
superstring revolution, where these theories were engineered via brane
systems, and made out of $\mathfrak{su}(m)\oplus\mathfrak{su}(n)$
bifundamental hypermultiplets \cite{Douglas:1996sw, Douglas:1997de,
Katz:1997eq, Kachru:1998ys, Park:1998zh, Lawrence:1998ja, Hanany:1998sd}, or
$\mathfrak{so}(m)\oplus\mathfrak{usp}(2n)$ in the presence of O-planes
\cite{Uranga:1998uj, Witten:1998xy, Sen:1996vd, Douglas:1996js, Ennes:2000fu}.
They follow finite and affine ADE classifications, see e.g. reference
\cite{Giveon:1998sr} for a review, and cover a large part of all possible
$\mathcal{N}=2$ Lagrangian SCFTs. More models can be constructed, including
some where matter fields do not transform in bifundamental representations. An
exhaustive enumeration of the possible outliers was achieved in reference
\cite{Bhardwaj:2013qia}. We stress that we are focussing here only on SCFTs
admitting a Lagrangian, which is but a small corner of the landscape of
$\mathcal{N}=2$ conformal theories. Myriad SCFTs are indeed non-perturbative in
nature, and can be obtained through deformations of a Lagrangian theory, via
compactification from six dimensions, or through geometric engineering in
string theory, see e.g. the reviews \cite{Tachikawa:2013kta, Akhond:2021xio,
Argyres:2022mnu} as well as references therein.

Lagrangian theories giving rise to $\mathcal{N}=2$ SCFTs are strongly
constrained by the vanishing of the $\beta$-function associated with their gauge
couplings, which severely restricts the possible gauge and flavour
representations under which the matter spectrum formed by hypermultiplets can transform. Following
reference \cite{Bhardwaj:2013qia}, in this work we will consider theories with
a semi-simple gauge symmetry:\footnote{For $\mathcal{N}=2$ theories, the
$\beta$-function of an Abelian gauge symmetry induces a Landau pole, and
can only appear in effective theories, rather than in \emph{bona fide} UV-complete SCFTs.}\footnote{We follow the convention for which
$\mathfrak{usp}(2r)\simeq\mathfrak{sp}(r)\simeq C_r^{(0)}$ has rank $r$, and
the dimension of its fundamental representation is $(2r)$.}
\begin{equation}
		\mathfrak{g} = \bigoplus_i \mathfrak{g}_i\,,\qquad \mathfrak{g}_i=\mathfrak{su}(k_i)\,,\mathfrak{so}(k_i)\,,\mathfrak{usp}(k_i)\,,
\end{equation}
and a matter spectrum consisting of hypermultiplets transforming in the
representations
\begin{equation}
		\bm{R}_a = \bigotimes_i \bm{R}_{a,i}\,.
\end{equation}

The possible representations $\bm{R}_a$ are constrained by supersymmetry.
Indeed, recall that in $\mathcal{N}=2$ supersymmetric theories, hypermultiplets
are endowed with a symplectic structure. In the presence of additional gauge or
flavour symmetries, the hypermultiplets must transform in representations
consistent with that structure. As a consequence, if a representation
$\bm{R}_a$ is real or complex, hypermultiplets must appear in pairs
$\bm{R}_a\oplus\overline{\bm{R}}_a$ to satisfy this requirement, and form a
\emph{full hypermultiplet}. By contradistinction, when $\bm{R}$ is pseudo-real
by itself, it is automatically compatible with the symplectic structure, and forms a
\emph{half hypermultiplet}. By definition, to obtain a superconformal theory,
the $\beta$-functions $\beta^i$ associated with each Yang--Mills coupling of
simple factors $\mathfrak{g}_i$ must vanish. For $\mathcal{N}=2$ theories,
these are one-loop exact, and fixed by representation theory:
\begin{equation}\label{beta-function}
		\begin{gathered}
				\beta^i = -\frac{g_i^3}{16\pi^2}\bigg(2h^\vee_i - \sum_{a}b_i(\bm{R}_a)\bigg)\,,\qquad
				b_i(\bm{R}_a) = a_a\,\,\textrm{T}(\bm{R}_{a,i})\,\prod_{j\neq i}\text{dim}(\bm{R}_{a,j})\,,
		\end{gathered}
\end{equation}
where $a_a = 1$ if $\bm{R}_{a}$ is pseudo-real, and $a_a=2$ otherwise so as to
take into account the associated full hypermultiplet; $h^\vee_i$ is the dual
Coxeter number of $\mathfrak{g}_i$, and $\textrm{T}(\bm{R})$ the Dynkin index of
the representation. 

From equation \eqref{beta-function}, we already see that the Dynkin index
$\textrm{T}(\bm{R})$ cannot exceed twice the dual Coxeter number $h^\vee$, and only a
handful of representations can appear in a Lagrangian SCFT as $\textrm{T}(\bm{R})$
grows with their dimensions. In addition, one can show that $\bm{R}_a$ can
involve at most three gauge groups, and this can be achieved for exactly three
gauge algebras, where the possibilities are the trifundamental of
$\mathfrak{su}(2)\oplus\mathfrak{su}(2)\oplus\mathfrak{su}(2)$ or
$\mathfrak{su}(2)\oplus\mathfrak{su}(2)\oplus\mathfrak{usp}(4)$. The complete
list of allowed representations can be found in e.g. reference 
\cite[Tables 1--3]{Bhardwaj:2013qia}.

All in all, up to a few outliers we will briefly discuss below, the
overwhelming majority of $\mathcal{N}=2$ quiver SCFTs---whether at large $N$ or not---involve only
hypermultiplets in the bifundamental representation of
$\mathfrak{su}(m)\oplus\mathfrak{su}(n)$,
$\mathfrak{so}(m)\oplus\mathfrak{usp}(n)$, $\mathfrak{su}(m)\oplus\mathfrak{so}(n)$, and
$\mathfrak{su}(m)\oplus\mathfrak{usp}(n)$ flavour or gauge symmetries. We will
use the quiver notation, which is the standard pictorial tool to represent
Lagrangian theories. We denote a gauge group by a circle, and a bifundamental
hypermultiplet by a line, so that all the relevant combinations above are depicted as:
\begin{equation}\label{quiver-definitions}
	\begin{matrix}
	\begin{tikzpicture}
		\node[node, label=above:{\footnotesize $\mathfrak{su}(k_{j})$}, fill=black] (A1)  {};
		\node[node, label=above:{\footnotesize $\mathfrak{su}(k_{i})$}, fill=black] (A0) [left=8mm of A1] {};
	    \draw (A0.east) -- (A1.west);
	\end{tikzpicture}
	\end{matrix}\qquad
	\begin{matrix}
	\begin{tikzpicture}
		\node[node, label=above:{\footnotesize $\mathfrak{usp}(k_{j})$}, fill=blue] (A1)  {};
		\node[node, label=above:{\footnotesize $\mathfrak{so}(k_{i})$}, fill=red] (A0) [left=8mm of A1] {};
	    \draw (A0.east) -- (A1.west);
	\end{tikzpicture}
	\end{matrix}\qquad 
	\begin{matrix}
	\begin{tikzpicture}
		\node[node, label=above:{\footnotesize $\mathfrak{usp}(k_i)$}, fill=blue] (A0)  {};
		\node[node, label=above:{\footnotesize $\mathfrak{su}(k_j)$}, fill=black] (A1) [right=8mm of A0] {};
		\node[yscale=1.4] (C) [right=1.6mm of A0] {$>$};
		\draw ([yshift=1.5pt]A0.east) -- ([yshift=1.5pt]A1.west);
	    \draw ([yshift=-1.5pt]A0.east) -- ([yshift=-1.5pt]A1.west);
	\end{tikzpicture}
	\end{matrix}\qquad
	\begin{matrix}
	\begin{tikzpicture}
		\node[node, label=above:{\footnotesize $\mathfrak{so}(k_i)$}, fill=red] (A0)  {};
		\node[node, label=above:{\footnotesize $\mathfrak{su}(k_j)$}, fill=black] (A1) [right=8mm of A0] {};
		\node[yscale=1.4] (C) [right=1.6mm of A0] {$>$};
		\draw ([yshift=1.5pt]A0.east) -- ([yshift=1.5pt]A1.west);
	    \draw ([yshift=-1.5pt]A0.east) -- ([yshift=-1.5pt]A1.west);
	\end{tikzpicture}
	\end{matrix}
\end{equation}
where we denote an $\textcolor{red}{\mathfrak{so}(k)}$ symmetry as a red node,
$\textcolor{blue}{\mathfrak{usp}(k)}$ in blue, and $\mathfrak{su}(k)$ in black.
Possible additional flavour symmetries are denoted by a square, and can only
attach to a single gauge node in the cases discussed in this work. For
bifundamental hypermultiplets of
$\textcolor{blue}{\mathfrak{usp}(k_i)}\oplus\mathfrak{su}(k_j)$ and
$\textcolor{red}{\mathfrak{so}(k_i)}\oplus\mathfrak{su}(k_j)$, we use a
double line with an arrow pointing to the $\mathfrak{su}(k_j)$. While this notation is
somewhat non-standard, we will see that it will be very helpful both to
emphasise that even though the fundamental representations of
$\mathfrak{usp}(k_i)$ and $\mathfrak{so}(k_i)$ are (pseudo-)real, we have a
full hypermultiplet due to the complex representation of $\mathfrak{su}(k_j)$,
and to see that the vanishing of the $\beta$-functions will be related to
a Dynkin diagram. 

In the sequel we will often label a node by the dimension $k_i$ of its
fundamental representation, as the colour is sufficient to know which type of
algebra it is associated with. Furthermore, imposing vanishing of the
$\beta$-function forbids the presence of hypermultiplets in the bifundamental
representations of $\mathfrak{so}(k_i)\oplus\mathfrak{so}(k_j)$, and we will
not consider this case. On the other hand, bifundamental representations of
$\mathfrak{usp}(k_i)\oplus\mathfrak{usp}(k_j)$ are possible and form full
hypermultiplets. They only appear in a handful of quivers
\cite{Bhardwaj:2013qia}, and for the cases relevant here can always be thought of
as a particular limit of certain infinite families. As such, we will only
briefly mention how they arise below.

Let us now give a concise argument to obtain the classification of
$\mathcal{N}=2$ quiver SCFTs with only bifundamental hypermultiplets. We use a
path that is slightly different from the traditional classifications, but will
make the relationship with group theory manifest. Consider the following
spectrum:
\begin{itemize}
		\item $\tilde{r}$ vector multiplets transforming in the adjoint representation of a simple algebra $\mathfrak{g}_i$ of classical type: $ \mathfrak{su}(k_i)$, $\mathfrak{so}(k_i)$, or $\mathfrak{usp}(k_i)$.
		\item Hypermultiplets in the bifundamental $(\overline{\bm{k}}_i,
				\bm{k}_j)$ of $\mathfrak{g}_i\oplus\mathfrak{g}_j$. The
				(symmetric) adjacency of the quiver is summarised in the matrix
				$B^{ij}$, where $B^{ij}=1$ for half-hypermultiplets, and
				$B^{ij}=2$ for full hypermultiplets. 
		\item Hypermultiplets in the bifundamental $(\overline{\bm{k}}_i, \bm{f}_i)$ of $\mathfrak{g}_i\oplus\mathfrak{f}_i$, where $\mathfrak{f}_i$ denotes the flavour symmetry. The adjacency of the gauge-flavour hypermultiplets is encoded in the diagonal matrix $D^{ij} = d_{i}\,\delta^{ij}$, with $d_i=2$ for full hypermultiplets and $d_i=1$ otherwise.
\end{itemize}
In our case, there will only be (at most) a single flavour per gauge node.
Furthermore, there are no self-loops in the quiver, and $B^{ii}=0$. These
would be associated with hypermultiplets transforming in the adjoint
representation; $\mathcal{N}=4$ super-Yang--Mills is the only $\mathcal{N}=2$
SCFT hosting such hypermultiplets considered in this work. Furthermore,
$\mathcal{N}=2$ quivers are not directed, since a full hypermultiplet comes in
a representation $\bm{R}\oplus\overline{\bm{R}}$, and half-hypermultiplets are in (pseudo-)real
representations, which implies that $B^{ij}=B^{ji}$ is symmetric. This is in
contradistinction to $\mathcal{N}=1$ quivers which can be chiral and
therefore are oriented graphs.

As an example, the following quiver
\begin{equation}\label{example-F4}
	\begin{tikzpicture}
		\node[node, label=above:{\footnotesize $\mathfrak{so}(k_{1})$}, fill=red] (A1) {};
	    \node[node, label=above:{\footnotesize $\mathfrak{usp}(k_{2})$}, fill=blue] (A2) [right=8mm of A1] {};
	    \node[node, label=above:{\footnotesize $\mathfrak{su}(k_{3})$}, fill=black] (A3) [right=8mm of A2] {};
	    \node[node, label=above:{\footnotesize $\mathfrak{su}(k_{4})$}, fill=black] (A4) [right=8mm of A3] {};
		\node[yscale=1.4] (C) [right=1.6mm of A2] {$>$};
	    \node[rectangle, label=below:{\footnotesize $\mathfrak{usp}(f^{1})$}, fill=blue] (F1) [below=4mm of A1] {};
	    \node[rectangle, label=below:{\footnotesize $\mathfrak{so}(f^{2})$}, fill=red] (F2) [below=4mm of A2] {};
	    \node[rectangle, label=below:{\footnotesize $\mathfrak{su}(f^{3})$}, fill=black] (F3) [below=4mm of A3] {};
	    \node[rectangle, label=below:{\footnotesize $\mathfrak{su}(f^{4})$}, fill=black] (F4) [below=4mm of A4] {};
	    \node (D) [right=6mm of C] {};
	    \draw (A1.east) -- (A2.west);
		\draw ([yshift=1.5pt]A2.east) -- ([yshift=1.5pt]A3.west);
	    \draw ([yshift=-1.5pt]A2.east) -- ([yshift=-1.5pt]A3.west);
	    \draw (A3.east) -- (A4.west);
		\draw (A1.south) -- (F1.north);
		\draw (A2.south) -- (F2.north);
		\draw (A3.south) -- (F3.north);
		\draw (A4.south) -- (F4.north);
	\end{tikzpicture}
\end{equation}
has its adjacency data given by 
\begin{equation}
		B = 
		\left(\begin{matrix}
				0& 1 &0 &0\\
				1& 0 &2 &0\\
				0& 2 & 0 & 2\\
				0& 0 & 2 &0
		\end{matrix}\right)\,,\qquad
		D = \text{diag}(1, 1, 2, 2)\,.
\end{equation}

We recall that the flavour algebra is fixed by the gauge algebra: $n_f$ \emph{full}
hypermultiplets transforming under a $\mathfrak{g}=\mathfrak{su}(k)$ are
rotated by an $\mathfrak{su}(n_f)$ flavour symmetry; if
$\mathfrak{g}=\mathfrak{so}(n)$, the vector representation is real and $n_f$
\emph{half} hypermultiplets have a $\mathfrak{usp}(2n_f)$ flavour symmetry; for
$\mathfrak{g}=\mathfrak{usp}(k)$, $n_f$ half hypermultiplets are rotated by a
$\mathfrak{so}(n_f)$ transformation. We stress that in our conventions, for an
$\mathfrak{so}(k)$ gauge symmetry, we have $f=2n_f$. For representations
beyond the (anti-)fundamental, the flavour symmetry will depend on whether
that representation is complex, pseudo-real, or real. However, up to a single
case involving the anti-symmetric representation of $\mathfrak{usp}(k)$---which is
real---we will only encounter the symmetric and anti-symmetric representations
of $\mathfrak{su}(k)$---which are both complex.

\paragraph{Allowed Quiver Shapes:} Given the matrices $D$ and $B$, we can
neatly summarise the hypermultiplet content, or in other words the shape of the
quiver in the following symmetric matrix:
\begin{equation}
		G^{ij} = 2D^{ij} - B^{ij}\,.
\end{equation}
In this form, $G^{ij}$ completely classifies the possible shapes of the quiver.
Indeed, it 1) is symmetric; 2) has only $2$ or $4$ on the diagonal; and 3) has only $0,-1,-2$
off-diagonal. These are nothing but the symmetrised version of
(generalised) Cartan matrices, which have been classified by Kac
\cite{Kac:1990gs}. For each quiver, we can therefore associate a Lie algebra
$\mathfrak{b}$ and a Cartan matrix $C^{ij}$:\footnote{We follow the same conventions used recently for a similar analysis of six-dimensional
		theories \cite{Ahmed:2025fmq}, where the shape of the quiver is
		associated with the \emph{base} of an elliptic fibration in the
F-theory construction, giving its name to the algebra $\mathfrak{b}$.} 
\begin{equation}
		\mathfrak{b}:\qquad C = D^{-1}G = 2\, \mathbf{1} - A\,,\qquad A = D^{-1}B\,.
\end{equation}
Contrary to $G$ and $B$, the matrices $C$ and $A$ need not be symmetric. 

The type of Lie algebra depends on the determinant of $C$: when
$\text{det}(C)>0$, it is the Cartan matrix of a finite/simple algebra, an
affine algebra when the determinant vanishes, or a Lie algebra of so-called
indeterminate type otherwise. To distinguish the algebras related to gauge
symmetries and the one describing the shape of the quiver, encoded in
$\mathfrak{b}$, we use Kac's notation \cite{Kac:1990gs} for the latter:
\begin{equation}
	\begin{aligned}
		\text{det}\,C>0\quad (\text{finite}):&\qquad \mathfrak{b}=X_r^{(0)}\,,\qquad \quad \tilde{r} = \text{rk}(\mathfrak{b}) = r\,;\\
		\text{det}\,C=0\quad (\text{affine}):&\qquad \mathfrak{b}=X_r^{(n>0)}\,,\qquad \tilde{r} = \text{rk}(\mathfrak{b}) + 1\,,
	\end{aligned}
\end{equation}
where we recall that $\tilde{r}$ is the number of gauge nodes in the quiver.
$X$ can be any choice of Dynkin type $ABCDEF$. In that notation, the integer
$n$ corresponds to the order of the twist performed on the original affine
algebra $X_r^{(1)}$.\footnote{We will not need to use the minutiae of twisted
		algebras---those that have $n>1$---in this work, and the reader
		unfamiliar with them should simply take $n$ as a labelling device so
		that $C$ is of affine type when $n>0$. A complete treatment can be
		found in e.g. reference \cite{Kac:1990gs}. Furthermore, the subscript
		for the twisted algebra $\mathfrak{b}= X^{(n>1)}_\ell$ refers to the rank
		of the \emph{untwisted} algebra $X^{(1)}_\ell$, and the rank of
		$\mathfrak{b}$ is usually lower. In all twisted cases, the convention
		is that if we have $\mathfrak{b}=X_{a\,r+b}^{(n>1)}$, then
$\text{rk}(\mathfrak{b})=r$ and therefore $\tilde{r}=r+1$.} 

In the example given in equation \eqref{example-F4}, we have four nodes and
$\tilde{r}=4$, and a shape related to $\mathfrak{b}=F_4^{(0)}$, i.e. a finite 
algebra. There are two affine versions that can be obtained by adding a simple
gauge node. If the flavour $\mathfrak{usp}(f^1)$ is gauged, one obtains a
$F_4^{(1)}$ quiver, or the twisted algebra $E_6^{(2)}$ if the flavour
$\mathfrak{su}(f^4)$ is gauged. In both cases, they are of affine type
with $\tilde{r}=5$. From this, we further see that the double arrow between the
nodes with $\mathfrak{usp}(k_2)$ and $\mathfrak{su}(k_3)$ arises naturally from
the adjacency, justifying the notation defined in equation
\eqref{quiver-definitions}.

Finally, observe that we cannot have $\mathfrak{b}= G_2^{(0)}$, $G_2^{(1)}$ or
$D_4^{(3)}$, since those algebras have an entry $G^{ij}=-3,-6$, which does not
make sense when $G$ is interpreted as the adjacency of either half- or full
hypermultiplets.

\paragraph{Vanishing of the $\beta$-Functions:} To fully classify all quivers
rather than only their shape, one must in addition classify the allowed gauge
algebras, or equivalently the dimensions of the fundamental representations of
gauge and flavour symmetries, denoted by $k_i$ and $f^i$, respectively. 

As mentioned before, those are constrained by demanding the vanishing of all
$\beta$-functions, which can be done node by node. For instance, considering
the following unitary quiver---that is, one with only $\mathfrak{su}(k)$
algebras---the $\beta$-function on the middle node is given by:
\begin{equation}\label{subquiver-example}
	\begin{matrix}
	\begin{tikzpicture}
		\node[node, label=above:{\footnotesize $k_{i}$}, fill=black] (A1)  {};
		\node[node, label=above:{\footnotesize $k_{i-1}$}, fill=black] (A0) [left=6mm of A1] {};
		\node[node, label=above:{\footnotesize $k_{i+1}$}, fill=black] (A2) [right=6mm of A1]  {};
		\node[rectangle, label=below:{\footnotesize $f^{i}$}, fill=black] (F1) [below=6mm of A1]  {};
	    \draw (A0.east) -- (A1.west);
	    \draw (A1.east) -- (A2.west);
	    \draw (A1.south) -- (F1.north);
	\end{tikzpicture}
	\end{matrix}\quad:
	\qquad 
	\beta^i = -\frac{g_i^3}{16\pi^2}\bigg( 2 k_i - 2\cdot\big(\frac{1}{2}\cdot k_{i-1} + \frac{1}{2}\cdot k_{i+1} + \frac{1}{2}\cdot f^i\big)\bigg),
\end{equation}
with the square depicting the $\mathfrak{su}(f^i)$ symmetry. We suppressed the
flavour symmetries of the other nodes as they do not participate in the
computation of $\beta^i$. We have three representations: $(\bm{k}_{i-1},
\overline{\bm{k}}_i)$, $(\bm{k}_{i}, \overline{\bm{k}}_{i+1})$, $(\bm{k}_{i},
\overline{\bm{f}}^i)$, and we have used that $k_i = \text{dim}(\bm{k}_i)$ and
$\textrm{T}(\bm{k}_i)=\frac{1}{2}$, taking into account that the fundamental of
$\mathfrak{su}(k)$ is complex. We have further collated some of the
Lie-algebraic quantities involved in the computation of $\beta$-functions in
Table \ref{tab:classical-quantities} for convenience.

\begin{table}
		\centering
		\begin{tabular}{cccc}
				\toprule
				$\mathfrak{g}$ & $\mathfrak{su}(k)$ & $\mathfrak{so}(k)$ & $\mathfrak{usp}(k)$\\\midrule
				$d$ & $2$ & $1$ & $1$ \\ 
				$\text{rk}(\mathfrak{g})$ & $k-1$ & $\lfloor\frac{k}{2}\rfloor$ & $\frac{k}{2}$\\
				$\text{dim}(\mathfrak{g})$ & $k^2-1$ & $\frac{1}{2}k(k-1)$ & $\frac{1}{2}k(k+1)$\\
				$h^\vee = \textrm{T}(\bm{adj})$ & $k$ & $k-2$ & $\frac{1}{2}(k+2)$ \\
				$\textrm{T}(\bm{F})$ & $\frac{1}{2}$ & $1$ & $\frac{1}{2}$ \\
				\bottomrule
		\end{tabular}
		\caption{\label{tab:classical-quantities}Quantities related to representations of classical algebras. The representation $\bm{F}$ denotes the fundamental representation, which includes the vector representation of $\mathfrak{so}(k)$ by abuse of notation. The floor function is denoted by $\lfloor \cdot\rfloor$. For $\mathfrak{su}(k)$, we will further need the Dynkin indices $\textrm{T}(\bm{S^2})=\frac{1}{2}(k+2)$ and $\textrm{T}(\bm{\Lambda^2})=\frac{1}{2}(k-2)$.}
\end{table}

The quiver given in equation \eqref{subquiver-example} can be seen as a
subquiver of a more involved Lagrangian SCFT, and used as a building block to constrain the gauge symmetry of these theories. In general, unitary quivers have only $\mathfrak{su}(k_i)$ algebras
by definition so that $D^{ij} = 2\delta^{ij}$. Since $G^{ij}=2\,C^{ij}$ in those
cases, the Cartan matrix itself must be symmetric and therefore associated with
a simply-laced, finite or affine, ADE algebra. The vanishing of $\beta^i$ in
equation \eqref{subquiver-example} is simply $f^i = -k_{i-1}+2k_i - k_{i+1}$,
which can be seen as arising from the row of a Cartan matrix. The
generalisation of equation \eqref{subquiver-example} to nodes with trivalent or
quadrivalent patterns is straightforward, and it is easy to convince oneself
that the vanishing of the $\beta$-function for any unitary quiver is given by
\begin{equation}
		\text{unitary quiver SCFTs: }\qquad C^{ij} k_j = f^i\,.
\end{equation}
This reasoning can be extended straightforwardly to ortho-symplectic
quivers---those with both $\mathfrak{usp}$ and $\mathfrak{so}$ gauge algebras,
and for which $D^{ij}=\delta^{ij}$---as well as those involving all three
classical algebras. This is achieved by noticing that while the dual Coxeter numbers
and the fundamental representations are different, for classical algebras they
can be related to the dimension of the fundamental representation and its
so-called Frobenius--Schur indicator $S=0,1,-1$ for $\mathfrak{su}(k)$,
$\mathfrak{so}(k)$ and $\mathfrak{usp}(k)$. From Table
\ref{tab:classical-quantities}, we see that the dual Coxeter number can be
rewritten as 
\begin{equation}
		h^\vee= d\,\textrm{T}(\bm{F})\big(\text{dim}(\bm{F})-2S(\mathfrak{g})\big)\,.
\end{equation}

By exploiting these relations, one finds that for all $\mathcal{N}=2$ SCFTs
with only bifundamental hypermultiplets, the $\beta$-function of the gauge
algebra $\mathfrak{g}_i$---corresponding to the $i$-th node of the quiver---is given by:
\begin{equation}
		\beta^i \propto C^{ij}k_j - (f^i - 4S^i)\,,
\end{equation}
where $S^i=S(\mathfrak{f}_i)$ is the Frobenius--Schur indicator of the \emph{flavour}
symmetry, see Table \ref{tab:classical-quantities}.\footnote{The quantity $S^i$ is defined with respect to the
		\emph{flavour} symmetry, as they are the more convenient quantities to
		label these SCFTs, and follow the same conventions as in reference
		\cite{Ahmed:2025fmq}---see also reference \cite{Blum:1997mm}---for all
group-theoretical quantities, where similar arguments were applied to
six-dimensional SCFTs and Little String Theories. For the quivers we consider here, since the flavour symmetry is fixed by the gauge algebra, we always have $S^i=S(\mathfrak{f}^i) = - S(\mathfrak{g}_i)$.}

To obtain a superconformal field theory, imposing the vanishing of all
$\beta$-functions constrains the dimension of the fundamental representations
of the gauge symmetries $k_i$ and those of the flavour algebras $f^i$, leading to
what we will refer to in the rest of this work as the \emph{SCFT condition}:
\begin{equation}\label{beta-function-bif}
		C^{ij}k_j = (f^i - 4S^i)\,,\qquad S^i=
		\begin{cases}
				\phantom{+}0\,,&\qquad\text{if }\mathfrak{f}_i= \mathfrak{su}(f^i)\,;\\
				+1\,,&\qquad\text{if }\mathfrak{f}_i= \mathfrak{so}(f^i)\,;\\
				-1\,,&\qquad\text{if }\mathfrak{f}_i= \mathfrak{usp}(f^i)\,.
		\end{cases}
\end{equation}
As mentioned above, the type of flavour symmetry is fixed by that of the gauge
algebra, and we use the convention where $S^i$ is set by $\mathfrak{f}_i$. If
there is no flavour symmetry at all, $S^i$ is set by the would-be flavour
symmetry.

We are ultimately interested in quivers which admit a large-$N$ limit for all
gauge symmetries, i.e. $k_i \propto N$ for $N\gg 1$. This condition promptly
dispenses with quivers that have a shape associated with an algebra of
indeterminate type. Indeed, when $\text{det}(C)<0$, in order to ensure the
vanishing of the $\beta$-function one will generically find negative values of
$k_i$ or $f^i$ due to the presence of negative eigenvalues of $C$. While there
are a few sporadic solutions with low values of $k_i$ \cite{Bhardwaj:2013qia},
those are not part of families admitting a large-$N$ limit and we will not
consider them in this work.

This means that any quiver that is part of a family admitting a large-$N$ limit
must be shaped like the Dynkin diagram of a finite or affine Lie algebra
$\mathfrak{b}$, and its flavour and gauge symmetries must satisfy the SCFT
condition \eqref{beta-function-bif}. The list of all such theories is collated
in Table \ref{tab:dynkin-bases-unitary} for unitary quivers, Table
\ref{tab:dynkin-bases-ortho-symplectic} for ortho-symplectic quivers, and Table
\ref{tab:dynkin-bases-unitary-ortho-symplectic} for those involving all three
algebras and shaped like non-simply-laced Dynkin diagrams.

\begin{table}[h]
    \centering
    \scriptsize 
    \begin{threeparttable}
    \resizebox{1\textwidth}{!}{
        \begin{tabular}[t]{cccc}
				\toprule
			$\mathfrak{b}$& unitary finite quiver & $\mathfrak{b}$ & unitary affine quiver\\\midrule
			$A^{(0)}_r$ & $\begin{matrix}\input{figures/Ak_flavor.tex}\end{matrix}$ & $A_r^{(1)}$ & $\begin{matrix}\input{figures/Ah.tex}\end{matrix}$ \\\midrule
			$D^{(0)}_r$ & $\begin{matrix}\input{figures/Dk_flavor.tex}\end{matrix}$ & $D_r^{(1)}$ & $\begin{matrix}\input{figures/Dh.tex}\end{matrix}$  \\\midrule
			$E_6^{(0)}$ & $\begin{matrix}\input{figures/E6_flavor.tex}\end{matrix}$ &$E_6^{(1)}$ & $\begin{matrix}\input{figures/E6h.tex}\end{matrix}$ \\\midrule
			$E_7^{(0)}$ & $\begin{matrix}\input{figures/E7_flavor.tex}\end{matrix}$ &$E_7^{(1)}$ & $\begin{matrix}\input{figures/E7h.tex}\end{matrix}$ \\\midrule
			$E_8^{(0)}$ & $\begin{matrix}\input{figures/E8_flavor.tex}\end{matrix}$ & $E_8^{(1)}$ & $\begin{matrix}\input{figures/E8h.tex}\end{matrix}$ \\
        \bottomrule
        \end{tabular}
        }
    \end{threeparttable}
	\caption{Lagrangian $\mathcal{N}=2$ SCFTs that can be described as a unitary quiver with only $\mathfrak{su}(k_i)$ gauge algebras, and admitting a large $k_i\sim N$ limit for all nodes. All are shaped like an ADE algebra of rank $r$. For finite quivers, $r$ counts the number $\tilde{r}$ of gauge nodes, and for affine we have $\tilde{r}=r+1$. For affine quivers, the cancellation of $\beta$-functions prohibits the presence of any flavour symmetry.}
    \label{tab:dynkin-bases-unitary}
\end{table}
\clearpage

\begin{table}[p]
    \centering
    \scriptsize 
    \begin{threeparttable}
			\resizebox{1\textwidth}{!}{
        \begin{tabular}[t]{cccc}
				\toprule
			$\mathfrak{b}$& ortho-symplectic finite quiver & $\mathfrak{b}$ & ortho-symplectic affine quiver\\\midrule
			$A_{r}^{(0)}$ & $\begin{matrix}\input{figures/Ak_os_flavor.tex} & \input{figures/Ak_os_flavor_3.tex}\\[3mm] & \text{+ inverted}\end{matrix}$ & $A_{r=2n-1}^{(1)}$ & $\begin{matrix}\input{figures/Ah_os.tex}\end{matrix}$\\\midrule
			$D^{(0)}_r$ & $\begin{matrix}\input{figures/Dk_os_flavor.tex}\\[3mm] \text{+ inverted} \\ \input{figures/Dk_os_flavor_2.tex} \\[3mm] \text{+ inverted} \end{matrix}$ & $D_{r=2n}^{(1)}$ &  $\begin{matrix}\input{figures/Dh_os_flavor_2.tex}\\[4mm]\input{figures/Dh_os_flavor.tex}\end{matrix}$ \\\midrule
			$E_6^{(0)}$ & $\begin{matrix}\input{figures/E6_os_flavor.tex}\\[3mm] \text{+ inverted}\end{matrix}$ & $E_6^{(1)}$ & $\begin{matrix}\input{figures/E6h_os.tex}\\[3mm] \text{+ inverted}\end{matrix}$ \\\midrule
			$E_7^{(0)}$ & $\begin{matrix}\input{figures/E7_os_flavor.tex}\\[3mm] \text{+ inverted}\end{matrix}$ & $E_7^{(1)}$ & $\begin{matrix}\input{figures/E7h_os_flavor.tex}\end{matrix}$ \\\midrule
			$E_8^{(0)}$ & $\begin{matrix}\input{figures/E8_os_flavor.tex}\\[3mm] \text{+ inverted}\end{matrix}$ & $E_8^{(1)}$ & $\begin{matrix}\input{figures/E8h_os_flavor.tex}\end{matrix}$ \\
        \bottomrule
        \end{tabular}
        }
    \end{threeparttable}
	\caption{Lagrangian $\mathcal{N}=2$ SCFTs that can be described as an ortho-symplectic quiver admitting a large $k_i\sim N$ limit for all nodes. Algebras of type $\textcolor{red}{\mathfrak{so}(k)}$ are denoted in red, while those of type $\textcolor{blue}{\mathfrak{usp}(k)}$ are in blue. All are shaped like an ADE algebra of rank $r$, counting the number of nodes, except for affine cases, where quivers associated with algebras $\mathfrak{b}=X_r^{(1)}$ have $r+1$ nodes due to the affine node. For affine quivers, the cancellation of $\beta$-functions either prohibits large flavour symmetries, or can forbid their existence altogether. Quivers with ``+ inverted'' indicate that a quiver of the same shape but with orthogonal and symplectic algebras exchanged is also possible.}
    \label{tab:dynkin-bases-ortho-symplectic}
\end{table}
\clearpage

\begin{table}[p]
    \centering
    \scriptsize 
    \begin{threeparttable}
			\resizebox{.85\textwidth}{!}{
        \begin{tabular}[t]{cccc}
				\toprule
			$\mathfrak{b}$& Unitary ortho-symplectic finite quiver & $\mathfrak{b}$ & Unitary ortho-symplectic affine quiver\\\midrule
			$B_{r}^{(0)}$ & $\begin{matrix}\input{figures/Bk_os_flavor.tex}\\[3mm]\text{+ inverted} \\[5mm] \input{figures/Bk_os_flavor_2.tex}\\[3mm]\text{+ inverted} \end{matrix}$ & $B_{r}^{(1)}$ & $\begin{matrix}\input{figures/Bh_os_flavor.tex}\\[3mm] \\ \input{figures/Bh_os_flavor_2.tex}\end{matrix}$\\\midrule
			$C^{(0)}_r$ & $\begin{matrix}\input{figures/Ck_os_flavor.tex}\\[3mm]\text{+ inverted}\end{matrix}$ & $C_{r}^{(1)}$ &  $\begin{matrix}\input{figures/Ch_os_flavor.tex}\\[3mm]\text{+ inverted} \\[5mm] \input{figures/Ch_os_flavor_2.tex}\\[3mm]\end{matrix}$ \\\midrule
			$F^{(0)}_4$ & $\begin{matrix} \input{figures/F4_flavor.tex} \\[3mm] \text{+ inverted}\end{matrix}$& $F_{4}^{(1)}$ & $\begin{matrix} \input{figures/F4h_flavor.tex}\end{matrix}$\\\midrule
			$\mathfrak{b}$& Quivers with twisted-affine shape \\\midrule
			$D^{(2)}_3$ & $\begin{matrix}\input{figures/D3_2_flavor.tex}\end{matrix}$ & $E_{6}^{(2)}$ &  $\begin{matrix}\input{figures/E6_2_flavor.tex}\\[3mm] \end{matrix}$ \\\midrule
			$\begin{matrix}\displaystyle D^{(2)}_{r+1}\\\displaystyle r>3\end{matrix}$ & $\begin{matrix}\input{figures/Dk_2_flavor.tex} \\[3mm] \text{+ inverted} \\[5mm] \input{figures/Dk_2_flavor_2.tex} \\[3mm] \text{+ inverted}\end{matrix}$ & $A_{2r-1}^{(2)}$ &  $\begin{matrix}\input{figures/Ak_2_flavor.tex} \\[3mm] \end{matrix}$ \\
        \bottomrule
        \end{tabular}
        }
    \end{threeparttable}
	\caption{Lagrangian $\mathcal{N}=2$ SCFTs that can be described by a unitary ortho-symplectic quiver---those involving all types of classical gauge symmetry---admitting a large $k_i\sim N$ limit for all nodes. The double line stresses the presence of a $\mathfrak{su}(k_1)\oplus\mathfrak{usp}(k_2)$ or $\mathfrak{su}(k_1)\oplus\mathfrak{so}(k_2)$ full hypermultiplet, with the arrow pointing in the direction of the $\mathfrak{su}(k_1)$ algebra, and indicating which type of Cartan matrix appears in the cancellation of $\beta$-functions. Algebras of type $\textcolor{red}{\mathfrak{so}(k)}$ are denoted in red, those of type $\textcolor{blue}{\mathfrak{usp}(k)}$ are in blue, and $\mathfrak{su}(k)$ in black. $r$ is the rank of the non-simply-laced algebra and counts the number of nodes in the finite case; for affine cases, where quivers associated with algebras $\mathfrak{b}=X_r^{(1)}$ have $r+1$ nodes due to the affine node; for twisted cases $X_r^{(2)}$, $r$ indicates the rank of the \emph{untwisted} algebra. For affine quivers, the cancellation of $\beta$-functions either prohibits large flavour symmetries, or forbids their existence altogether. Quivers with ``+ inverted'' indicate that a quiver of the same shape but with orthogonal and symplectic algebras exchanged is also possible.}
    \label{tab:dynkin-bases-unitary-ortho-symplectic}
\end{table}

\paragraph{Flavour Symmetry and Root Systems:} In the literature of
three-dimensional quiver theory, the SCFT condition given in equation
\eqref{beta-function-bif} is also referred to as the \emph{balancing
condition}. All four-dimensional large-$N$ $\mathcal{N}=2$ quiver theories must
be balanced, and therefore by definition ``good'' in the parlance of Gaiotto
and Witten \cite{Gaiotto:2008ak}. 

Like the shape of the quiver, this opens a natural Lie-algebraic interpretation
of the dimension of both gauge and flavour data: the dimensions $k_i$ can be
interpreted as the coefficients of a coroot expressed in terms of the simple
coroots $\alpha^{i\,\vee}$ of the algebra $\mathfrak{b}$: $\mu^\vee = k_i
\alpha^{i\,\vee}$. As shown in Appendix \ref{app:lie-algebras}, from that point
of view, the SCFT condition can be thought of as requiring that $\mu^\vee$
is both a dominant coweight and a positive coroot.

While for brevity we will not delve into the details of this
interpretation---see e.g. references \cite{Bourget:2019aer, Bourget:2021siw,
Fazzi:2023ulb, Lawrie:2024zon, Ahmed:2025fmq} for quiver SCFTs in various
dimensions---it enables the use of standard results of Lie theory and
properties of the underlying algebra $\mathfrak{b}$ to bypass certain
gauge-theory computations. For instance, the flavour-gauge adjacency $D^{ij}$
is set by the length of the simple roots $D^{ii} \propto 1/|\alpha^i|^2$, and we once
again see that the Lie algebra knows both about the shape of the quiver and the
allowed gauge symmetries.

Furthermore, it will enable us to find many results in this work in terms of
group-theoretical invariants. For instance, if the Lie algebra $\mathfrak{b}$
defining the shape of the quiver is of affine type, as the Cartan matrix is not
invertible, the possible flavour symmetries---if allowed at all---cannot be
large. Indeed, since in the affine case $C^{ij}$ has a one-dimensional
null space, substituting this constraint in equation \eqref{beta-function-bif}
we obtain
\begin{equation}\label{affine-level}
		\mathfrak{b}\,\text{ affine}:\qquad \theta_i f^i = 4 \,\theta_i S^i\,,\qquad \theta_i C^{ij} = 0= C^{ij}(D^{-1}\theta)_j\,,
\end{equation}
where we have used the unique left null vector $\theta_i$---the null root---whose coefficients are positive
integers and $\text{gcd}(\theta_i)=1$, called the ``Dynkin marks'' of the
Lie algebra $\mathfrak{b}$.

In practice, we observe that we can only have $4\theta_i S^i=0, 4, 8$. As unitary quivers always
have $S^i=0$, they cannot have flavour, and one finds that for those of affine
type $k_i = \theta_i N$. All other quivers always have at least one flavour node, which leads to the constraint 
$k_i\propto (D^{-1}\theta)_i \,N+ \dots$ In terms of Lie algebra, this means
that the coweight $\mu^\vee$ defined above must have level $4 \theta_i
S^i$, and the parameter $N$ is the so-called scaling element of the coweight.
We therefore see that all consistent flavour symmetries are neatly determined
by the root system of $\mathfrak{b}$, and one can show that the solutions are
classified by certain $SO(8)$ and $SU(4)$ instanton
moduli spaces \cite{Blum:1997mm, Ahmed:2025fmq}.

Anticipating a more thorough discussion in Section \ref{sec:orientifolds}, all affine quivers can be
realised as the worldvolume theory of a stack of $D3$-branes probing an
orbifold singularity, and possibly an orientifold, and the affine root system also arises naturally in
that picture: the flavour symmetries are due to the presence of eight (half)
D7-branes, and different choices of coweights/flavour symmetries correspond to
the boundary data for these branes.

For SCFTs related to a finite algebra $\mathfrak{b}$, the Cartan matrix is
invertible and the flavour symmetry does not have additional constraints: every
coweight $\mu^\vee$ that is both dominant (its coefficients $f^i-4S^i$ are non-negative integers) and a
positive coroot (all $k_i=C^{-1}_{ij}(f^j-4S^{j})$ are also positive integers)
leads to a consistent $\mathcal{N}=2$ quiver SCFT. Furthermore, contrary to the
affine case, the absence of the constraint in equation \eqref{affine-level}
means that there can be flavour growing like $N$, $f^i\sim N$. We will come
back to the flavour symmetry in Section \ref{sec:orientifolds}.

We have therefore found that for large-$N$ quiver SCFTs, it is convenient to
label the gauge data in Lie-algebraic terms. One can then go over all
combinations of $(\mathfrak{b}, S^i, f^i)$ leading to consistent solutions of
the SCFT condition. The results are collated in Tables
\ref{tab:dynkin-bases-unitary}--\ref{tab:dynkin-bases-unitary-ortho-symplectic},
and are summarised as follows:
\begin{classification}[Large-$N$ Quiver SCFTs \cite{Douglas:1996sw, Douglas:1997de, Katz:1997eq, Kachru:1998ys, Park:1998zh, Lawrence:1998ja, Hanany:1998sd, Uranga:1998uj, Witten:1998xy, Sen:1996vd, Douglas:1996js, Ennes:2000fu, Giveon:1998sr, Bhardwaj:2013qia}] \label{classification-scft}
		All four-dimensional $\mathcal{N}=2$ SCFTs admitting a quiver
		description and a large-$N$ limit, and involving only bifundamental hypermultiplets are described by a triplet $(\mathfrak{b}, S^i, f^i)$ consisting of:
		\begin{itemize}
				\item A finite or (possibly-twisted) affine Lie algebra $\mathfrak{b}$ describing the shape of the quiver.
				\item A vector $S^i \in \{0, 1, -1\}$ labelling the type of classical flavour algebras, $\mathfrak{su}(f^i)$, $\mathfrak{so}(f^i)$, $\mathfrak{usp}(f^i)$.
				\item The dimension of the fundamental representation $f^i$ of
						the flavour symmetries, subject to the SCFT condition:
					\begin{equation}\label{cft-condition}
						C^{ij}k_j = f^i - 4S^i\,,
					\end{equation}
					where $k_i$ is the dimension of the fundamental representation
					of the gauge algebras, and $C^{ij}$ the Cartan matrix of
					$\mathfrak{b}$. Demanding that $f^i$ and $k_i$ are integers
					imposes that they are related to the root system of
					$\mathfrak{b}$. 
		\end{itemize}
		Up to a few sporadic cases discussed below and involving matter
		transforming in representations beyond the bifundamental, this exhausts
		all possibilities.
\end{classification}
For $\mathfrak{usp}(k_i)$, $k_i$ must be even, and similarly for flavour
symmetries. Furthermore, if $\mathfrak{b}$ is of affine type, there is a
one-parameter family of solutions to the SCFT condition \eqref{cft-condition}
labelled by $N$, see discussion around equation \eqref{affine-level}.

\paragraph{Sporadic Quivers:} As mentioned above, there are a few cases that do
not admit only hypermultiplets transforming in bifundamental representations.
They appear mostly for theories with a single gauge node due to the relatively
small value of the Dynkin index $\textrm{T}(\bm{R})$ of certain
representations. We will discuss only families admitting a large-$N$ limit for
all gauge groups; for an exhaustive discussion of all other cases, see reference
\cite{Bhardwaj:2013qia}.

The simplest example is $\mathcal{N}=4$ super-Yang--Mills, which has two
hypermultiplets transforming in the adjoint representation:
\begin{equation}\label{N=4-quivers}
	\begin{matrix}
	\begin{tikzpicture}
		\node[node, label=left:{\footnotesize $k$}, fill=black] (A1)  {};
		\draw (A1.north east) to[out=60, in=120, looseness=10] (A1.north west);
		\draw (A1.south east) to[out=-60, in=-120, looseness=10] (A1.south west);
	\end{tikzpicture}
	\end{matrix}\qquad\qquad
	\begin{matrix}
	\begin{tikzpicture}
		\node[node, label=left:{\footnotesize $k$}, fill=red] (A1)  {};
		\draw (A1.north east) to[out=60, in=120, looseness=10] (A1.north west);
		\draw (A1.south east) to[out=-60, in=-120, looseness=10] (A1.south west);
	\end{tikzpicture}
	\end{matrix}\qquad\qquad
	\begin{matrix}
	\begin{tikzpicture}
		\node[node, label=left:{\footnotesize $k$}, fill=blue] (A1)  {};
		\draw (A1.north east) to[out=60, in=120, looseness=10] (A1.north west);
		\draw (A1.south east) to[out=-60, in=-120, looseness=10] (A1.south west);
	\end{tikzpicture}
	\end{matrix}\qquad\qquad
\end{equation}
The self-loop indicates the adjoint representation. In those cases, there is
an extra $\mathfrak{su}(2)_L$ flavour symmetry rotating the two
hypermultiplets, which can be understood from the decomposition of the
$\mathcal{N}=4$ R-symmetry,
$\mathfrak{su}(4)_R\to\mathfrak{su}(2)_R\oplus\mathfrak{u}(1)_r\oplus\mathfrak{su}(2)_L$.

For completeness, we also give quivers corresponding to $\mathcal{N}=2$
superconformal Quantum Chromodynamics (SQCD):
\begin{equation}\label{SQCD-quivers}
	\begin{matrix}
	\begin{tikzpicture}
		\node[node, label=above:{\footnotesize $k$}, fill=black] (A1)  {};
		\node[rectangle, label=below:{\footnotesize $2k$}, fill=black] (F1) [below=6mm of A1]  {};
		\draw (A1.south) -- (F1.north);
	\end{tikzpicture}
	\end{matrix}\qquad\qquad
	\begin{matrix}
	\begin{tikzpicture}
		\node[node, label=above:{\footnotesize $k$}, fill=blue] (A1)  {};
		\node[rectangle, label=below:{\footnotesize $2k+4$}, fill=red] (F1) [below=6mm of A1]  {};
	    \draw (A1.south) -- (F1.north);
	\end{tikzpicture}
	\end{matrix}\qquad\qquad
	\begin{matrix}
	\begin{tikzpicture}
		\node[node, label=above:{\footnotesize $k$}, fill=red] (A1)  {};
		\node[rectangle, label=below:{\footnotesize $2k-4$}, fill=blue] (F1) [below=6mm of A1]  {};
	    \draw (A1.south) -- (F1.north);
	\end{tikzpicture}
	\end{matrix}
\end{equation}
We recall that $f$ half hypermultiplets with a $\mathfrak{so}(k)$ gauge
symmetry are rotated by a $\mathfrak{usp}(f)$ flavour symmetry, and vice-versa.
In the notation discussed above, these quivers have a shape associated with an
algebra $\mathfrak{b}=A_1^{(0)}$, and can be understood as the building blocks
of more complicated quivers through a gauging of common flavour factors.

The remaining sporadic quivers arising in a large-$N$ family have been
discussed in references \cite{Bhardwaj:2013qia, Razamat:2020pra}, and more
recently in the context of the CFT Distance Conjecture
\cite{Calderon-Infante:2024oed}. They involve two-symmetric $\bm{S}^2$ and
two-anti-symmetric $\bm{\Lambda}^2$ representations, which with the adjoint are
the only allowed representations that are not bifundamental. They can appear for
quivers with an arbitrary number of nodes, but only in specific circumstances.
At one node, we have:
\begin{equation}\label{sporadic-cases-1}
	\begin{matrix}
	\begin{tikzpicture}
		\node[node, label=above:{\footnotesize $k$}, fill=black] (A1)  {};
		\node[rectangle, label=above:{\footnotesize $1$}, fill=black] (F1) [right=8mm of A1]  {};
		\node[rectangle, label=below:{\footnotesize $k+2$}, fill=black] (F2) [below=6mm of A1]  {};
		\node[label=below:{\footnotesize $\bm{\Lambda^2}$}] (C) [right=4mm of A1]  {};
		\draw[decorate, decoration={snake}] (A1.east) -- (F1.west);
	    \draw (A1.south) -- (F2.north);
	\end{tikzpicture}
	\end{matrix}\qquad\qquad
	\begin{matrix}
	\begin{tikzpicture}
		\node[node, label=above:{\footnotesize $k$}, fill=black] (A1)  {};
		\node[rectangle, label=above:{\footnotesize $1$}, fill=black] (F1) [right=8mm of A1]  {};
		\node[rectangle, label=below:{\footnotesize $k-2$}, fill=black] (F2) [below=6mm of A1]  {};
		\node[label=below:{\footnotesize $\bm{S^2}$}] (C) [right=4mm of A1]  {};
		\draw[decorate, decoration={snake}] (A1.east) -- (F1.west);
	    \draw (A1.south) -- (F2.north);
	\end{tikzpicture}
	\end{matrix}\qquad\qquad
	\begin{matrix}
	\begin{tikzpicture}
		\node[node, label=above:{\footnotesize $k$}, fill=black] (A1)  {};
		\node[rectangle, label=above:{\footnotesize $1$}, fill=black] (F1) [right=8mm of A1]  {};
		\node[rectangle, label=below:{\footnotesize $1$}, fill=black] (F2) [below=6mm of A1]  {};
\node[label=below:{\footnotesize $\bm{\Lambda^2}$}] (C) [right=4mm of A1]  {};
		\node[label=left:{\footnotesize $\bm{S^2}$}] (D) [below=2mm of A1]  {};
		\draw[decorate, decoration={snake}] (A1.east) -- (F1.west);
		\draw[decorate, decoration={snake}] (A1.south) -- (F2.north);
	\end{tikzpicture}
	\end{matrix}\qquad\qquad
	\begin{matrix}
	\begin{tikzpicture}
		\node[node, label=above:{\footnotesize $k$}, fill=black] (A1)  {};
		\node[rectangle, label=above:{\footnotesize $2$}, fill=black] (F1) [right=8mm of A1]  {};
		\node[rectangle, label=below:{\footnotesize $4$}, fill=black] (F2) [below=6mm of A1]  {};
		\node[label=below:{\footnotesize $\bm{\Lambda^2}$}] (C) [right=4mm of A1]  {};
		\draw[decorate, decoration={snake}] (A1.east) -- (F1.west);
	    \draw (A1.south) -- (F2.north);
	\end{tikzpicture}
	\end{matrix}\qquad\qquad
	\begin{matrix}
	\begin{tikzpicture}
		\node[node, label=above:{\footnotesize $k$}, fill=blue] (A1)  {};
		\node[rectangle, label=above:{\footnotesize $2$}, fill=red] (F1) [right=8mm of A1]  {};
		\node[rectangle, label=below:{\footnotesize $8$}, fill=red] (F2) [below=6mm of A1]  {};
		\node[label=below:{\footnotesize $\bm{\Lambda^2}$}] (C) [right=4mm of A1]  {};
		\draw[decorate, decoration={snake}] (A1.east) -- (F1.west);
	    \draw (A1.south) -- (F2.north);
	\end{tikzpicture}
	\end{matrix}
\end{equation}
For $\mathfrak{so}(k)$, the adjoint is the anti-symmetric representation,
$\bm{adj}=\bm{\Lambda}^2$, while it is the symmetric for $\mathfrak{usp}(k)$,
$\bm{adj}=\bm{S}^2$, and these cases correspond to $\mathcal{N}=4$ theories.

For the last three quivers, the flavour does not scale like $k$ and when gauged
does not lead to another large-$N$ family. For the first two, using that for
$\mathfrak{su}(k)$, we have $\textrm{T}(\bm{S^2}) = \frac{1}{2}(k+2)$ and
$\textrm{T}(\bm{\Lambda^2}) = \frac{1}{2}(k-2)$, a short computation reveals that, at
the level of the SCFT condition, the presence of a symmetric or anti-symmetric
representation of $\mathfrak{su}(k)$ is equivalent to replacing the
corresponding node by one involving only a fundamental representation of
$\mathfrak{so}(k)$ or $\mathfrak{usp}(k)$, respectively. More generally, when
they are part of a longer quiver, we have the equivalence:
\begin{equation}\label{sporadic-cases-2}
	\begin{aligned}
	\begin{matrix}
	\begin{tikzpicture}
		\node (A0)  {$\cdots$};
		\node[node, label=above:{\footnotesize $k_{i-1}$}, fill=black] (A1) [right=4mm of A0] {};
		\node[node, label=above:{\footnotesize $k_{i}$}, fill=black] (A2)  [right=8mm of A1] {};
		\node[rectangle, label=below:{\footnotesize $f^{i-1}$}, fill=black] (F1) [below=6mm of A1]  {};
		\node[rectangle, label=above:{\footnotesize $1$}, fill=black] (F2) [right=8mm of A2]  {};
		\node[rectangle, label=below:{\footnotesize $f^i$}, fill=black] (F3) [below=6mm of A2]  {};
		\node[label=below:{\footnotesize $\bm{S^2}$}] (C) [right=4mm of A2]  {};
	    \draw (A0.east) -- (A1.west);
	    \draw (A1.east) -- (A2.west);
		\draw[decorate, decoration={snake}] (A2.east) -- (F2.west);
	    \draw (A2.south) -- (F3.north);
	    \draw (A1.south) -- (F1.north);
	\end{tikzpicture}
	\end{matrix}\qquad&\longleftrightarrow\qquad
	\begin{matrix}
	\begin{tikzpicture}
		\node[node, label=above:{\footnotesize $k_i$}, fill=red] (A1)  {};
		\node[rectangle, label=below:{\footnotesize $(2f^i)$}, fill=blue] (F2) [below=6mm of A1]  {};
		\node[node, label=above:{\footnotesize $k_{i-1}$}, fill=black] (A0) [left=6mm of A1] {};
		\node[rectangle, label=below:{\footnotesize $f^{i-1}$}, fill=black] (F3) [below=6mm of A0]  {};
		\node (B) [left=6mm of A0] {\dots};
		\node[yscale=1.4] (D) [right=.2mm of A0] {$<$};
	    \draw (A1.south) -- (F2.north);
	    \draw (A0.south) -- (F3.north);
	    \draw (B.east) -- (A0.west);
		\draw ([yshift=1.5pt]A1.west) -- ([yshift=1.5pt]A0.east);
	    \draw ([yshift=-1.5pt]A1.west) -- ([yshift=-1.5pt]A0.east);
	\end{tikzpicture}
	\end{matrix}\\
	\begin{matrix}
	\begin{tikzpicture}
		\node (A0)  {$\cdots$};
		\node[node, label=above:{\footnotesize $k_{i-1}$}, fill=black] (A1) [right=4mm of A0] {};
		\node[node, label=above:{\footnotesize $k_{i}$}, fill=black] (A2)  [right=8mm of A1] {};
		\node[rectangle, label=below:{\footnotesize $f^{i-1}$}, fill=black] (F1) [below=6mm of A1]  {};
		\node[rectangle, label=above:{\footnotesize $1$}, fill=black] (F2) [right=8mm of A2]  {};
		\node[rectangle, label=below:{\footnotesize $f^i$}, fill=black] (F3) [below=6mm of A2]  {};
		\node[label=below:{\footnotesize $\bm{\Lambda^2}$}] (C) [right=4mm of A2]  {};
	    \draw (A0.east) -- (A1.west);
	    \draw (A1.east) -- (A2.west);
		\draw[decorate, decoration={snake}] (A2.east) -- (F2.west);
	    \draw (A2.south) -- (F3.north);
	    \draw (A1.south) -- (F1.north);
	\end{tikzpicture}
	\end{matrix}\qquad&\longleftrightarrow\qquad
	\begin{matrix}
	\begin{tikzpicture}
		\node[node, label=above:{\footnotesize $k_i$}, fill=blue] (A1)  {};
		\node[rectangle, label=below:{\footnotesize $(2f^i)$}, fill=red] (F2) [below=6mm of A1]  {};
		\node[node, label=above:{\footnotesize $k_{i-1}$}, fill=black] (A0) [left=6mm of A1] {};
		\node[rectangle, label=below:{\footnotesize $f^{i-1}$}, fill=black] (F3) [below=6mm of A0]  {};
		\node (B) [left=6mm of A0] {\dots};
		\node[yscale=1.4] (D) [right=.2mm of A0] {$<$};
	    \draw (A1.south) -- (F2.north);
	    \draw (A0.south) -- (F3.north);
	    \draw (B.east) -- (A0.west);
		\draw ([yshift=1.5pt]A1.west) -- ([yshift=1.5pt]A0.east);
	    \draw ([yshift=-1.5pt]A1.west) -- ([yshift=-1.5pt]A0.east);
	\end{tikzpicture}
	\end{matrix}
	\end{aligned}
\end{equation}
Observe that while the original quiver has $f^i$ full hypermultiplets rotated
by an $\mathfrak{su}(f^i)$ flavour, the would-be quiver is understood as having
$2f^i$ half-hypermultiplets with an $\mathfrak{usp}(2f^i)$ or
$\mathfrak{so}(2f^i)$ symmetry.

\paragraph{Summary:} We therefore conclude that only six sporadic cases,
namely $\mathcal{N}=4$ super-Yang--Mills with the three different classical
gauge algebras, and the last three quivers in equation \eqref{sporadic-cases-1}
cannot be captured by a Lie algebra dictating their quiver description. For
all other large-$N$ Lagrangian SCFTs, we have Classification
\ref{classification-scft} which we have obtained from a bottom-up perspective
via a study of finite and affine Cartan matrices. 

The quivers involving (anti-)symmetric representations of $\mathfrak{su}(k)$
also fall into that classification after trading that representation for a
non-simply-laced diagram. In the following sections, we will see that this is
not an accident, and they are related to orbifolds and orientifolds of
$\mathcal{N}=4$ super-Yang--Mills and their decoupling limits. This
classification will moreover prove itself very useful, and most of our results
will be stated directly in terms of the Lie-algebraic data of the
quiver---namely its Lie algebra $\mathfrak{b}$, the vector $S$ and the flavour
content $f$---rather than in terms of a specific quiver presentation. In
particular, we will see that the algebra $\mathfrak{b}$ enables one to define
universality classes dictated by their Hagedorn temperature, and imposes bounds
on the exponential rates predicted by the CFT Distance Conjecture.

\section{The Thermal Partition Function of \texorpdfstring{$\mathcal{N}=2$}{N=2} Quiver SCFTs}\label{sec:thermal}

In this section we study the thermal partition function of four-dimensional
$\mathcal{N}=2$ quiver SCFTs on $S^3\times S^1$ in the large-$N$ limit. Our
goal here is to study the divergence leading to a Hagedorn temperature. More
precisely, we will show that as the ranks of the gauge symmetries are taken to
be large, the gauge information is washed away and the partition function can
be recast as a Gaussian integral given in terms of an adjacency matrix of the
quiver. The Hagedorn temperature is then obtained by solving a simple algebraic
equation depending only on the largest eigenvalue of this matrix, which as we
have seen in the previous section is related to a Cartan matrix. Our main
finding is that the Hagedorn temperature is the same for all affine quivers,
and despite finite quivers being possibly non-simply-laced, their Hagedorn
temperature nonetheless follows an ADE classification. 

Let us briefly review how to compute the thermal partition function in the
large-$N$ limit, and how its divergences set the Hagedorn temperature. In four
dimensions, the thermal partition function on $S^{1}_T \times S^3$ is a trace
over the Hilbert space quantised on $S^3$:
\begin{equation}
		Z[x] = \text{Tr}_{\mathcal{H}_{S^3}} x^{E}\,,\qquad x = e^{-\frac{1}{T}}\,,
\end{equation}
where $\frac{1}{T}$ sets the size of the thermal circle. We work throughout at
the overall free point of the conformal manifold where all gauge couplings
vanish. In this limit, the Hilbert space factorises into bosonic and fermionic
Fock spaces and the thermal partition function can be reconstructed from the
corresponding one-particle states, giving rise to a plethystic exponential:
\begin{equation}
		Z[x] = \text{exp}\bigg[\sum_{n>0}\frac{1}{n}\big(z_B(x^n) + (-1)^{n+1}z_F(x^n)\big)\bigg]\,.
\end{equation}
The functions $z_B(x)$ and $z_F(x)$ correspond to the single-letter partition
functions of bosonic and fermionic states respectively.

In conformal field theories, the single-letter partition function encodes
contributions from the full conformal multiplet, although one must take into account the
presence of null states, which in the free-field limit follow from the
equations of motion, as well as gauge invariance for vector fields. For scalars
$\phi$, Weyl fermions $\psi$, and gauge bosons $A_\mu$, one finds \cite{Aharony:2003sx}:
\begin{equation}
		z_\phi(x) = \frac{x(1+x)}{(1-x)^3}\,,\qquad
		z_\psi(x) = \frac{4x^{\frac{3}{2}}}{(1-x)^3}\,,\qquad
		z_A(x) = \frac{6x^2-2x^3}{(1-x)^3}\,.
\end{equation}
As our focus is on $\mathcal{N}=2$ SCFTs, we define the contributions
associated with the complete half hyper- and vector multiplets. In terms of
their free-field content, we have:
\begin{equation}\label{Eq:sl_pfs}
	\begin{aligned}
			z_\text{V}(x, n) &= \, z_A(x^n) + 2\,z_\phi(x^n) + (-1)^{n+1}2z_\psi(x^n)\,,\\
			z_\text{H}(x, n) &= 2\, z_\phi(x^n) + (-1)^{n+1}\,\,z_\psi(x^n)\,,\\
	\end{aligned}
	\qquad t(x,n) = \frac{1-z_\text{V}(x,n)}{z_\text{H}(x,n)}\,.
\end{equation}
We stress that $z_\text{H}$ defines the contribution of a \emph{half}
hypermultiplet. The combination $t$ is defined for later convenience.

If the theory is endowed with gauge and flavour symmetries $\mathfrak{g}$ and
$\mathfrak{f}$, we can turn on fugacities $U_a$ and $V_b$, one for each
symmetry eigenvalue. They act as chemical potentials counting states of a given
charge, and by standard representation-theoretic arguments, see Appendix
\ref{app:weyl-characters}, the thermal partition function reorganises into Weyl
characters $\chi_{\bm{R}}(U)$ and $\chi_{\bm{R}'}(V)$ of irreducible
representations of these symmetries. Multi-particle states can then be
rewritten in terms of products of vector multiplets transforming in the adjoint
representation of the gauge symmetries, and the hypermultiplets transforming in
(possibly reducible or even trivial) representations $(\bm{R}, \bm{R}')$ of
$\mathfrak{g}\oplus\mathfrak{f}$:
\begin{equation}\label{partition-function-as-haar-integral}
		Z[x; U, V] = \text{exp}\bigg[\sum_{n>0}\frac{1}{n}\bigg(z_\text{V}(x,n) \sum_{i=1}^{\tilde{r}}\chi_{\bm{adj}_i}(U_i^n) + z_\text{H}(x,n)\sum_{(\bm{R}, \bm{R}')}\chi_{\bm{R}}(U^n)\chi_{\bm{R}'}(V^n) \bigg)\bigg]\,.
\end{equation}
The partition function must of course be gauge invariant and one must project
$Z[x; U, V]$ onto the gauge-singlet sector. This can be achieved via Weyl
integration, and the partition function is given as a matrix model:
\begin{equation}
		Z[x;V] = \int_G d\mu(U)\, Z[x; U, V]\,,
\end{equation}
with $d\mu$ the Haar measure associated with the Lie group $G$ of the gauge
algebra $\mathfrak{g}$. 

\paragraph{Weyl Integration at Large-$N$:} To perform the integral, we will
resort to a large-$N$ limit. This was first achieved for $G=SU(N)$ in reference
\cite{Aharony:2003sx}, and can be extended to $G=SO(2N), USp(2N)$ as well
\cite{Calderon-Infante:2024oed}, see also references \cite{Imamura:2016abe,
Calderon-Infante:2026rkj}. The quivers we consider involve \emph{a priori} all
three classical groups, and we therefore need to have a uniform way to treat
all cases at the same time. 

To do so, one can see that up to factors $(-1)^{n+1}$ coming from the
single-letter partition function of fermions, equation
\eqref{partition-function-as-haar-integral} is the Haar integration of a
plethystic exponential over a group $G$:
\begin{equation}
		Z = \int_G d\mu\,\text{PE}[f(U)]\,,\qquad \text{PE}[f(U)] = \text{exp}\bigg[\sum_{n>0}\frac{f(U^n)}{n}\bigg]\,,
\end{equation}
which can be recast into an integral over torus angles $\theta_a$, $a=1,\dots,N$
using Weyl's integration formula
\begin{equation}
		Z[x] = \frac{1}{|W|} \int_{-\pi}^\pi\prod_{a=1}^{N}\frac{d\theta_a}{2\pi} ~\text{PE}[\text{rk}(G)-\chi_{\bm{adj}}(U) + f(U)]\,,\qquad
         U_a = \text{exp}(i\theta_a)\,,
\end{equation}
where the term $\text{rk}(G)-\chi_{\bm{adj}}(U)$ comes from the Haar measure.
To evaluate the integral, one utilises properties of the Weyl character in the
large-$N$ limit. While one often writes Weyl characters $\chi_{\bm{R}}$ using
the basis of fundamental weights $\omega_i$---which we briefly encountered
in the previous section, and discuss further in the appendix---it is more
convenient to use the so-called orthogonal basis to find the thermal
partition function in the large-$N$ limit. There, the character of the fundamental representation is expressed
as a power sum:\footnote{\label{fn:weyl}Technically, we have
		given the polynomial of $U(N)$ rather than $SU(N)$, where the component
		associated with the trace must be removed by imposing
		$\prod_{a=1}^NU_a=1$. Moreover, for $SO(2N)$, one needs to supplement the power sums with another independent
		polynomial corresponding to the character of the spinor
		representation, and for $SO\left( 2N+1 \right)$ we have
		$p_n^{SO(2N+1)}= p_n^{SO(2N)}+1$. These subtleties are not relevant to
us, as we ultimately take the stable limit $N\to\infty$, and the $p_n$
become independent integration variables.}
\begin{equation}\label{def:p_n}
		\chi_{\bm{F}}(U^n) = p_n(U) = 
		\begin{cases}
				~\sum_{a=1}^N U_a^n\,,&\qquad \text{for }G=SU(N)\,;\\
				~\sum_{a=1}^N (U_a^n + U_a^{-n})\,,&\qquad \text{for }G=SO(2N)\,,USp(2N)\,.
		\end{cases}
\end{equation} 
Using the properties of Weyl characters, one finds that for representations
relevant to this work:
\begin{equation}\label{def-characters}
		\chi_{\bm{F}}(U^n) = p_n(U)\,,\qquad \chi_{\bm{adj}}(U^n) =\frac{d}{2}\big( p_n(U) \overline{p}_{n}(U) + S\, p_{2n}(U)\big) - (d-1)\,,
\end{equation}
where we have used the quantities defined in Table
\ref{tab:classical-quantities} to write a single formula valid for all
classical groups. Furthermore, recall that we are using conventions where $S_i=S(\mathfrak{f}^i)$ is
defined in terms of the \emph{flavour} symmetry, and we have $S(\mathfrak{g}_i)
= -S(\mathfrak{f}^i)$. For reference, the adjoint character of $SU(N)$ is $\chi_{\bm{adj}}(U^n)=|p_n|^2-1$, for $SO(2N)$ we have
$\chi_{\bm{adj}}(U^n) = (p_n^2-p_{2n})/2$, and for $USp(2N)$ the sign is reversed in the previous equation.

When $n>N$, the polynomials $p_n$ satisfy certain relations; for instance, for
$SU(N)$ the power sums obey Newton's identities. However, in the limit
$N\to\infty$ these relations are washed away. In the language of symmetric
functions, this is called the stable limit, and we are able to treat the $p_n$ as
independent variables, giving a basis for all Weyl characters 
$\chi_{\bm{R}}(U^n)$ \cite{macdonald1998symmetric}.

Using the eigenvalue-density method \cite{Aharony:2003sx}, see also references
\cite{Imamura:2016abe, Calderon-Infante:2024oed}, the
Haar measure therefore becomes an integral over $p_n$, and amounts to the
following replacement:
\begin{equation}\label{haar-replacement}
		\int_G d\mu\quad \longrightarrow \quad \int \prod_{n>0}d^{d}p_n\, \left( \frac{d}{2\pi n} \right)^{d/2} \text{exp}\bigg[-\sum_{n>0}\frac{d}{2n}(|p_n|^2 +S\, p_{2n} )\bigg]\,.
\end{equation}
The interested reader can find additional details in Appendix \ref{app:Haar}.

If $G = SU(N)$, we have a complex integration with the relation $\overline{p}_n
= p_{-n}$, while $G = SO(2N)$ or $USp(2N)$ leads to real integration variables
satisfying $\overline{p}_n = p_{n}$. In all three cases, this correctly
reproduces that the large-$N$ limit of Haar integrals is associated with
Gaussian ensembles \cite{Collin:2005, Anninos:2020ccj, Eynard:2015aea}.

Finally, we note that for $SU(N)$, a method using orthogonality properties of
Schur polynomials in the stable regime $n<N$ leads to the same result
\cite{Dolan:2007rq, Dolan:2008qi, macdonald1998symmetric}, although we are not
aware of a simple argument for the other classical groups. 

The thermal partition function of Lagrangian theories in the large-$N$ and
overall free-field limits therefore considerably simplifies and is given by a
Gaussian integral. In the sequel, we explicitly compute it for all quivers
discussed in Section \ref{sec:classification-quivers}. We will find that the
pole structure is universally fixed by the adjacency matrix of the quiver,
which will lead us to a simple algebraic equation from which the Hagedorn
temperature can be extracted.

\paragraph{$\mathcal{N}=4$ Super-Yang--Mills:} As a warm-up, let us compute the
thermal partition function of $\mathcal{N}=4$ super-Yang--Mills, corresponding
to the quivers given in equation \eqref{N=4-quivers}. It has a single gauge
node, and in terms of $\mathcal{N}=2$ single-letter partition functions, we
have:
\begin{equation}
		Z_{\mathcal{N}=4} = \int_G\,d\mu\, \text{exp}\bigg[\sum_{n>0}\frac{1}{n}\chi_{\bm{adj}}(U^n)\,z_{\mathcal{N}=4}(x,n)\bigg]\,,\qquad z_{\mathcal{N}=4} = z_\text{V} +\chi_{\bm{2}}\, z_\text{H}\,,
\end{equation}
where $\chi_{\bm{2}}(V)$ is associated with the $\mathfrak{su}(2)_L$ flavour
symmetry rotating the two $\mathcal{N}=2$ hypermultiplets in the
$\mathcal{N}=4$ vector multiplet. Using the definition of the character of the
adjoint given in equation \eqref{def-characters} and the replacement rule for
the Haar measure above, one finds \cite{Aharony:2003sx,
Calderon-Infante:2024oed}:
\begin{equation}\label{n=4-thermal-partition}
		Z_{\mathcal{N}=4}\simeq
		\prod_{n>0}\frac{\text{exp}\big(\frac{S^2}{4n}\frac{[1-z_{\mathcal{N}=4}(x,n)]^2}{[1-z_{\mathcal{N}=4}(x,2n)]}-\frac{(1-S^2)}{n}z_{\mathcal{N}=4}(x,n)\big)}{\big[1-z_\text{V}(x,n) - \chi_{\bm{2}}(V^n)\, z_\text{H}(x,n)\big]^{d/2}}\,,
\end{equation}
where we have cleared factors of $d$ in the exponential, as $S=0$ for
$\mathfrak{su}(N)$. We see that while the cases with
$\mathfrak{g}=\mathfrak{so}(2N)$ and $\mathfrak{usp}(2N)$ have different
quivers, the partition function solely depends on $S^2 = (\pm 1)^2$ and is the
same in both cases.

Furthermore, observe that the denominator does not depend on the type of gauge
symmetry, but only on the Weyl character of the extra $\mathfrak{su}(2)_L$
symmetry. As we will find below, the appearance of the character of that
symmetry is the reason that all $\mathcal{N}=2$ SCFTs obtained as orientifolds
of $\mathcal{N}=4$ super-Yang--Mills share the same Hagedorn
temperature.\footnote{In the rest of this work, we will use orientifold as a
shorthand to also include the possible presence of orbifolds.}
 
\paragraph{Simply-Laced Quivers:} Quivers with more gauge nodes can be
computed in a similar fashion. In the previous section, we have seen that for
SCFTs with only bifundamental hypermultiplets, the full content of the theory
is encoded by a triplet $(\mathfrak{b}, S^i, f^i)$, from which one easily
obtains the adjacency matrices $D^{ij}$ and $B^{ij}$, as well as the types of
gauge nodes.
For these theories, the thermal partition function is therefore given by
\begin{equation}
		Z_{(\mathfrak{b}, S, f)}[x;V] = \int_{G_1 \times \dots \times G_{\tilde{r}}}\,d\mu \,\text{exp}\bigg[\sum_{n>0} \frac{1}{n}\bigg(z_\text{V}(x,n)\sum_{i=1}^{\tilde{r}}\chi_{\bm{adj}_i}(U^n) + z_\text{matter}[x; U, V; n]\bigg)\bigg]\,,
\end{equation}
with the contribution of the matter spectrum given by
\begin{equation}\label{N=2-thermal}
		z_\text{matter}[x; U, V; n] =  \frac{z_\text{H}(x,n)}{2}\bigg(B^{ij}\overline{\chi}_{\bm{k}_i}(U^n)\chi_{\bm{k}_j}(U^n) + \big(D^{ij}\overline{\chi}_{\bm{f}^i}(V^n)\chi_{\bm{k}_j}(U^n)+ \text{c.c}\big)\bigg)\,,
\end{equation}
where we have suppressed the indices of the gauge and flavour fugacities for ease
of reading. The generalisation to theories involving (anti-)symmetric
representations is immediate. Note that our definition of $D$ automatically
takes into account double counting, e.g. for $\mathfrak{su}(k)$, we have
$\overline{\chi}_{\bm{k}_i}\neq\chi_{\bm{k}_i}$ and $D^{ii}=2$ since
$z_\text{H}$ counts half-hypermultiplets.

By abuse of notation, for a semi-simple group $G= G_1\times\dots\times
G_{\tilde{r}}$, we write the measure for the fundamental characters $p_n$
obtained from the replacement given in equation \eqref{haar-replacement} as
\begin{equation}
		\mathcal{D}p = \prod_{i=1}^{\tilde{r}}\prod_{n>0}d^{d_i}p^{(i)}_n\left( \frac{d_i}{2\pi n} \right)^{d_i/2}\,,
\end{equation}
where we have not included the Gaussian term coming from the Weyl determinant,
and $d^{d_i}p_n$ with $d_i=2$ refers to a complex integration for
$\mathfrak{g}_i=\mathfrak{su}(k_i)$, and real variables for the other two
cases, with $d_i=1$. Taking into account the differences between the three classical gauge
symmetries, for any simply-laced quiver, the partition function reduces to a
multivariate Gaussian integral:
\begin{equation}
		\begin{aligned}
		Z = \int \mathcal{D}p\, \text{exp}\bigg[-\sum_{n>0}\frac{1}{n}\bigg( 
						& \frac{1}{2}z_\text{H}(x,n)\,\overline{p}_n\cdot \big( t(x,n)D -  B\big)\cdot p_n -\frac{z_\text{H}(x,n)}{2}\big(\overline{\chi}_{\bm{f}}(V^n)\cdot D\cdot p_n + \text{c.c.}\big) \\
		& +\frac{1}{2}  S\cdot \big(1-z_\text{V}(x,n)\big)\cdot p_{2n} + (\tilde{r}-(S_i)^2)z_\text{V}(x,n)\bigg)\bigg]\,,
		\end{aligned}
\end{equation}
where we have used that $(D\cdot S)^i = S^i$ by virtue of their respective
definitions and $t(x,n)$ was defined in equation \eqref{Eq:sl_pfs}. The result
depends on whether the quiver is unitary or ortho-symplectic, as we have either
complex or real integration variables. For unitary quivers $B=2\,A\,,
D=2\,\mathbf{1}$, and we find
\begin{equation}\label{thermal-partition-function-unitary}
	\begin{aligned}
		Z_{(\mathfrak{b},S=0, f)} & \simeq 
		\prod_{n>0}\frac{\text{exp}\big[\frac{1}{n}\big(-\tilde{r}\,z_\text{V}(x,n) + z_\text{H}(x,n)^2\, \overline{\chi}_{\bm{f}}(V^n) \cdot M^{-1}(x,n) \cdot\chi_{\bm{f}}(V^n)\big)\big]}{\det\big(M(x,n)\big)}\,,
	\end{aligned}
\end{equation}
with
\begin{equation}\label{1-A-def}
		M(x,n) =  \big(1-z_\text{V}(x,n)\big)\,\mathbf{1} - z_\text{H}(x,n)\, A\,,
\end{equation}
where $C = 2\,\mathbf{1} - A$ is the Cartan matrix of $\mathfrak{b}$. We recall
that for affine cases, the Cartan matrix $C$ is not invertible, but $M$ is for
generic values of the temperature $x=e^{-1/T}$. 

Moving on to ortho-symplectic quivers, we have $S^i=\pm1$ for all nodes, and
using that $ B=A\,,D=\mathbf{1}$, we obtain a similar result:
\begin{equation}\label{thermal-partition-function-ortho-symplectic}
		Z_{(\mathfrak{b},S=\pm1, f)} \simeq 
\prod_{n>0}\frac{\text{exp}\big[\frac{1}{2n} J(x,n)\cdot M^{-1}(x,n)\cdot J(x,n)\big]}{\sqrt{\det\big(M(x,n)\big)}}\,,
\end{equation}
where $M(x,n)$ is as in equation \eqref{1-A-def}. The exponential is now
slightly more complicated due to the presence of variables $p_{2n}$ coming from
the expansion of the character of the adjoint, but is still quadratic in the
single-letter partition functions:
\begin{equation}
		J_i(x,n) = z_\text{H}(x,n)\chi_{\bm{f^i}}(V^n) - \delta_{2|n}\,\big(1-z_\text{V}(x,\frac{n}{2})\big)S^i\,,\qquad 
		\delta_{2|n} = \begin{cases}
				1 & \qquad \text{if $n$ even}\,;\\
				0 & \qquad \text{if $n$ odd}\,.
			\end{cases}
\end{equation}

All in all, the thermal partition function of quivers with a simply-laced shape
has a similar form, up to a square root in the ortho-symplectic case, and
possibly a more involved exponential term. The denominator, however, depends
solely on the Cartan adjacency $A$ and not on the rest of the gauge data. This
is unsurprising, as we are taking the large-$N$ limit, and the denominator is
not affected by how this scaling is done. 

For affine quivers, this is expected, as $k_i\propto (D^{-1}\theta)_i N +\dots$ is
fixed, see discussion around equation \eqref{affine-level}. On the other hand,
the $k_i$ are unconstrained when $\mathfrak{b}$ is finite, as long as the SCFT
condition is satisfied. But in both cases the partition function is blind to
particular choices of scaling $k_i = m_i N + \dots$. The only remnant of that
data is through the Weyl characters of the flavour symmetry. As we will find in
all cases, and further discuss in Section \ref{sec:Hagedorn}, this is a generic
result: the denominator of the partition function is always set by a Cartan
adjacency, and the Hagedorn temperature only depends on the shape of the
quiver, not its gauge content. 

\paragraph{Non-Simply-Laced Quivers:} For the quivers above, we have made use
of the fact that the quiver had either only $\mathfrak{su}(k_i)$ gauge
algebras, and the partition function corresponds to a complex Gaussian
integral, or ortho-symplectic with only alternating $\mathfrak{so}(k_i)$ and
$\mathfrak{usp}(k_i)$ symmetries, and therefore real integration variables.

We have however seen in the previous section that in the presence of
$\mathfrak{so}(k_i)\oplus\mathfrak{su}(k_j)$ or
$\mathfrak{usp}(k_i)\oplus\mathfrak{su}(k_j)$ bifundamental, the quiver has a
shape set by a non-simply-laced algebra $\mathfrak{b}$. In the partition
function this is reflected by having both real and complex integration
variables. To perform the integral, the simplest route is to pass to real
variables for all gauge algebras. At the level of the quiver, this corresponds
to a doubling of unitary nodes and their adjacency. In terms of Lie algebras,
this is known as the \emph{unfolding} of the Dynkin diagram. For
instance, for $F_4^{(1)}$ and $E_{6}^{(2)}$ which have the same topology but
different values of $S_i$, unfolding leads to $E_6^{(1)}$ and $E_7^{(1)}$,
respectively:
\begin{equation}
		\begin{aligned}
				F_{4}^{(1)}:\quad
	\begin{matrix}
	\begin{tikzpicture}
	    \node[node, fill=blue] (A0)  {};
	    \node[node, fill=red] (A1) [right=6mm of A0] {};
	    \node[node, fill=blue] (A2) [right=6mm of A1] {};
	    \node[node, fill=black] (A3) [right=6mm of A2] {};
	    \node[node, fill=black] (A4) [right=6mm of A3] {};
		\node[yscale=1.4] (C) [right=.2mm of A2] {$>$};
	    \node (D) [right=6mm of C] {};
	    \draw (A0.east) -- (A1.west);
	    \draw (A1.east) -- (A2.west);
		\draw ([yshift=1.5pt]A2.east) -- ([yshift=1.5pt]A3.west);
	    \draw ([yshift=-1.5pt]A2.east) -- ([yshift=-1.5pt]A3.west);
	    \draw (A3.east) -- (A4.west);
	\end{tikzpicture}
	\end{matrix}
	&\qquad\longrightarrow\qquad
			\mathcal{U}(F_4^{(1)}) = E_{6}^{(1)}:\quad
		\begin{matrix}
		\begin{tikzpicture}
		    \node[node, fill=blue] (A1) [right=6mm of A0] {};
		    \node[node, fill=red]   (A2) [right=6mm of A1] {};
		    \node[node, fill=blue] (A3) [right=6mm of A2] {};
		    \node[node, fill=white]   (A4) [right=6mm of A3] {};
		    \node[node, fill=white] (A5) [right=6mm of A4] {};
		    \node[node, fill=white]   (A6) [above=6mm of A3] {};
		    \node[node, fill=white]   (A0) [above=6mm of A6] {};
			\node[draw=none]  [left=6mm of A1] {};
			\node[draw=none]  (B1) [left=1mm of A5] {};
			\node[draw=none]  (B2) [below=1mm of A0] {};
		    \draw (A0.south) -- (A6.north);
		    \draw (A1.east) -- (A2.west);
		    \draw (A2.east) -- (A3.west);
		    \draw (A3.east) -- (A4.west);
		    \draw (A4.east) -- (A5.west);
		    \draw (A3.north) -- (A6.south);
			\draw[<->, dashed, bend right=30, shorten >=6pt, shorten <=6pt]  (B1) to node [auto] {} (B2);
		\end{tikzpicture}
		\end{matrix}\\
		& \\
				E_{6}^{(2)}:\quad
	\begin{matrix}
	\begin{tikzpicture}
	    \node[node, fill=black] (A0)  {};
	    \node[node, fill=black] (A1) [right=6mm of A0] {};
	    \node[node, fill=black] (A2) [right=6mm of A1] {};
	    \node[node, fill=blue] (A3) [right=6mm of A2] {};
	    \node[node, fill=red] (A4) [right=6mm of A3] {};
		\node[yscale=1.4] (C) [right=.2mm of A2] {$<$};
	    \draw (A0.east) -- (A1.west);
	    \draw (A1.east) -- (A2.west);
		\draw ([yshift=1.5pt]A2.east) -- ([yshift=1.5pt]A3.west);
	    \draw ([yshift=-1.5pt]A2.east) -- ([yshift=-1.5pt]A3.west);
	    \draw (A3.east) -- (A4.west);
	\end{tikzpicture}
	\end{matrix}
		&\qquad\longrightarrow\qquad
		\mathcal{U}(E_6^{(2)}) = E_{7}^{(1)}:\quad
	\begin{matrix}
	\begin{tikzpicture}
	    \node[node, fill=white] (A0)  {};
	    \node[node, fill=white]   (A1) [right=6mm of A0] {};
	    \node[node, fill=white] (A2) [right=6mm of A1] {};
	    \node[node, fill=blue]   (A3) [right=6mm of A2] {};
	    \node[node, fill=white] (A4) [right=6mm of A3] {};
		\node[node, fill=white]   (A5) [right=6mm of A4] {};
	    \node[node, fill=white] (A6) [right=6mm of A5] {};
	    \node[node, fill=red]   (A7) [above=6mm of A3] {};
	    %
		\coordinate (B1) at ([xshift=1mm, yshift=-2mm]A1.east);
		\coordinate (B2) at ([xshift=-1mm, yshift=-2mm]A5.west);
		\draw[<->, dashed, bend right=30, shorten >=6pt, shorten <=6pt]  (B1) to node [auto] {} (B2);
	    \draw (A0.east) -- (A1.west);
	    \draw (A1.east) -- (A2.west);
	    \draw (A2.east) -- (A3.west);
	    \draw (A3.east) -- (A4.west);
	    \draw (A4.east) -- (A5.west);
	    \draw (A5.east) -- (A6.west);
	    \draw (A3.north) -- (A7.south);
	\end{tikzpicture}
	\end{matrix}
	\end{aligned}
\end{equation}
where $\mathcal{U}$ denotes the unfolding operation, we have omitted the
flavour nodes for ease of reading, and indicated unfolded gauge nodes with
white circles. The inverse operation, folding, consists of identifying the
invariant subalgebras under outer-automorphisms, and is the standard operation
to construct non-simply-laced Lie algebras \cite{Kac:1990gs}.

In terms of the quiver, the unfolding procedure is natural, and can be
understood as using half-hypermultiplets rather than full hypermultiplets. We
then have a node for both representations $\bm{k}_i$ and $\overline{\bm{k}}_i$
of $\mathfrak{su}(k_i)$. Doing so at the level of the partition function and
expressing everything in terms of real variables $q_n$, we have
\begin{equation}
	\begin{aligned}
		Z_{(\mathfrak{b},S, f)} & \simeq
		 \int \mathcal{D}p\, \text{exp}\bigg[-\sum_{n>0}\frac{1}{n}\bigg( 
		 \frac{1}{2}\,\overline{p}_n\cdot \big( (1-z_\text{V})\,D -  z_\text{H}\,B_{\mathfrak{b}}\big)\cdot p_n + \cdots\big)\bigg] \\
		& = \int \mathcal{D}q\, \text{exp}\bigg[-\sum_{n>0}\frac{1}{n}\bigg( 
			\frac{1}{2}\,q_n\cdot \big( (1-z_\text{V}) -  z_\text{H}\,A_{\mathcal{U}(\mathfrak{b})}\big)\cdot q_n + \cdots\big)\bigg] \\
			& =  \prod_{n>0}\frac{\text{exp}\big[\cdots\big]}{\sqrt{\text{det}\big[z_\text{H}(x,n)\,\big(t(x,n)\,\mathbf{1} - A_{\mathcal{U}(\mathfrak{b})}\big)\big]}} \,.
	\end{aligned}
\end{equation}
The result is then given in terms of the adjacency $A = 2\,\mathbf{1} - C$ of
the unfolded algebra $\mathcal{U}(\mathfrak{b})$. For instance,
$\mathcal{U}(F_4^{(1)}) = E_6^{(1)}$ in the example above.

We therefore see that the thermal partition function is of the same form as
for ortho-symplectic quivers, see equation
\eqref{thermal-partition-function-ortho-symplectic}, but replacing the adjacency
matrix $A = 2\mathbf{1} - C$ by its unfolded version. This means that the
singularity structure of all large-$N$ quivers with only hypermultiplets is set
by an ADE algebra, even in non-simply-laced cases.

\paragraph{(Anti-)Symmetric Representations:} The rest of the possible
large-$N$ quiver SCFTs are those with (anti-)symmetric representations of
$\mathfrak{su}(k)$, up to the outlier with an anti-symmetric representation of
$\mathfrak{so}(k)$ shown in equation \eqref{sporadic-cases-1}. For the latter,
the thermal partition function was computed in reference
\cite{Calderon-Infante:2024oed} and we therefore only focus on
cases with $\mathfrak{su}(k)$ symmetries.

When performing the Gaussian integral, a feature akin to unfolding occurs
here as well. Indeed, in terms of power sums, the characters of the two representations
are given by:
\begin{equation}
		\chi_{\bm{S^2}}(U^n) = \frac{1}{2}(p_n^2 + p_{2n})\,,\qquad
		\chi_{\bm{\Lambda^2}}(U^n) = \frac{1}{2}(p_n^2 - p_{2n})\,.
\end{equation}
This means that---recalling that these representations are complex---if the quiver
has such a hypermultiplet on e.g. the first node, the integrand takes the
following form in real coordinates $p_n = x_n + i y_n$:
\begin{equation}
	\begin{aligned}
		Z & \simeq
		 \int \mathcal{D}p\, \text{exp}\bigg[-\sum_{n>0}\frac{1}{n}\bigg( 
				 x_n\cdot M(x,n)\cdot x_n + y_n\cdot M(x,n)\cdot y_n - z_\text{H}\big( (x_n^1)^2-(y^1_n)^2 \pm x^1_{2n}\big) + \cdots
		\bigg)\bigg] \,,
	\end{aligned}
\end{equation}
where we have ignored other terms for brevity, as we are ultimately interested in
the Hagedorn temperature. We see that the presence of either a
symmetric or anti-symmetric representation induces the same shift to the entry
of $M$ of the corresponding node, and we obtain:
\begin{equation}
		Z \simeq \prod_{n>0}\frac{\text{exp}[\cdots]}{\sqrt{\text{det}\big( M_+\, M_-\big)}}\,, \qquad 
		M_\pm^{ij}(x,n) = (1-z_\text{V}(x,n))\delta^{ij} - A^{ij}_\pm\,z_\text{H}(x,n)\,,
\end{equation}
where we defined $A_{\pm}^{ij} = A^{ij}\pm \delta^{1i}\delta^{1j}$. The
generalisation to the same representations at another node is immediate. 

By inspection, it is easy to see that the determinant of these two matrices
recombines into that of an adjacency matrix $A$ of a Dynkin diagram obtained by
doubling the graph:
\begin{equation}\label{determinant-symmetric-reps}
		\text{det}\big((t- A_+)(t-A_-)\big) = \text{det}(t - A_\text{doubled})\,.
\end{equation}
This is proved easily by writing $A_\text{doubled}$ in block-diagonal form, and
we find the following replacement rules:
\begin{equation}
	\begin{aligned}
		A_{r}^{(0)}\text{ with 1 } \bm{S}^{2}\text{ or }\bm{\Lambda}^{2} & \qquad\longrightarrow\qquad A_\text{doubled} = A^{(0)}_{2r}\,,\\
		A_{r}^{(0)}\text{ with }2\text{ of } \bm{S}^{2}\text{ or }\bm{\Lambda}^{2} & \qquad\longrightarrow\qquad A_\text{doubled} = A^{(1)}_{2r-1}\,,\\
		D_{r}^{(0)}\text{ with 1 } \bm{S}^{2}\text{ or }\bm{\Lambda}^{2} & \qquad\longrightarrow\qquad A_\text{doubled} = D^{(1)}_{2r-1}\,,\\
		B_{r}^{(0)}, C_{r}^{(0)}\text{ with 1 } \bm{S}^{2}\text{ or }\bm{\Lambda}^{2} & \qquad\longrightarrow\qquad A_\text{doubled} = A^{(1)}_{2r-1}\,,
	\end{aligned}
\end{equation}
where in each case the (anti-)symmetric representation is located on the side
of the Dynkin diagram that does not have a trivalent (for D-type diagrams) or
non-simply-laced (for BC-type) node. For type A, they can be attached at either
end of the linear chain.

As for the non-simply-laced quivers discussed above, the fact that we obtain
these replacements can also be understood in terms of unfolding. In Section
\ref{sec:classification-quivers}, we have seen that in the presence of these
representations, the SCFT condition can be rewritten in terms of the Cartan
matrix where the node hosting the representation is replaced by an
$\mathfrak{so}(k)$ or $\mathfrak{usp}(k)$ spurious symmetry, and the adjoining
edge becomes non-simply-laced, see discussion around equation
\eqref{sporadic-cases-2}. From Tables \ref{tab:dynkin-bases-ortho-symplectic}
and \ref{tab:dynkin-bases-unitary-ortho-symplectic}, one can see that these
correspond to $\mathfrak{b}=A_{2r}^{(2)}, A_{2r-1}^{(2)}, C^{(0)}_{r},
C^{(1)}_{r}$, and $A_\text{doubled}$ then corresponds to their respective
unfolded algebras. By abuse of notation, in the presence of such
representations, we will write $A_\text{doubled} = A_{\mathcal{U}(\mathfrak{b})}$.
Since quivers with (anti-)symmetric representations are uniquely associated
with a non-simply-laced algebra, this operation is unambiguous.

\paragraph{Summary for all Quivers:} We have therefore shown that in the
large-$N$ limit, the thermal partition function takes a particularly simple form that
depends only on the Lie-algebraic data. To summarise the results above, for
all $\mathcal{N}=2$ quivers with only bifundamental hypermultiplets, in the
purely unitary case, we have:
\begin{equation}
		Z^{\text{unitary}}_{\mathfrak{b}, 0, f} \simeq \prod_{n>0}\frac{\text{exp}[\dots]}{\text{det}\big[z_\text{H}(x,n)\,\big(t(x,n)\,\mathbf{1} - A_{\mathfrak{b}}\big)\big]}\,,\qquad t = \frac{1-z_\text{V}}{z_\text{H}}\,,
\end{equation}
where $C=2\,\mathbf{1}- A$ is the Cartan matrix of the ADE algebra $\mathfrak{b}$
defining the shape of the quiver. 

For all other quivers, including those with (anti-)symmetric representations, it
takes a similar form, but is instead given in terms of the adjacency matrix of
the possibly-unfolded algebra $\mathcal{U}(\mathfrak{b})$: 
\begin{equation}
		Z \simeq \prod_{n>0}\frac{\text{exp}[\dots]}{\sqrt{\text{det}\big[z_\text{H}(x,n)\,\big(t(x,n)\,\mathbf{1} - A_{\mathcal{U}(\mathfrak{b})}\big)\big]}}\,,\qquad t = \frac{1-z_\text{V}}{z_\text{H}}\,.
\end{equation}

The difference between this and the unitary cases is due to the presence of complex
fundamental representations for $\mathfrak{su}(k)$. We conclude that for all
large-$N$ $\mathcal{N}=2$ quiver SCFTs, the denominator of the thermal partition
function in the free-field limit is universally set by the adjacency matrix of
an ADE algebra.

\subsection{A Beautiful Cosine Formula for the Hagedorn Temperature}\label{sec:Hagedorn}

We have seen that in the large-$N$ limit the Haar measure becomes Gaussian,
leading to a very simple form of the partition function. Perhaps
unsurprisingly, it does not depend on the particular scalings of the gauge
data $k_i \sim m_i N$ as this information is lost in the large-$N$ limit, and
its only remnant is through the flavour symmetry. 

Our goal is now to find the smallest value of the temperature
$x_\text{Hag.} = e^{-1/T_\text{Hag.}}$ for which the thermal partition function develops a
Hagedorn behaviour, that is when the density of states defined in equation
\eqref{density-thermal} becomes exponential $\rho(E)\sim e^{E/T_\text{Hag.}}$, up
to polynomial contributions. It is easy to convince oneself that this occurs
when the partition function develops a pole or a branch cut around
$x_\text{Hag.}$:
\begin{equation}
		Z\sim \int_0^\infty dE\, E^{p-1}\, e^{E(\frac{1}{T_\text{Hag.}}- \frac{1}{T})} \propto \frac{1}{[\log(x_\text{Hag.}/x)]^p}\sim \frac{1}{(1-\frac{x}{x_\text{Hag.}})^p} \,,\qquad \text{as }x \to x_\text{Hag.}\,,
\end{equation}
where we have ignored constant prefactors for simplicity. This means that all
$\mathcal{N}=2$ quivers will exhibit a Hagedorn behaviour, as we have found
that the singularity structure of the partition function is determined by the
characteristic polynomial of the Cartan adjacency $A= 2\,\mathbf{1}-C$
\begin{equation}
		Z\propto \prod_{n>0}\frac{1}{\text{det}\big[t(x,n)\,\mathbf{1} - A\big]^{p}}\,,
\end{equation}
where $p=1$ for all unitary quivers with only bifundamental representations,
and $p=\frac{1}{2}$ for all others. The Hagedorn behaviour is therefore set by
the largest eigenvalue of $A$:
\begin{equation}
		t(x_\text{Hag.}, 1) = \lambda_\text{max}\quad \Leftrightarrow \quad 1-z_\text{V}(x_\text{Hag.},1) = \lambda_\text{max}\, z_\text{H}(x_\text{Hag.},1)\,,\qquad \lambda_\text{max} = \text{max}\big[\text{Spec}(A)\big]\,.
\end{equation}
Note that the presence or absence of the square root does not impact the
Hagedorn temperature, and therefore unitary and ortho-symplectic quivers will
have the same behaviour at large $N$.
\begin{table}
		\centering
		\begin{tabular}{clccc}
			\toprule
			$\mathfrak{b}$ & Exponents $m_i$ & $h$ & $h^\vee$ & $R_{\mathfrak{b}}^\text{max}$\\\midrule
			$A_r^{(0)}$ & $1,2,3,\dots,r$& $r+1$ & $r+1$ & $\frac{6}{2r+1}$\\
			$B_r^{(0)}$ & $1,3,5,\dots,2r-1$ & $2r$& $2r-1$ & $\frac{2}{2r-1}$\\
			$C_r^{(0)}$ & $1,3,5,\dots,2r-1$ & $2r$& $r+1$ & $\frac{2}{2r-1}$\\
			$D_r^{(0)}$ & $1,3,5,\dots,2r-3, r-1$ & $2r-2$ & $2r-2$ & $\frac{2}{2r-3}$\\
			$E_6^{(0)}$ & $1,4,5,7,8,11$ & $12$ & $12$ & $\frac{6}{53}$\\
			$E_7^{(0)}$ & $1,5,7,9,11,13,17$ & $18$ & $18$ & $\frac{6}{115}$\\
			$E_8^{(0)}$ & $1,7,11,13,17,19,23,29$ & $30$ & $30$ & $\frac{2}{119}$\\
			$F_4^{(0)}$ & $1,5,7,11$ & $12$ & $9$ & $\frac{2}{23}$\\
			$G_2^{(0)}$ & $1,5$ & $6$ & $4$ & $\frac{2}{7}$\\
			\bottomrule

		\end{tabular}
		\caption{\label{tab:exponents}Exponents of finite algebras $\mathfrak{b}=X_r^{(0)}$ and their Coxeter numbers. For
		non-twisted affine algebras $X_r^{(1)}$, one supplements the
affine exponent $m_0=0$. $R_{\mathfrak{b}}^\text{max}$ is related to a bound on $n_\text{V}/n_\text{H}$, see equation \eqref{Rmax}. $G_2^{(0)}$ is not associated with an $\mathcal{N}=2$ quiver and is given for completeness.}
\end{table}

It turns out that the eigenvalues of the adjacency matrix $A$ have a
particularly nice form when they are related to Lie algebras, in particular for
those of finite type. Indeed, it was shown by Damianou that its spectral
properties satisfy the following theorem:
\begin{theorem}[The beautiful sine formula \cite{MR3149030}]\label{thm:sin-formula}
	Let $\mathfrak{b}$ be a simple Lie algebra $X_r^{(0)}$ with Cartan
	matrix $C$, Coxeter number $h$, and Weyl-group exponents
	$m_1, m_2, \cdots, m_r$. The characteristic polynomial of $C$ is given by
	\begin{equation}
			\textup{det}\big(t\,\mathbf{1} - C\big) = \prod_{i=1}^r\bigg(t-4\sin^2(\frac{m_i\, \pi}{2\,h})\bigg)\,.
	\end{equation}
\end{theorem}
The Weyl-group exponents $m_i$ are collated in Table \ref{tab:exponents}, and
it immediately follows that the characteristic polynomial of the adjacency
$A=2\,\mathbf{1}-C$ satisfies
\begin{equation}\label{eqn:cos-formula}
		\mathfrak{b}\text{ finite}:\qquad 
		\textrm{det}\big(t\,\mathbf{1} - A\big) = \prod_{i=1}^r\bigg(t-2\cos(\frac{m_i \pi}{h})\bigg)\,.
\end{equation}

For affine algebras, the general form of equation \eqref{eqn:cos-formula} holds
\cite{damianou2014characteristic}, but the $m_i$ are not directly related to
the affine Weyl group. However, due to the fact that $\text{det}(C)=0$ in the
affine case, we always have $\lambda_\text{max}=2$. As it turns out,
$\lambda_\text{max}$ is always set by $m_1=1$ for finite algebras, and $m_0=0$
for affine algebras.

Furthermore, we need only to consider unfolded algebras
$\mathcal{U}(\mathfrak{b})$, which are always of ADE type and simply laced.
Using the fact that for ADE algebras, the Coxeter number and its dual are the
same $h^\vee = h$, we therefore find that a direct corollary of Theorem
\ref{thm:sin-formula} is that the Hagedorn temperature is always set by solving
a simple algebraic equation determined by its shape. It gives a physics avatar
of the beautiful sine formula of Theorem \ref{thm:sin-formula}:
\paragraph{The Beautiful Cosine Formula:} \emph{Given an $\mathcal{N}=2$ quiver
	SCFT defined by the triplet $(\mathfrak{b},S,f)$, the Hagedorn
	temperature in the large-$N$ and overall free-field limits is uniquely fixed by
	Lie-algebraic quantities of the Dynkin diagram of the unfolded
	algebra $\mathcal{U}(\mathfrak{b})$, and obtained as the smallest
	value of $x= e^{-1/T}$ solving the following equation:
	\begin{equation}\label{beautiful-cos-formula}
			1-z_\text{V}(x) = \lambda_\text{max}\, z_\text{H}(x)\,,\qquad
		\lambda_\text{max}=
		\begin{cases}
			2\,,\quad &\text{if }\mathcal{U}(\mathfrak{b})\text{ affine}\,;\\
			2\cos\big(\frac{\pi}{h^\vee_{\mathcal{U}(\mathfrak{b})}}\big)\,,\quad &\text{if }\mathcal{U}(\mathfrak{b})\text{ finite}\,;\\
		\end{cases}
	\end{equation}
	$\lambda_\text{max}$ is the largest eigenvalue of the Cartan adjacency $A=2\,\mathbf{1}-C$.
}

This result generalises recent findings \cite{Calderon-Infante:2024oed,
Calderon-Infante:2026rkj} to all $\mathcal{N}=2$ quiver theories, including
unitary-ortho-symplectic quivers, and following the discussion around equation
\eqref{determinant-symmetric-reps}, it is also valid in the presence of
(anti-)symmetric representations. The single-letter partition functions
$z_\text{H}$ and $z_\text{V}$ are those associated with free hyper- and vector
multiplets respectively, and defined in equation \eqref{Eq:sl_pfs}.

We therefore find that all these SCFTs fall into universality classes
corresponding to a string-like spectrum whose Hagedorn temperature is set by
$\lambda_\text{max}$. We stress that while there is an infinite number of
classical algebras and therefore an infinite number of universality classes, as
unfolding must be taken into account, we have two main sectors: all affine
quivers have the same Hagedorn temperature as $\mathcal{N}=4$ super-Yang--Mills, and those
associated with finite algebras follow an ADE classification that is not
sensitive to folding, nor to the presence of other representations.

We remind the reader that we have assumed throughout that we were in the strict
$N\to\infty$ limit. Only this regime exhibits a Hagedorn behaviour, which
follows from the fact that single-trace operators are independent. At large but
finite $N$ on the other hand, there is conversely only a fixed number of
independent Casimir operators, which form a basis for single-trace operators.
This means that the exponential growth of states cannot hold indefinitely, and
is valid only in a large but finite window of energy, after which one expects
only sub-exponential growth \cite{Aharony:2003sx}. Note that this is precisely
the same reason that allowed us to perform the integral in the first place,
whereby the large-$N$ limit enables us to ignore any relation between the power
sums $p_n$, and the Haar integral is then Gaussian.

\paragraph{Hagedorn Behaviour and Flavour Symmetries:} We close this section by
commenting on possible stronger singularities in the presence of flavour
symmetries, or more generally when the integrand of the thermal partition
function has linear terms in the power sums $p_n$. The simplest examples are
unitary quivers associated with finite algebras, see equation
\eqref{thermal-partition-function-unitary}. This appears generically for all
quivers that are not affine unitary, the partition function taking roughly the
form 
\begin{equation}
		Z\sim \prod_{n>0}\frac{\text{exp}[\frac{1}{n}z_\text{H}(x,n)^2 Q\cdot M^{-1}(x,n)\cdot Q]}{\sqrt{\text{det}\big[M(x,n)\big]}}\,,\qquad M(x,n)=1-z_\text{V}(x,n) - z_\text{H}(x,n)\,A\,,
\end{equation}
where $Q$ is a vector that does not depend on the temperature. As $M$ develops a
zero eigenvalue, the partition function exhibits an essential singularity
rather than a pole or a branch cut. However, its location is still set by the
largest eigenvalue of $A$, and the Hagedorn temperature is the same. By similar
arguments as above, going to an eigenbasis of $M$, we (very) schematically
have:
\begin{equation}
		Z\sim \frac{\text{exp}(\frac{1}{x_\text{Hag.}-x})}{(x_\text{Hag.}-x)^p} \sim \int_0^\infty dE\, E^{\frac{p}{2}-\frac{3}{4}}\, e^{E(\frac{1}{T_\text{Hag.}}- \frac{1}{T})} e^{\sqrt{E/T_\text{Hag.}}}\,,
\end{equation}
where we have ignored all order-one factors. We conclude that while the density
of states is enhanced by a sub-exponential factor, the Hagedorn behaviour
remains unchanged at large energies.

Finally, as we take the large-$N$ limit, finite quivers can also have formally
infinite flavour and the thermal partition function is divergent for any
temperature. As noted in reference \cite{Calderon-Infante:2024oed} this should
not be understood as a genuine Hagedorn behaviour, but rather as a signal that
the theory does not have a sparse spectrum in that limit.

\section{Universality Classes, Orientifolds, and Their Partial Decoupling}\label{sec:orientifolds}

In the previous section, we have seen that while the full thermal partition
function depends on the particular data of the gauge and flavour symmetries and
the shape of the quiver, the Hagedorn temperature in the overall-free limit is
solely set by the largest eigenvalue $\lambda_\text{max}$ of its Cartan
adjacency $A$---or an unfolding thereof, see equation
\eqref{beautiful-cos-formula}. The Hagedorn temperature therefore defines
universality classes that group SCFTs with the same growth of states at high
energy. In this section, we describe some of the properties of these
universality classes, their string-theoretical realisations, and some of their
holographic descriptions.

As we have seen in Section \ref{sec:Hagedorn}, the Hagedorn temperature
is set by the Lie algebra $\mathfrak{b}$ defining the shape of the quiver
rather than its gauge and flavour data, but several algebras share the same
Hagedorn temperature. They can furthermore be separated into two main groups:
the universality class formed by all affine quivers, and a finite ADE
classification of all the others. For the former, all quivers have
$\lambda_\text{max} = 2$, and share their Hagedorn temperature with
$\mathcal{N}=4$ super-Yang--Mills, thereby forming the largest grouping of
SCFTs.

This is of course not an accident: all $\mathcal{N}=2$ affine quivers admitting
a large-$N$ limit can be obtained from orientifold projections of $\mathcal{N}=4$
super-Yang--Mills, which is well known to be the low-energy description of a
stack of $N$ D3-branes probing a singularity. In the sequel, we will always split the
ten-dimensional flat spacetime as $\mathbb{R}^{1,3}\times \mathbb{R}^2\times
\mathbb{C}^2$ or orientifolds thereof, with the D3-branes filling the first four
directions.

When the D3-branes probe an orbifold singularity, we obtain an SCFT with
unitary gauge group whose number of conserved supercharges depends on the
associated discrete group. For instance, to preserve $\mathcal{N}=2$
supersymmetry, the extra dimensions must be of the form
$\mathbb{R}^2\times(\mathbb{C}^2/\Gamma)$, with $\Gamma$ a discrete subgroup of
$SU(2)$. For $\mathcal{N}=1$ theories, we have
$\mathbb{C}^3/\Gamma_{\mathcal{N}=1}$ with $\Gamma_{\mathcal{N}=1}\subset
SU(3)$ instead.

Discrete subgroups of $SU(2)$ famously follow an affine ADE classification,
which we denote by $\widehat{A}_r\simeq\mathbb{Z}_{r+1}, \widehat{D}_r,
\widehat{E}_6, \widehat{E}_7, \widehat{E}_8$. They correspond to the cyclic
groups, dicyclic groups, and the three exceptional binary tetrahedral, binary
octahedral, and binary icosahedral groups, respectively. To each discrete group
$\Gamma=\widehat{X}_r$, we can therefore associate an ADE affine algebra
$\mathfrak{b}=X_r^{(1)}$ via McKay duality \cite{McKay:1981aap}. When the
$N$ D3-branes probe the $\mathbb{C}^2/\Gamma$ singularity, one then obtains the
unitary affine quiver $(\mathfrak{b},S,f)=(X^{(1)}_r,0,0)$ in the notation of
Classification \ref{classification-scft}. For instance, with the $E_8$
singularity \cite{Douglas:1996sw, Katz:1997eq, Kachru:1998ys, Hanany:1998sd}:
\begin{equation}
		\begin{matrix}\begin{tikzpicture}
		\node[node, label=left:{\footnotesize $N$}, fill=black] (A1)  {};
		\draw (A1.north east) to[out=60, in=120, looseness=10] (A1.north west);
		\draw (A1.south east) to[out=-60, in=-120, looseness=10] (A1.south west);
	\end{tikzpicture}\end{matrix}
	\quad\xrightarrow{\mathbb{C}\times \mathbb{C}^2/\widehat{E}_8}\quad
	\begin{matrix}\begin{tikzpicture}
		\node[node, fill=black, label=below:{\footnotesize $N$}] (A0)  {};
    	\node[node, fill=black, label=below:{\footnotesize $2N$}]   (A1) [right=6mm of A0] {};
    	\node[node, fill=black, label=below:{\footnotesize $3N$}] (A2) [right=6mm of A1] {};
    	\node[node, fill=black, label=below:{\footnotesize $4N$}]   (A3) [right=6mm of A2] {};
    	\node[node, fill=black, label=below:{\footnotesize $5N$}] (A4) [right=6mm of A3] {};
    	\node[node, fill=black, label=below:{\footnotesize $6N$}]   (A5) [right=6mm of A4] {};
    	\node[node, fill=black, label=below:{\footnotesize $4N$}] (A6) [right=6mm of A5] {};
    	\node[node, fill=black, label=below:{\footnotesize $2N$}] (A7) [right=6mm of A6] {};
    	\node[node, fill=black, label=above:{\footnotesize $3N$}]   (A8) [above=6mm of A5] {};
    	\draw (A0.east) -- (A1.west);
    	\draw (A1.east) -- (A2.west);
    	\draw (A2.east) -- (A3.west);
    	\draw (A3.east) -- (A4.west);
    	\draw (A4.east) -- (A5.west);
    	\draw (A5.east) -- (A6.west);
    	\draw (A6.east) -- (A7.west);
    	\draw (A5.north) -- (A8.south);
	\end{tikzpicture}\end{matrix}
\end{equation}
More generally, we can also consider \emph{orientifolds}, associated with the
presence of $O3$- or $O7$-planes, see e.g. references \cite{Ahn:1998ku,
Uranga:1998uj, Witten:1998xy, Sen:1996vd, Douglas:1996js, Ennes:2000fu}. There
are four types of $O3$-planes: $O3^{\pm}$, $\widetilde{O3}^{\pm}$-planes, where
the signs correspond to the action of worldsheet parity on the Chan--Paton
factors. $\widetilde{O3}^\pm$-planes can be thought of as $O3^\pm$-planes with
a half-D3-brane stuck on them. These give rise to $\mathfrak{so}$ and
$\mathfrak{usp}$ algebras, depending on the chosen type of O3-plane. For instance, in
the presence of an $\widetilde{O3}^-$-plane, we obtain $\mathfrak{so}(2N+1)$ $\mathcal{N}=4$ super-Yang--Mills
\begin{equation}
	\begin{matrix}\begin{tikzpicture}
		\node[node, label=left:{\footnotesize $N$}, fill=black] (A1)  {};
		\draw (A1.north east) to[out=60, in=120, looseness=10] (A1.north west);
		\draw (A1.south east) to[out=-60, in=-120, looseness=10] (A1.south west);
	\end{tikzpicture}\end{matrix}
	\xrightarrow{\quad\widetilde{O3}^-\quad}
		\begin{matrix}\begin{tikzpicture}
		\node[node, label=right:{\footnotesize $2N+1$}, fill=red] (A1)  {};
		\draw (A1.north east) to[out=60, in=120, looseness=10] (A1.north west);
		\draw (A1.south east) to[out=-60, in=-120, looseness=10] (A1.south west);
	\end{tikzpicture}\end{matrix}
\end{equation}

We can also introduce $O7^\pm$-planes, which must be accompanied by a stack of
eight (half) D7-branes to cancel the RR charge. The prototypical
example is a stack of D3-branes in the presence of an $O7^-$-plane and eight
D7-branes \cite{Douglas:1996js}. The resulting theory is a quiver with a
single $\mathfrak{usp}(2N)$ gauge symmetry, two anti-symmetric representations, and
eight fundamentals. This corresponds to the last quiver in equation
\eqref{sporadic-cases-1}, which was shown to be in the same universality class
as $\mathcal{N}=4$ super-Yang--Mills \cite{Calderon-Infante:2024oed}.

If in addition to O-planes an additional orbifold is performed, the resulting
quiver will be non-simply-laced or will have (anti-)symmetric representations.
For instance, with an $\widehat{A}_{k-1}\simeq\mathbb{Z}_k$ orbifold, when $k$
is even, we can obtain two different quivers, depending on the location of the
$O7^-$-plane:
\begin{equation}
		\begin{matrix}\resizebox{1.1cm}{!}{\begin{tikzpicture}
		\node[node, label=left:{\footnotesize $N$}, fill=black] (A1) {};
		\draw (A1.north east) to[out=60, in=120, looseness=10] (A1.north west);
		\draw (A1.south east) to[out=-60, in=-120, looseness=10] (A1.south west);
	\end{tikzpicture}}\end{matrix}
	\quad\xrightarrow{~\mathbb{Z}_{k}}\quad
	\begin{matrix}\resizebox{4cm}{!}{\begin{tikzpicture}
	    \node[node, label=below:{\footnotesize $N$}, fill=black] (A1) [below right=6mm of A0] {};
	    \node (A2) [right=6mm of A1] {\dots};
	    \node[node, label=south west:{\footnotesize $N$}, fill=black] (A3) [right=6mm of A2] {};
	    \node (A4) [right=6mm of A3] {\dots};
	    \node[node, label=below:{\footnotesize $N$}, fill=black] (A5) [right=6mm of A4] {};
	    \node[node, label=above:{\footnotesize $N$}, fill=black] (A6) [above right=6mm of A0] {};
	    \node (A7) [right=6mm of A6] {\dots};
	    \node[node, label=north west:{\footnotesize $N$}, fill=black] (A8) [right=6mm of A7] {};
	    \node (A9) [right=6mm of A8] {\dots};
	    \node[node, label=above:{\footnotesize $N$}, fill=black] (A10) [right=6mm of A9] {};
	    \draw (A1.south) -- (A6.north);
	    \draw (A1.east) -- (A2.west);
	    \draw (A2.east) -- (A3.west);
	    \draw (A3.east) -- (A4.west);
	    \draw (A4.east) -- (A5.west);
	    \draw (A6.east) -- (A7.west);
	    \draw (A7.east) -- (A8.west);
	    \draw (A8.east) -- (A9.west);
	    \draw (A9.east) -- (A10.west);
	    \draw (A10.south) -- (A5.north);

		\draw[dashed] ($(A3.south)+(0,-5mm)$) -- ($(A8.north)+(0,5mm)$);

		\draw[dotted] ($(A1)!0.5!(A6)+(-5mm,0)$) -- ($(A5)!0.5!(A10)+(5mm,0)$);
	\end{tikzpicture}}\end{matrix}
	\quad
	\begin{matrix}\begin{tikzpicture}
		\node (A0) at (0,0.1) {};
		\node (A1) at (0,-0.1) {};
		\node (A2) at (1.1,0.7) {};
		\node (A3) at (1.1,-0.7) {};
		\node (A4) at (.8,0) {\footnotesize $O7^-$};
		\draw[->, dashed] (A0.east) -- (A2.west);
		\draw[->, dotted] (A1.east) -- (A3.west);
	\end{tikzpicture}\end{matrix}
	\quad
	\begin{gathered}
		\begin{matrix}\resizebox{4.7cm}{!}{\begin{tikzpicture}
			\node[node, label=above:{\footnotesize $2N$}, fill=blue]   (A1) [right=6mm of A0] {};
    		\node[node, label=above:{\footnotesize $2N$}, fill=black] (A2) [right=6mm of A1] {};
    		\node[node, label=above:{\footnotesize $2N$}, fill=black]   (A3) [right=6mm of A2] {};
    		\node (A4) [right=6mm of A3] {\dots};
    		\node[node, label=above:{\footnotesize $2N$}, fill=black]   (A5) [right=6mm of A4] {};
    		\node[node, label=above:{\footnotesize $2N$}, fill=black] (A6) [right=6mm of A5] {};
    		\node[node, label=above:{\footnotesize $2N$}, fill=blue] (A7) [right=6mm of A6] {};
    		\node[rectangle, label=below:{\footnotesize $4$}, fill=red] (F1) [below=4mm of A1] {};
    		\node[rectangle, label=below:{\footnotesize $4$}, fill=red] (F7) [below=4mm of A7] {};
			\node[yscale=1.4] (C) [right=.2mm of A1] {$>$};
			\node[yscale=1.4] (D) [right=.2mm of A6] {$<$};
			\draw ([yshift=1.5pt]A1.east) -- ([yshift=1.5pt]A2.west);
    		\draw ([yshift=-1.5pt]A1.east) -- ([yshift=-1.5pt]A2.west);
    		\draw (A2.east) -- (A3.west);
    		\draw (A3.east) -- (A4.west);
    		\draw (A4.east) -- (A5.west);
    		\draw (A5.east) -- (A6.west);
    		\draw (A1.south) -- (F1.north);
    		\draw (A7.south) -- (F7.north);
			\draw ([yshift=1.5pt]A6.east) -- ([yshift=1.5pt]A7.west);
    		\draw ([yshift=-1.5pt]A6.east) -- ([yshift=-1.5pt]A7.west);
		\end{tikzpicture}}\end{matrix}\\
		\begin{matrix}\resizebox{4.5cm}{!}{\begin{tikzpicture}
		 	\node[node, label=above:{\footnotesize $N$}, fill=black]   (A1) {};
	   	 	\node (A2) [right=6mm of A1] {\dots};
	   	 	\node[node, label=above:{\footnotesize $N$}, fill=black]   (A3) [right=6mm of A2] {};
	   	 	\node[node, label=above:{\footnotesize $N$}, fill=black] (A4) [right=6mm of A3] {};
	   	 	\node[rectangle, label=below:{\footnotesize $2$}, fill=black] (F1) [below=4mm of A1] {};
	   	 	\node[rectangle, label=below:{\footnotesize $2$}, fill=black] (F4) [below=4mm of A4] {};
	   	 	\node[rectangle, label=below:{\footnotesize $\bm{\Lambda^2}$}, fill=black] (AS1) [left=6mm of A1] {};
	   	 	\node[rectangle, label=below:{\footnotesize $\bm{\Lambda^2}$}, fill=black] (AS4) [right=6mm of A4] {};
	   	 	\draw (A1.east) -- (A2.west);
	   	 	\draw (A2.east) -- (A3.west);
	   	 	\draw (A3.east) -- (A4.west);
	   	 	\draw (A1.south) -- (F1.north);
	   	 	\draw (A4.south) -- (F4.north);
	   	 	\draw[decorate, decoration={snake}] (AS1.east) -- (A1.west);
	   	 	\draw[decorate, decoration={snake}] (A4.east) -- (AS4.west);
		\end{tikzpicture}}\end{matrix}\\
	\end{gathered}
\end{equation}
with the dashed lines expressing how the folding is performed.

Although we have classified affine quivers from a bottom-up perspective in
Section \ref{sec:classification-quivers} and found that in those cases the
flavour symmetry cannot be large, see equation \eqref{affine-level}, this is
consistent with the brane point of view, and explained by the fact that there
must be only eight flavour branes. Furthermore, we have seen that the flavour
configurations are classified by instanton moduli spaces---or equivalently by
affine coweights---which is understood here as the boundary data of the
D7-branes, see e.g. references \cite{Douglas:1996sw, Blum:1997mm}. Similarly,
the brane construction gives a top-down explanation as to why the cancellation
of the $\beta$-functions involving (anti-)symmetric representations can be
understood as equivalent to replacing the associated $\mathfrak{su}(k)$ node by
an $\mathfrak{usp}(k)$ or $\mathfrak{so}(k)$ symmetry: the O-plane leads to a
folding of the orbifold.

All in all, every large-$N$ quiver associated with an affine algebra
$\mathfrak{b}=X_r^{(n>0)}$---including quivers hosting (anti-)symmetric
representations mimicking an affine quiver---can be obtained by D3-branes
probing orientifolds, and therefore descend from $\mathcal{N}=4$
super-Yang--Mills. We defer to e.g. reference \cite{Abajian:2024rjq} for a
complete survey of these constructions in four dimensions. This explains why
while the partition function itself depends on the full shape of the quiver,
the pole structure is insensitive to the folding: the presence of O-planes does
not significantly affect the spectrum and thus does not change the Hagedorn
temperature.

Furthermore, for theories associated with an orbifold and without
O-planes, the full partition function can be obtained via group theory through
McKay duality. This duality can be concisely explained as follows: let $R_i$ be
any irreducible representation of the discrete group $\Gamma\subset SU(2)$. If
$g\in \Gamma$, the character of the representation is given by $\chi_i =
\text{Tr}_{R_i}(g)$, and it can then be shown that \cite{McKay:1981aap}:
\begin{equation}\label{mcKay-decomp}
		\chi_{\bm{2}}(g)\,\chi_i(g) = A^{ij}\chi_j(g)\,,\qquad A = 2\mathbf{1} - C\,,
\end{equation}
where $C$ is the Cartan matrix of the algebra dual to $\Gamma$, which is always
symmetric in the ADE case. It follows that the dimensions of irreducible
representations of $\Gamma$ arrange into an eigenvector of the Cartan matrix:
$C^{ij}\text{dim}(R_j)=0$. 

As shown in references \cite{Lawrence:1998ja, Hanany:1998sd}, this relation is
at the heart of the orbifolding procedure: fields in the resulting theory must
form invariants of $\Gamma$. At the level of the single-letter partition
function of $\mathcal{N}=4$ super-Yang--Mills, we observe that this amounts to
the substitution:
\begin{equation}
		z_{\mathcal{N}=4} = z_\text{V} + \chi_{\bm{2}}\,z_\text{H} \qquad\xrightarrow{~\mathbb{C}^2/\Gamma ~}\qquad z_{\Gamma}^{ij} = \delta^{ij}z_\text{V} + A^{ij}\,z_\text{H}\,,
\end{equation}
with the extra $\mathfrak{su}(2)_L$ symmetry understood from the
$\mathcal{N}=2$ point of view by decomposing the $\mathcal{N}=4$ R-symmetry:
$\mathfrak{su}(4)\to\mathfrak{su}(2)_R\oplus\mathfrak{u}(1)\oplus\mathfrak{su}(2)_L$.
The full thermal partition function of unitary quivers then descends from that of
$SU(N)$ $\mathcal{N}=4$ super-Yang--Mills in a simple way:
\begin{equation}\label{replacement-orbifolds}
		Z_{\mathcal{N}=4} \simeq \prod_{n>0}\frac{\text{exp}[-\frac{1}{n}z_{\mathcal{N}=4}(x,n)]}{1-z_{\mathcal{N}=4}(x,n)}
	\qquad\xrightarrow{~\mathbb{C}^2/\Gamma ~}\qquad 
	Z_{\Gamma}\simeq\prod_{n>0}\frac{\text{exp}[-\frac{1}{n}\text{Tr}(z_{\Gamma}(x,n))]}{\text{det}[\mathbf{1}-z_\Gamma(x,n)]}\,.
\end{equation}
Using that $\text{Tr}(A)=0$ in all cases, we recover the correct result for
unitary affine quivers, see equation \eqref{thermal-partition-function-unitary}.

We are not aware of a similar procedure that applies to all other unitary
quivers, as there are complications due to the flavour. It would however be
interesting to see if this perspective can be used to uniformly describe the
thermal partition function, including the effect of O-planes. We expect that
such a group-theoretic relation between the partition functions in the affine
and finite cases would lead to a better understanding of the extra towers of
states discussed in e.g. reference \cite{Mantegazza:2026spd}.

\subsection{Decoupling the Affine Node}\label{sec:affine-decoupling}

In Section \ref{sec:classification-quivers}, we have shown that every
$\mathcal{N}=2$ quiver SCFT admitting a large-$N$ limit has the shape of a
finite or affine Dynkin diagram, up to the subtleties involving non-fundamental
representations. Furthermore, we have seen above that those with an affine
shape can always be thought of as arising from a stack of D3-branes probing an
orientifold singularity. 

Those of finite shape however do not admit such a description, although they
can often be engineered in terms of type-IIA string theory via
Hanany--Witten-like setups \cite{Hanany:1996ie}, see e.g. reference
\cite{Giveon:1998sr} for a review of early constructions.
Other---equivalent---possibilities include F-theory geometric engineering, or
class-S constructions from six dimensions, see below.

From Tables
\ref{tab:dynkin-bases-unitary}--\ref{tab:dynkin-bases-unitary-ortho-symplectic}
one can see that finite quivers can be obtained by deleting a single node from
an affine quiver, up to some freedom with the flavour symmetry. While this can
often be achieved in multiple ways, the canonical choice is of course to
decouple the affine node. Starting with a theory associated with an affine
algebra $\mathfrak{b}=X^{(1)}_r$, there is always a unique node whose
Yang--Mills coupling can be sent to zero, leading to a quiver with shape
$X_r^{(0)}$ accompanied by a free vector multiplet. For instance with
$\mathfrak{b}=E_7^{(1)}$:
\begin{equation}\label{example-affine-decoupling}
	\begin{matrix}\begin{tikzpicture}
	    \node[node, fill=black, label=below:{\footnotesize $N$}] (A0)  {};
	    \node[node, fill=black, label=below:{\footnotesize $2N$}]   (A1) [right=6mm of A0] {};
	    \node[node, fill=black, label=below:{\footnotesize $3N$}] (A2) [right=6mm of A1] {};
	    \node[node, fill=black, label=below:{\footnotesize $4N$}]   (A3) [right=6mm of A2] {};
	    \node[node, fill=black, label=below:{\footnotesize $3N$}] (A4) [right=6mm of A3] {};
	    \node[node, fill=black, label=below:{\footnotesize $2N$}]   (A5) [right=6mm of A4] {};
	    \node[node, fill=black, label=below:{\footnotesize $N$}] (A6) [right=6mm of A5] {};
	    \node[node, fill=black, label=above:{\footnotesize $2N$}]   (A7) [above=6mm of A3] {};
	    \draw (A0.east) -- (A1.west);
	    \draw (A1.east) -- (A2.west);
	    \draw (A2.east) -- (A3.west);
	    \draw (A3.east) -- (A4.west);
	    \draw (A4.east) -- (A5.west);
	    \draw (A5.east) -- (A6.west);
	    \draw (A3.north) -- (A7.south);
	\end{tikzpicture}\end{matrix}
		\qquad\longrightarrow\qquad
	\begin{matrix}\begin{tikzpicture}
	    \node[node, fill=black, label=below:{\footnotesize $N$}]   (A0) {};
	    \node[node, fill=black, label=below:{\footnotesize $2N$}]   (A1) [right=16mm of A0] {};
	    \node  (B) [left=9mm of A1] {$\oplus$};
	    \node[node, fill=black, label=below:{\footnotesize $3N$}] (A2) [right=6mm of A1] {};
	    \node[node, fill=black, label=below:{\footnotesize $4N$}]   (A3) [right=6mm of A2] {};
	    \node[node, fill=black, label=below:{\footnotesize $3N$}] (A4) [right=6mm of A3] {};
	    \node[node, fill=black, label=below:{\footnotesize $2N$}]   (A5) [right=6mm of A4] {};
	    \node[node, fill=black, label=below:{\footnotesize $N$}] (A6) [right=6mm of A5] {};
	    \node[node, fill=black, label=above:{\footnotesize $2N$}]   (A7) [above=6mm of A3] {};
	    \node[rectangle, fill=black, label=below:{\footnotesize $N$}] (F1) [left=4mm of A1] {};
	    \draw (A1.west) -- (A2.west);
	    \draw (A2.east) -- (A3.west);
	    \draw (A3.east) -- (A4.west);
	    \draw (A4.east) -- (A5.west);
	    \draw (A5.east) -- (A6.west);
	    \draw (A3.north) -- (A7.south);
		\draw (A1.west) -- (F1.east);
	\end{tikzpicture}\end{matrix}
\end{equation}
For twisted algebras, i.e. those of the form $X_r^{(n>1)}$, decoupling the
affine node will lead to a quiver of different type, e.g.  $E_6^{(2)}\to
F_4^{(0)}$. The same reasoning applies to all quivers, although when the decoupling node
hosts an additional flavour symmetry, there might also be a free hypermultiplet
in addition to the free vector multiplet.

Furthermore, with this prescription the decoupled theory always has a fixed
flavour symmetry dictated by the SCFT condition given in equation
\eqref{beta-function-bif}. However all other configurations of flavour
symmetries can be reached through Higgs-branch deformations. This is guaranteed
by the properties of the coroot lattice of $\mathfrak{b}$, see e.g. 
\cite{Bourget:2021siw, Ahmed:2025fmq, Fazzi:2023ulb, Bourget:2026ono} and
references therein for their relation to the weight system of $\mathfrak{b}$ in
various dimensions.

This decoupling	is possible in all but three cases, namely with
$\mathfrak{b}=F_4^{(0)}, E_7^{(0)}, E_8^{(0)}$. For those quivers, there are
two possible assignments of gauge algebras, and only one choice can be
obtained by decoupling. This is because its affine parent must satisfy
$\theta_i\, f^i = 4 \theta_i\,S^i$, see equation \eqref{affine-level}. For
$\mathfrak{b}=E_7^{(1)}, E_8^{(1)}$, we have either a unitary quiver (in which
case the affine decoupling is possible), or ortho-symplectic, in which case the
only consistent choice has $\theta_i\, f^i = 8$. The would-be affine quiver
can then be obtained by exchanging $\mathfrak{so}$ and $\mathfrak{usp}$
algebras. However, this results in the constraint $\theta_i\, f^i = -8$ on the
flavour symmetry, which is not a consistent choice since $\theta_i>0$. The same
argument applies to the remaining outliers. It is quite peculiar that out of
all the possible $\mathcal{N}=2$ quiver shapes, only three cannot be
obtained through affine decoupling. However, we have only used the vanishing of
$\beta$-functions, and it could be possible that these SCFTs suffer from other
pathologies that render them inconsistent. Due to their exceptional shapes,
they cannot be easily constructed via a brane system. Furthermore, when lifted
to a six-dimensional theory, they are part of the frozen phase of F-theory
\cite{Tachikawa:2015wka, Bhardwaj:2018jgp, Bhardwaj:2019hhd}, which can be
difficult to realise explicitly. 

Finally, as the node becoming ungauged admits a large-$N$ limit, so will the
resulting flavour node. The decoupling also leads to an infinite tower of
new BPS states: as the anomalous dimension of certain long multiplets vanishes,
and superconformal recombination rules show that they split into various short
multiplets whose presence can be detected via a superconformal-index
computation \cite{Gadde:2009dj, Mantegazza:2026spd}. 

A consequence of the absence of these states in the decoupled theory is that
the large-energy growth of states is affected, and the Hagedorn temperature is
never preserved when performing such a partial-decoupling limit.

Even though the thermal partition function is computed at the overall-free
point in both cases, the absence of a gauge node in the resulting theory
changes its singularity structure. The group-theoretical explanation is that we
have lost the null eigenvalue of the Cartan matrix, and therefore we do not
have $\lambda_\text{max}=2$ anymore. 
 Similar arguments should also
apply more generally when the decoupling is performed for arbitrary nodes also
in finite cases, and could help characterise the decoupled tower of states. 

\subsection{Generalisations to \texorpdfstring{$\mathcal{N} \leq 1$}{N ≤ 1}}\label{sec:N=1}

When computing the partition function of quiver SCFTs, we have primarily used
our knowledge of $\mathcal{N}=2$ theories to find all possible adjacency
matrices. Relaxing this condition to $\mathcal{N}=1$ theories, we are opening
up a larger landscape of possibilities no longer solely classified by group
theory. However, most of the techniques we have developed in Section
\ref{sec:thermal} to perform the Gaussian integral easily extend to large-$N$
$\mathcal{N}=1$ quiver SCFTs, and the Hagedorn temperature is also set by the
spectral properties of the corresponding adjacency matrix.

For $\mathcal{N}=2$ we have relied heavily on the fact that the matrix $G^{ij}
= 2D^{ij} - B^{ij}$ was symmetric and is associated with a Dynkin diagram. We
are unfortunately not aware of a similar simple classification for the
adjacency matrix of $\mathcal{N}=1$ theories. We can nonetheless make a
statement for orbifolds of $\mathcal{N}=4$ super-Yang--Mills preserving four
supercharges. In the same way that $\mathcal{N}=2$ orbifolds are classified by a
finite subgroup of $SU(2)$, those preserving four supercharges correspond to a
choice of finite subgroups $\Gamma_{\mathcal{N}=1}\subset SU(3)$. 

In terms of $\mathcal{N}=1$ supermultiplets, the single-letter partition
function of $\mathcal{N}=4$ is given by
\begin{equation}\label{zN=4-N=1}
	\begin{gathered}
			z_{\mathcal{N}=4}(x,n) = z_\text{v}(x,n) + \chi_{\bm{3}}(V^n) z_\text{c}(x,n)\,,\\
			z_\text{v}(x,n)=z_A(x^n)+(-1)^{n+1}z_\psi(x^n)\,,\qquad z_\text{c} = 2z_\phi(x^n)+(-1)^{n+1}z_\psi(x^n)\,,
	\end{gathered}
\end{equation}
where $z_\text{v}$ and $z_\text{c}$ are the single-letter partition functions
associated with $\mathcal{N}=1$ vector and chiral multiplets, respectively.
Similarly as for the $\mathcal{N}=2$ case, from the $\mathcal{N}=1$ point of
view there is an extra $\mathfrak{su}(3)_L$ symmetry after decomposing the
R-symmetry $\mathfrak{su}(4)_R\to \mathfrak{u}(1)_r\oplus\mathfrak{su}(3)_L$.
The three complex scalars in the $\mathcal{N}=4$ vector multiplet are rotated
by this symmetry, explaining the presence of the Weyl character
$\chi_{\bm{3}}(V)$ in equation \eqref{zN=4-N=1}.

One can then use similar arguments as with the McKay correspondence, see
equation \eqref{mcKay-decomp}, to go from a discrete group
$\Gamma_{\mathcal{N}=1}$ to the quiver: its adjacency matrix satisfies
\cite{Lawrence:1998ja, Hanany:1998sd}
\begin{equation}\label{su3-relation}
		\chi_{\bm{3}}(g)\,\chi_i(g) = A^{ij}\chi_j(g)\,,
\end{equation}
with $\chi_i(g)$ the characters of irreducible representations of the finite
subgroup $\Gamma_{\mathcal{N}=1}\subset SU(3)$. Contrary to $SU(2)$ subgroups,
which fall in an ADE classification, the adjacency matrix need not be
symmetric. We can nonetheless go through the same procedure for $\mathcal{N}=1$
orbifolds, and using the properties of group characters, it follows from
equation \eqref{su3-relation} that $\text{det}(3\,\mathbf{1}-A)=0$.

Given an $\mathcal{N}=1$ theory with adjacency $A$, in the absence of O-planes
the thermal partition function will be of the same form as for $\mathcal{N}=2$
orbifolds, and given by the replacement rule in equation
\eqref{replacement-orbifolds}. We stress that since $A$ is not symmetric, one
will obtain the determinant of $(A+A^T)/2$ instead, which can be understood as
counting the contribution of chiral multiplets in a representation $\bm{R}$
together with their CPT conjugate \cite{Calderon-Infante:2024oed}, and we find
the following result:
\paragraph{Hagedorn Temperature of $\mathcal{N}=1$ Orbifolds:} \emph{For a
		$\mathbb{C}^3/\Gamma_{\mathcal{N}=1}$ orbifold of $\mathcal{N}=4$
		super-Yang--Mills with $\Gamma_{\mathcal{N}=1}\subset SU(3)$, the
		Hagedorn temperature is set by solving the following constraint in the
		large-$N$ limit:
\begin{equation}\label{N=1-tmp}
		1- z_\text{v}(x,1) = \lambda_\text{max}^{\mathcal{N}=1} \,z_\text{c}(x,1)\,,\qquad \lambda_\text{max}^{\mathcal{N}=1} = 3\,,
\end{equation}
and these $\mathcal{N}=1$ SCFTs are therefore all in the same universality
class as all $\mathcal{N}=2$ affine quivers.
}

Indeed, the large-$N$ computation of the partition function is completely
analogous to that of Section \ref{sec:thermal}, replacing the single-letter
partition functions of hyper- and vector multiplets by those of $\mathcal{N}=1$
vector and chiral multiplets, $z_\text{v}$, $z_\text{c}$, respectively. The
equation above is the same as that of $\mathcal{N}=4$ or equivalently to that
of affine quivers, see equation \eqref{beautiful-cos-formula} when rewritten in
terms of $\mathcal{N}=1$ contributions. Furthermore, finite subgroups of
$SU(3)$ are therefore ``affine'', in the sense that the matrix $C=3\,\mathbf{1}
- A$ has determinant zero, and we indeed find that the largest eigenvalue
of $A$ is $\lambda_\text{max}^{\mathcal{N}=1}=3$.
Observe that if we start with $\mathcal{N}=2$ supersymmetry,
expanding in terms of chiral multiplets, we find 
\begin{equation}
		\mathcal{N}=2 \text{ quivers:}\qquad \lambda_\text{max}^{\mathcal{N}=1} = \lambda_\text{max}^{\mathcal{N}=2} + 1\,.
\end{equation}

The discussion above goes beyond $\mathcal{N}\geq1$ and applies to
non-supersymmetric orbifolds in the overall-free limit as well. One then has a
stack of D3-branes probing $\mathbb{R}^6/\Gamma_{\mathcal{N}=0}$ with
$\Gamma_{\mathcal{N}=0}\subset SU(4)\simeq Spin(6)$, which preserves no
supercharges. The resulting theories only have vanishing $\beta$-function at
leading order in $N$ and therefore are ``quasi-conformal''. Furthermore, in the
supergravity regime their bulk duals are plagued by tachyonic instabilities
\cite{Pomoni:2009joh}. We can nonetheless compute their thermal partition
function in the overall free-field limit, where they are trivially conformal.
These theories also admit a quiver description, although the absence of
supersymmetry forces one to consider two adjacency matrices: one for bosons and
the other for fermions, again given in terms of the discrete group and the
R-symmetry of the $\mathcal{N}=4$ theory:
\begin{equation}
		\chi_{\bm{6}}(g)\,\chi_i(g) = A_\varphi^{ij}\,\chi_j(g)\,,\qquad
		\chi_{\bm{4}}(g)\,\chi_i(g) = A_\psi^{ij}\,\chi_j(g)\,,
\end{equation}
where $\bm{6}$ and $\bm{4}$ relate to the $\mathfrak{su}(4)_R$ R-symmetry
representations of the six real scalars and four fermions of the $\mathcal{N}=4$
vector multiplet, respectively. Through the arguments we have used many times by
now, albeit applied to the two adjacency matrices separately, one finds that
$\lambda^\varphi_\text{max}=6$ and $\lambda^\psi_\text{max}=4$ in the obvious
notation, and their Hagedorn temperature is set by
\begin{equation}
		\mathcal{N}=0 \text{ orbifolds:}\qquad		1-z_A(x) = 6\, z_\phi(x) + 4\, z_\psi(x)\,,
\end{equation}
which is the same as equations \eqref{beautiful-cos-formula} and
\eqref{N=1-tmp} when expressed in terms of the single-letter partition functions
of fundamental fields. This shows that $\mathcal{N}=4$ super-Yang--Mills and
all its orbifolds, supersymmetric or not, share the same Hagedorn temperature,
and fall into the same universality class.

While these orbifold projections are all in the same universality class, the
landscape of $\mathcal{N}=1$ large-$N$ quiver theories is much richer and
more subtle than those with $\mathcal{N}=2$ supersymmetry. Demanding vanishing of
the $\beta$-functions may relate different gauge couplings, or they can develop
an anomalous dimension and are no longer exactly marginal. Additionally, these
theories also admit exactly-marginal superpotential deformations, which further
complicates the structure of the conformal manifold.

To illustrate these points, consider the CFT obtained as the low-energy
worldvolume theory of a stack of D3-branes probing the conifold singularity
$Y_6 = \{(x,y,u,v)\in\mathbb{C}^4\,|\, xy=uv\}$. The associated $\mathcal{N}=1$
quiver theory is then given by \cite{Klebanov:1998hh}:
\begin{equation}\label{conifold-quiver}
		\begin{matrix}\begin{tikzpicture}
		    \node[node, label=left:{\footnotesize $\mathfrak{su}(N)$}, fill=black] (A1) {};
		    \node[node, label=right:{\footnotesize $\mathfrak{su}(N)$}, fill=black] (A2) [right=10mm of A1] {};
		    \draw[doublemidarrow]
		        (A1) to[bend left=35]
		        node[above] {$\footnotesize X$}
		        (A2);
		    \draw[doublemidarrow]
		        (A2) to[bend left=35]
		        node[below] {$\footnotesize Y$}
		        (A1);
		\end{tikzpicture}\end{matrix}\qquad
		A =\left(\begin{matrix}
			0&2\\2&0
		\end{matrix}\right) 
\end{equation}
The double arrows each denote two $\mathcal{N}=1$ chiral multiplets in the
bifundamental representation, with a total of four chiral multiplets $X_a,
Y_a$, $a=1,2$. Since $\mathcal{N}=1$ theories are chiral, we now have an
oriented graph depicting the quiver. This is the same matter content as the
$\mathbb{C}\times\mathbb{C}^2/\mathbb{Z}_2$ orbifold; however, we do not have an
$\mathcal{N}=2$ SCFT, as we are lacking adjoint chiral multiplets on each node
to form a \emph{bona fide} $\mathcal{N}=2$ multiplet.

Computing the overall-free limit of the thermal partition function by performing the
large-$N$ integral in terms of $\mathcal{N}=1$ single-letter partition
functions, one easily finds that the Hagedorn temperature for the quiver above
is set by
\begin{equation}
		1- z_\text{v}(x,1) = 2 \,z_\text{c}(x,1)\,,
\end{equation}
where $\lambda^{\mathcal{N}=1}_\text{max}=2$ corresponds to the largest eigenvalue of the
adjacency matrix given in equation \eqref{conifold-quiver}.

One might then expect that it shares the same universality class as the
$\mathcal{N}=2$ quiver with $\mathfrak{b}=A_2^{(0)}$. However, the quiver of
the conifold above is not, as it is, conformal. To obtain an $\mathcal{N}=1$ SCFT,
one must turn on a superpotential:
\begin{equation}
		W = \frac{g_W}{2}\,\varepsilon^{ab}\varepsilon^{cd}\,\text{Tr}\,X_aY_cX_bY_d\,.
\end{equation}
It is then possible to show that to be conformal, the chiral fields $X,Y$ must
have anomalous dimensions satisfying \cite{Klebanov:1998hh}:
\begin{equation}
		\gamma_X(g_1,g_2,g_W) + \gamma_Y(g_1,g_2,g_W) + \frac{1}{2} = 0\,.
\end{equation}
Taking into account the symmetry exchanging the two gauge nodes, the conformal
manifold is two-dimensional. However, since conformality demands that
$\gamma_X,\gamma_Y\neq 0$, that SCFT does not admit an overall-free point, and
the Hagedorn temperature associated with the free thermal partition function
gives no information about the structure of its conformal manifold, as the
overall-free theory is not part of it.

This example can be generalised to D3-branes probing a local Calabi--Yau
singularity $Y_6$. As for the conifold, they have $a\sim c$ in the conformal
regime and a weakly-coupled $\text{AdS}_5\times X_5$ dual, where $Y_6$ is then
the cone over $X_5$ \cite{Klebanov:1998hh, Morrison:1998cs, Acharya:1998db,
Gauntlett:2004yd}. While there is a procedure to find the quiver and its
adjacency matrix, turning on a superpotential will generically be necessary to
ensure conformality, and the resulting conformal manifold will not have
an overall-free point, except in the case of orbifold projections.

This demonstrates that the question of the nature of tensionless strings in
$\mathcal{N}=1$ theories is much more subtle, and the free Hagedorn temperature
is not always a good tool to study it. In particular, while we have found that
almost all $\mathcal{N}=2$ finite quivers can be obtained by affine decoupling
of an orbifold of $\mathcal{N}=4$ super-Yang--Mills, without a classification
of $\mathcal{N}=1$ quivers and their adjacency, it is not clear whether a
similar result holds with fewer supercharges, even when their conformal
manifold pass through an overall-free point.

It would therefore be interesting to explore whether and when similar arguments
as in Section \ref{sec:classification-quivers} can be applied to
$\mathcal{N}=1$, and if it is possible to classify all quivers admitting a
non-trivial conformal manifold with an overall-free point. In particular, we
have not checked whether all seven $\mathcal{N}=1$ theories in the
``mini-landscape'' discussed below are decoupling points of the conformal
manifold of orientifolds of $\mathcal{N}=4$ super-Yang--Mills.

When a quiver does admit a conformal manifold with an overall-free point, given
its adjacency $A$ the Hagedorn temperature will be set by its largest
eigenvalue $\lambda_\text{max}$. While we do not expect O-planes to have an
influence on the Hagedorn temperature as they do not influence the singularity
structure of the partition function, it is difficult to estimate the total
number of universality classes without a full classification of quiver shapes.
Furthermore, while certain $\mathcal{N}=1$ SCFTs have a quiver description,
their conformal manifolds do not always have an infinite-distance point.

\subsection{Explaining the One-Node \texorpdfstring{``Mini-Landscape''}{"Mini-Landscape"}}\label{sec:mini-landscape}

A complete analysis of four-dimensional large-$N$ $\mathcal{N}=1$ quiver SCFTs
with a single gauge node and their Hagedorn temperatures has been performed in
reference \cite{Calderon-Infante:2024oed}. It was found that, out of 21
different consistent theories, only three distinct universality classes share
the same Hagedorn temperature, which the authors referred to as a
``mini-landscape'' of SCFTs. Our analysis gives an elegant explanation for the
presence of three universality classes when the quiver is made out of only one
simple gauge factor: at least for cases with eight supercharges, it is either
associated with a stack of $N$ D3-branes probing an orientifold singularity, or
is the deformation of an SCFT that has the shape of a higher-rank algebra. In
each case, we can associate the largest eigenvalue $\lambda_\text{max}$ of the
corresponding Cartan adjacency matrix $A = 2\,\mathbf{1}-C$ to label the
universality class.

Restricting ourselves to $\mathcal{N}=2$ theories, only eleven remain, and as
we have seen above, one of the universality classes is given by $\mathcal{N}=4$
super-Yang--Mills and its orbifold projections. While it contains an infinite
number of quiver shapes for all possible affine algebras, those that have a
single gauge algebra are as follows: 
\begin{equation}
		\lambda_\text{max} = 2:\qquad 
	\begin{matrix}
	\begin{tikzpicture}
		\node[node, label=left:{\footnotesize $N$}, fill=black] (A1) {};
		\draw (A1.north east) to[out=60, in=120, looseness=10] (A1.north west);
		\draw (A1.south east) to[out=-60, in=-120, looseness=10] (A1.south west);
	\end{tikzpicture}
	\end{matrix}\qquad
	\begin{matrix}
	\begin{tikzpicture}
		\node[node, label=left:{\footnotesize $N$}, fill=red] (A1)  {};
		\draw (A1.north east) to[out=60, in=120, looseness=10] (A1.north west);
		\draw (A1.south east) to[out=-60, in=-120, looseness=10] (A1.south west);
	\end{tikzpicture}
	\end{matrix}\qquad
	\begin{matrix}
	\begin{tikzpicture}
		\node[node, label=left:{\footnotesize $2N$}, fill=blue] (A1)  {};
		\draw (A1.north east) to[out=60, in=120, looseness=10] (A1.north west);
		\draw (A1.south east) to[out=-60, in=-120, looseness=10] (A1.south west);
	\end{tikzpicture}
	\end{matrix}\qquad
	\begin{matrix}
	\begin{tikzpicture}
		\node[node, label=above:{\footnotesize $N$}, fill=black] (A1)  {};
		\node[rectangle, label=above:{\footnotesize $2$}, fill=black] (F1) [right=8mm of A1]  {};
		\node[rectangle, label=below:{\footnotesize $4$}, fill=black] (F2) [below=6mm of A1]  {};
		\node[label=below:{\footnotesize $\bm{\Lambda^2}$}] (C) [right=4mm of A1]  {};
		\draw[decorate, decoration={snake}] (A1.east) -- (F1.west);
	    \draw (A1.south) -- (F2.north);
	\end{tikzpicture}
	\end{matrix}\qquad
	\begin{matrix}
	\begin{tikzpicture}
		\node[node, label=above:{\footnotesize $N$}, fill=black] (A1)  {};
		\node[rectangle, label=above:{\footnotesize $1$}, fill=black] (F1) [right=8mm of A1]  {};
		\node[rectangle, label=below:{\footnotesize $1$}, fill=black] (F2) [below=6mm of A1]  {};
\node[label=below:{\footnotesize $\bm{\Lambda^2}$}] (C) [right=4mm of A1]  {};
		\node[label=left:{\footnotesize $\bm{S^2}$}] (D) [below=2mm of A1]  {};
		\draw[decorate, decoration={snake}] (A1.east) -- (F1.west);
		\draw[decorate, decoration={snake}] (A1.south) -- (F2.north);
	\end{tikzpicture}
	\end{matrix}\qquad
	\begin{matrix}
	\begin{tikzpicture}
		\node[node, label=above:{\footnotesize $2N$}, fill=blue] (A1)  {};
		\node[rectangle, label=above:{\footnotesize $2$}, fill=red] (F1) [right=8mm of A1]  {};
		\node[rectangle, label=below:{\footnotesize $8$}, fill=red] (F2) [below=6mm of A1]  {};
		\node[label=below:{\footnotesize $\bm{\Lambda^2}$}] (C) [right=4mm of A1]  {};
		\draw[decorate, decoration={snake}] (A1.east) -- (F1.west);
	    \draw (A1.south) -- (F2.north);
	\end{tikzpicture}
	\end{matrix}
\end{equation}
Apart from the $\mathcal{N}=4$ theories themselves, all others have
(anti-)symmetric representations, and the number of extra bifundamental
hypermultiplets is small, which is consistent with the fact that in string
theory, they are obtained via the orientifold construction where $O7$-planes
are accompanied by eight flavour half D7-branes.

The second universality class is made out of the second-simplest
$\mathcal{N}=2$ theory: superconformal QCD for the three classical gauge
symmetries, with $\mathfrak{b}=A_1^{(0)}$:
\begin{equation}
		\lambda_\text{max} = 2 \cos(\frac{\pi}{2}) = 0:\qquad
	\begin{matrix}
	\begin{tikzpicture}
		\node[node, label=above:{\footnotesize $N$}, fill=black] (A1)  {};
		\node[rectangle, label=below:{\footnotesize $2N$}, fill=black] (F1) [below=6mm of A1]  {};
		\draw (A1.south) -- (F1.north);
	\end{tikzpicture}
	\end{matrix}\qquad
	\begin{matrix}
	\begin{tikzpicture}
		\node[node, label=above:{\footnotesize $N$}, fill=red] (A1)  {};
		\node[rectangle, label=below:{\footnotesize $2N-4$}, fill=blue] (F1) [below=6mm of A1]  {};
	    \draw (A1.south) -- (F1.north);
	\end{tikzpicture}
	\end{matrix}\qquad
	\begin{matrix}
	\begin{tikzpicture}
		\node[node, label=above:{\footnotesize $2N$}, fill=blue] (A1)  {};
		\node[rectangle, label=below:{\footnotesize $2(2N)+4$}, fill=red] (F1) [below=6mm of A1]  {};
	    \draw (A1.south) -- (F1.north);
	\end{tikzpicture}
	\end{matrix}
\end{equation}
They can be understood as decoupling limits of a quiver of shape
$\mathfrak{b}=A^{(1)}_1$, i.e. of a $\mathbb{Z}_2$ orbifold. Equivalently,
there is no gauge-gauge adjacency matrix for $A_1^{(0)}$, so that
$\lambda_\text{max}=0$ trivially.

Finally, the last class of theories has (anti-)symmetric representations. A
quiver with a single node, i.e. of shape $A_1^{(0)}$, and an (anti-)symmetric
representation unfolds to $\mathcal{U}(\mathfrak{b})=A_2^{(0)}$, so that
$\lambda_\text{max}=1$:
\begin{equation}\label{lambda_max=1-one-node}
		\lambda_\text{max} = 2\,\text{cos}(\frac{\pi}{3}) = 1:\qquad 
	\begin{matrix}
	\begin{tikzpicture}
		\node[node, label=above:{\footnotesize $N$}, fill=black] (A1)  {};
		\node[rectangle, label=above:{\footnotesize $1$}, fill=black] (F1) [right=8mm of A1]  {};
		\node[rectangle, label=below:{\footnotesize $N+2$}, fill=black] (F2) [below=6mm of A1]  {};
		\node[label=below:{\footnotesize $\bm{\Lambda^2}$}] (C) [right=4mm of A1]  {};
		\draw[decorate, decoration={snake}] (A1.east) -- (F1.west);
	    \draw (A1.south) -- (F2.north);
	\end{tikzpicture}
	\end{matrix}\qquad
	\begin{matrix}
	\begin{tikzpicture}
		\node[node, label=above:{\footnotesize $N$}, fill=black] (A1)  {};
		\node[rectangle, label=above:{\footnotesize $1$}, fill=black] (F1) [right=8mm of A1]  {};
		\node[rectangle, label=below:{\footnotesize $N-2$}, fill=black] (F2) [below=6mm of A1]  {};
		\node[label=below:{\footnotesize $\bm{S^2}$}] (C) [right=4mm of A1]  {};
		\draw[decorate, decoration={snake}] (A1.east) -- (F1.west);
	    \draw (A1.south) -- (F2.north);
	\end{tikzpicture}
	\end{matrix}
\end{equation}
This is the universality class related to affine decoupling of the
$\mathbb{Z}_3$ orientifold of $\mathcal{N}=4$ super-Yang--Mills. In the last
two cases, we therefore see that the classes correspond to decoupling limits of
a rank-one algebra. In the latter case, the presence of (anti-)symmetric
representations in the orbifold theory is associated with an O-plane. As we
have seen at the beginning of this section, the Hagedorn temperature is not
sensitive to those; these theories belong to the same universality class as the
decoupled unfolded theory.

The remaining seven theories studied in reference
\cite{Calderon-Infante:2024oed} have only $\mathcal{N}=1$ supersymmetry, and as we
have seen above, finding the temperature is done similarly for such theories. They further admit a Hanany--Witten realisation in terms of 1, 2, or 3 NS5-branes \cite{Calderon-Infante:2026zmd}, which correlates with the rank of the algebras $\mathfrak{b}$ in the $\mathcal{N}=2$ cases.
One then straightforwardly finds that they all have adjacencies mimicking
$\lambda_\text{max}\in\{2,0,1\}$. In the same reference, it was shown that the Hagedorn temperature sets the exponential rate $\kappa$ of the
Distance Conjecture, see equation \eqref{CDC-decay}. As we will see in Section
\ref{sec:bound-rate}, this is a particular feature of the fact that there is a
single gauge group. Generically, $\kappa$ will depend on all the eigenvalues of
the adjacency matrix, not only the largest.

\subsection{Holography}\label{sec:holography}

An $\mathcal{N}=2$ SCFT obtained as the worldvolume theory of a stack of
D3-branes probing a $\mathbb{C}\times \mathbb{C}^2/\Gamma$ orbifold is well
known to have a weakly-coupled Einstein gravitational dual given by type-IIB
string theory on $\text{AdS}_5\times S^5/\Gamma$ with $N$ units of $F_5$-flux; in
the presence of O-planes, one obtains
$\text{AdS}_5\times\mathbb{R}\text{P}^5/\Gamma$ instead \cite{Kachru:1998ys,
Aharony:1998rm, Witten:1998xy, Kakushadze:1998tr, Kakushadze:1998tz}.

More generally, for a four-dimensional CFT to admit a weakly-coupled Einstein
supergravity description on an $\text{AdS}_5$ background, it must satisfy
various consistency conditions. Some of these constraints are related to the
central charges $a$ and $c$, corresponding to the coefficients of the Weyl
anomaly \cite{Duff:1993wm}. For $\mathcal{N}=2$ Lagrangian theories, they are
given in terms of the numbers $n_\text{H}, n_\text{V}$ of $\mathcal{N}=2$ \emph{full} hyper-
and vector multiplets:
\begin{equation}\label{central-charges-def}
		a = \frac{1}{24}(5n_\text{V} + n_\text{H})\,,\qquad
		c = \frac{1}{24} (4 n_\text{V}  + 2n_\text{H})\,,\qquad
		24(a-c) = n_\text{V} - n_\text{H}\,.
\end{equation}
Via the standard AdS/CFT dictionary, the central charge $c$ is related to the
ratio between the AdS radius $L_\text{AdS}$ and the five-dimensional Newton
constant, $c\sim L^3/G_N$ up to order-one factors. To have a weakly-coupled
gravity dual $G_N\ll L_\text{AdS}^3$, the central charge must therefore be large.
Furthermore, the quantity $(a-c)$ can be shown to control higher-order
curvature terms, which must be suppressed in order to obtain Einstein gravity.
A candidate dual must therefore have $a\sim c\gg1$, which, as can be seen above,
corresponds to a large-$N$ limit where $n_\text{H}\sim n_\text{V}$.

In this section we briefly reformulate those constraints in Lie-algebraic
terms, and discuss some of the holographic consequences of the universality
classes defined by the Hagedorn temperature. For quivers with only
bifundamental hypermultiplets, the number of hyper- and vector multiplets is
given by  
\begin{equation}
		n_\text{V} = \sum_{i}\text{dim}(\mathfrak{g}_i) \simeq \frac{1}{2}k_i D^{ij} k_{j}\,,\qquad 
		n_\text{H} = \frac{1}{4}B^{ij}k_ik_j + \frac{1}{2}D^i_j k_if^j\simeq n_\text{V}+\frac{1}{4}k_iD^i_jf^j\,,
\end{equation}
where indices are raised and lowered with $\delta^{ij}$, and we have used the
SCFT condition \eqref{beta-function-bif} and the quantities defined in Table
\ref{tab:classical-quantities} to write an expression working for all types of
gauge and flavour algebras. We have ignored terms proportional to $S^i$ as
they will always be subleading, but they can be re-established easily to find
finite-$N$ expressions, albeit slightly more complicated ones. We also focus on
bifundamental hypermultiplets for simplicity, but the inclusion of
(anti-)symmetric representations is straightforward.

Let us now consider affine and finite quivers separately. We will closely
follow the arguments made in reference \cite{Calderon-Infante:2026rkj} for
unitary quivers. As we will see, the presence of non-unitary gauge algebras
does not significantly change their results, and will generalise it to all
large-$N$ $\mathcal{N}=2$ quiver SCFTs.

\paragraph{Affine Quivers:} As we have seen around equation
\eqref{affine-level}, affine quivers can have very little flavour symmetry, and
the vector $f^i$ is associated with an affine coweight at fixed level. In the
brane construction, this is explained through the presence of the eight half
D7-branes. A corollary is that, labelling the affine node by $i=0$, the
SCFT condition can be inverted up to an integer parameter $N$:
\begin{equation}
		k_i = N\, K_i + X_{ij}(f^j-4S^j)\,,\qquad X_{0i}=0\,, \quad X_{ij}=(C^{-1})_{ij}\,,\,i,j\neq0\,,
\end{equation}
where $K_i$ is the right null vector of the Cartan matrix, $C^{ij}K_j=0$. The
integers $K_i$ are the so-called ``Dynkin comarks'' of $\mathfrak{b}$
\cite{Kac:1990gs}. We therefore find that at leading order in $N$,
\begin{equation}\label{affine-nV}
		n_\text{V} \simeq \frac{1}{2}(K\cdot D\cdot K)\,N^2\,.
\end{equation}
For ADE algebras, it is well known that the sum of the squared comarks is equal to the
order of the McKay dual group, $\sum_iK_i^2=|\Gamma|$ and we have $n_\text{V}=
|\Gamma| N^2$ for fully unitary quivers \cite{Lawrence:1998ja}, and
$n_\text{V}= \frac{1}{2} |\Gamma| N^2$ otherwise. For non-simply-laced
algebras, by inspection one finds that $n_\text{V}$ is also set by the order of
an ADE discrete group corresponding to an unfolded or untwisted algebra.

As a result, since the flavour symmetry cannot be large, we have
$n_\text{V}\sim n_\text{H}$ and we conclude that affine quivers are always
holographic:
\begin{equation}\label{affine-nh-nv}
		\frac{n_\text{H}}{n_\text{V}} = 1 + \mathcal{O}(\frac{1}{N})\,.
\end{equation}
As mentioned above already, this is a well-known result, and unsurprising as
affine quivers arise from a stack of $N$ D3-branes probing an
orientifold singularity $\mathbb{C}\times\mathbb{C}^2/\Gamma$, whose dual is
described by type-IIB string theory on an $\text{AdS}_5\times S^5/\Gamma$
background.

All affine quivers are in the same universality class and share the same
Hagedorn temperature as $\mathcal{N}=4$ super-Yang--Mills. In the bulk, we
therefore expect an exponential growth of single-particle states corresponding
to closed-string modes. In the supergravity regime, the spectrum can be found
through a sigma model associated with a supercoset \cite{Metsaev:1998it}. This
description breaks down in the overall-free limit, which is equivalent to
$g_s\to0$, where we instead have a tensionless string propagating in
$\text{AdS}_5\times S^5$ \cite{Gaberdiel:2021jrv, Gaberdiel:2021qbb}. The
orbifold procedure will remove a finite fraction of these states, while adding
finitely many twisted modes at a given excitation level
\cite{Gaberdiel:2022iot}. As we are considering the large-$N$ limit, these
changes will only affect the sub-leading growth of states, the leading Hagedorn
behaviour remaining the same.

\paragraph{Finite Quivers:} Conversely, the flavour symmetry is much less
constrained for quivers with $\mathfrak{b}$ of finite type, and can be of order
$N$. The simplest example is the affine-node decoupling: since the gauge
symmetry of the affine theory scales like $N$, so will the flavour of the
resulting quiver, see for example equation \eqref{example-affine-decoupling}.
Using the SCFT condition, up to subleading terms we have
\begin{equation}\label{finite-nh-nv}
		\frac{n_\text{H}}{n_\text{V}} \simeq 1 + \frac{1}{2}R_\mathfrak{b}\,,\qquad R_\mathfrak{b} = \frac{k_i (DC)^{ij}k_j}{k_iD^{ij}k_j}\,.
\end{equation}
As was already observed in reference \cite{Calderon-Infante:2026rkj} for linear
quivers, the quantity $R_\mathfrak{b}$ is the Rayleigh quotient of the Cartan
matrix $C$, which can be used to test whether finite quivers are holographic.
It satisfies $R_{\mathfrak{b}}(a\,k) = R_{\mathfrak{b}}(k)$, and we can see
that, rotating in an eigenbasis, it is bounded by the largest and smallest
eigenvalues. Using the \emph{beautiful sine formula} of Theorem
\ref{thm:sin-formula} we then have 
\begin{equation}
		4\sin^2(\frac{\pi}{2h})\leq R_{\mathfrak{b}} \leq 4\sin^2(\frac{(h-1)\pi}{2h})\,,
\end{equation}
where we have used that the smallest and largest exponents are $m_1=1$ and
$m_r=h-1$, respectively. Observe that contrary to the temperature where we had
unfolded ADE algebras, here the Cartan matrix can be that of a non-simply-laced
algebra, and one must use the Coxeter number $h$ rather than its dual $h^\vee$,
see Table \ref{tab:exponents}. Using trigonometric identities, we can rewrite
this bound in terms of the largest eigenvalue $\lambda_\text{max}$ of the
adjacency $A = 2\,\mathbf{1}-C$:
\begin{equation}\label{bound-rayleigh}
		2-\lambda_\text{max} \leq R_{\mathfrak{b}} \leq 2+\lambda_\text{max}\,,
		\qquad \lambda_\text{max}=2\,\text{cos}(\frac{\pi}{h_{\mathfrak{b}}})\,.
\end{equation}
This bound will be very useful when we consider partial-decoupling limits in
Section \ref{sec:bound-rate}, and it has two important properties. First, it is
also valid in the affine case, where $\lambda_\text{max}=2$ and equation
\eqref{finite-nh-nv} trivially saturates the lower bound. More importantly
we have $R_{\mathfrak{b}}\geq 2-\lambda_\text{max}$ for all Lie algebras.
Furthermore, in the presence of an (anti-)symmetric representation on e.g. the
first node, we have
\begin{equation}
		n_\text{H} = \frac{1}{4} B^{ij}k_ik_j + \frac{1}{2}D^i_j k_i f^j + \frac{1}{2}k_1(k_1\pm1) \simeq \frac{1}{4}(DA^+)^{ij}k_ik_j + \frac{1}{2}D^i_j k_if^j\,.
\end{equation}
This means that these representations can once again be traded for a quiver
with only bifundamental hypermultiplets, but a different shape. As the largest
eigenvalue of the adjacency matrix $A^+$ is the same as that of the doubled
algebra, see equation \eqref{determinant-symmetric-reps}, we have
$\lambda_\text{max}(A^+) = \lambda_\text{max}(A_\text{doubled})$ and we can use
the bound on the Rayleigh quotient in equation \eqref{bound-rayleigh} also when
(anti-)symmetric representations are involved.

As we alluded to above, equation \eqref{bound-rayleigh} generalises the bound
found in reference \cite{Calderon-Infante:2026rkj} for unitary quivers to all
$\mathcal{N}=2$ quivers, as the Rayleigh quotient is not affected by the
diagonal matrix $D$. There, it was used to show that $\lambda_\text{max}\leq
2$. However, from our perspective, this bound is reached directly from the
properties of Cartan matrices.

From this bound, we see that large-$N$ finite quivers do not admit a
weakly-coupled gravity dual. The simplest example is $\mathcal{N}=2$
$\mathfrak{su}(N)$ superconformal QCD, which has $n_\text{H}\sim 2n_\text{V}$
since conformality imposes $f = 2N$. On the other hand, for a large number of
gauge nodes we can get theories with $a\sim c$. Indeed, a better upper bound
can be found using the SCFT condition, $k_i = (C^{-1})_{ij}\mu^j$ with $\mu^i =
f^i-4S^i$. Since $DC$ is symmetric one finds,
\begin{equation}\label{Rmax}
	R_{\mathfrak{b}} = \frac{\mu\cdot D \cdot C^{-1}  \cdot \mu}{\mu \cdot D \cdot C^{-2} \cdot \mu}\,\leq\, R_{\mathfrak{b}}^\text{max} = \text{max}_{i,j} \frac{C^{-1}_{ij}}{C^{-2}_{ij}}\,,
\end{equation}
where we used that since $\mu^i\geq0$ and $C^{-1}$ has only non-negative
entries, then $R_{\mathfrak{b}}^\text{max} DC^{-2}- DC^{-1}\geq0$ entrywise.
The values of $R_{\mathfrak{b}}^\text{max}$ are collated in Table
\ref{tab:exponents}, as we are not aware of a simple closed-form formula, but
by inspection we see that the dual Coxeter number gives a moderately weaker
bound $R_{\mathfrak{b}}\leq\frac{4}{h^\vee}$. This means that when the shape of
the quiver is of classical type, we can take a limit where it becomes very long
and $r\gg1$, in which case we have $n_\text{H}\simeq n_\text{V}$.
 In that limit the
Hagedorn temperature approaches that of $\mathcal{N}=4$, as the maximal
eigenvalue is given by
\begin{equation}
		\lambda_\text{max} = 2\cos(\frac{\pi}{h})\quad\xrightarrow{r\to\infty}\quad 2 = \lambda_\text{max}^{\mathcal{N}=4}\,.
\end{equation}

This can be thought of as the fact that, as the quiver gets longer and longer,
it approximately looks like a long necklace quiver, and the information about
its endpoints (such as BCD-type trivalent or non-simply-laced ends) becomes
negligible. 

We however caution against interpreting the associated tensionless
string as the same as that of $\text{AdS}_5\times S^5$. While they have the
same temperature, the duals are different. For the $\mathfrak{b}=A_{r}^{(0)}$
unitary quiver for instance, type-IIA realisation of the CFT via Hanany--Witten
setups has been discussed in the literature
\cite{Aharony:2012tz}---see also reference \cite{Nunez:2019gbg} for details
about the holographic dictionary of various quantities and marginal
deformations---but the supergravity approximation can only be trusted when
$r\gg1$. These backgrounds are however very different from $\text{AdS}_5\times S^5$ and
its orbifolds, and it is not clear whether in the free-field limits these
will lead to the same type of tensionless string in the bulk. 

As has been noted in reference \cite{Calderon-Infante:2026rkj} the Hagedorn
temperature is bounded from below by the value of $\mathcal{N}=4$
super-Yang--Mills, following from $\lambda_\text{max}\leq2$, a value that is also saturated
for theories with $a\sim c$ and therefore have a weakly-coupled holographic
dual. Furthermore, as alluded to above, if a large number of
flavour symmetries is present, the spectrum will not be sparse, a necessary
condition for controlled holography. While, as noted in references
\cite{Gadde:2009dj, Calderon-Infante:2024oed}, a sparse spectrum can be obtained
by projecting on the flavour-singlet sector, considering these qualitative
differences with $\mathcal{N}=4$ super-Yang--Mills and its orbifolds, we expect
the bulk tensionless strings to be different.

\subsubsection{Interlude: Little String Theories and Double-Scaling Limits}

As we have seen, quivers associated with orientifold projections of
$\mathcal{N}=4$ super-Yang--Mills have $a\sim c$, and therefore have a clear
interpretation in terms of a fundamental critical string in the holographic
dual. On the other hand, finite quivers have different large-$N$ central
charges, and the bulk strings are harder to identify. Here, we briefly discuss
how both classes of quivers can be engineered in terms of F-theory on the
field-theory side, and how this framework provides a unifying picture that
generalises the double-scaling limits that have been used to put forward a
``little-string holography'' interpretation of the bulk duals of finite
quivers~\cite{Gadde:2009dj}.

We have seen in Section \ref{sec:affine-decoupling} that finite quivers can be
obtained by decoupling a node from an affine quiver. One possible
interpretation for the resulting string can be that close to the
partial-decoupling point of the conformal manifold, the fundamental string
splits into a collection of non-critical or field-theory strings. This is a
known phenomenon: for instance the $E_8\times E_8$ heterotic string can be
thought of as the combination of two field-theory E-strings
\cite{Haghighat:2014pva}.

This mechanism is particularly striking when lifting the SCFTs considered in
this work to six dimensions, where they can be understood as deformations of
Little String Theories (LSTs). LSTs are six-dimensional, UV-complete,
non-local, non-gravitational theories, and can be realised as the worldvolume
theories of NS5-branes \cite{Witten:1995gx}. 

LSTs are indeed closely related to four-dimensional $\mathcal{N}=2$ quiver
SCFTs. As six-dimensional theories, the former can be engineered via F-theory
compactifications on an elliptically-fibred Calabi--Yau three-fold $Y_3\to B_2$
\cite{Bhardwaj:2015oru}, where the base $B_2$ is taken to be non-compact in
order to decouple gravity. Wrapping D3-branes on curves $\Sigma_i\subset B_2$
leads to the celebrated 6d BPS string, which becomes tensionless when
$\Sigma^i$ shrinks to zero volume. The eponymous little string is then
characterised by a curve of self-intersection zero:
\begin{equation}
		\text{Little string: D3-brane on}\quad \Sigma_\text{LST} = \ell_i\, \Sigma^i\,,\qquad \Sigma_\text{LST}\cdot\Sigma_\text{LST} = 0\,,
\end{equation}
where $\Sigma^i$ form a basis of curves in the base $B_2$ and $\ell_i$ is the null
vector of the intersection matrix:
\begin{equation}
		\eta^{ij}=\Sigma^i\cdot\Sigma^j\,,\qquad \eta^{ij}\,\ell_j =0\,.
\end{equation}
When the volume of all other curves in the base can be sent to zero, one
obtains a singular Calabi--Yau resulting in a Little String Theory in six
dimensions, and where the volume of the curve $\Sigma_\text{LST}$ sets the
tension of the little string. If there are no curves of self-intersection zero
in the base, one instead obtains a 6d SCFT at the singular point, as the limit
of vanishing volume removes every dimensionful parameter in that case. The
gauge data is then dictated by the type of the elliptic fiber over each curve,
and matter arising from fibers-singularity enhancements at their intersection.
We defer to e.g. the reviews \cite{Heckman:2018jxk, DelZotto:2023ahf} for an
introduction to the geometric construction.

It turns out that demanding consistency of the Calabi--Yau---or equivalently
cancellation of gauge anomalies of the six-dimensional field theory---is
closely related to the SCFT condition \eqref{cft-condition} in four dimensions
\cite{Blum:1997mm, Blum:1997fw, Bhardwaj:2015xxa, Baume:2024oqn,
Ahmed:2025fmq}. In fact, the intersection pairing $\eta^{ij}$ is itself related
to a Cartan matrix, and the curves associated with six-dimensional gauge
theories must also intersect like a Dynkin diagram: those of affine type lead
to an LST, while finite Dynkin diagrams correspond to 6d SCFTs.

Away from the singular point---i.e. a generic point of the tensor branch where
all curves have finite volume---we have a weakly-coupled gauge theory, and
further compactifying on a torus $T^2$, one then obtains the four-dimensional
$\mathcal{N}=2$ quiver SCFTs discussed in this work.\footnote{In six
		dimensions, demanding cancellation of gauge anomalies for theories with
		classical algebras and an intersection pairing related to a Cartan
		matrix $C^{ij}$ gives the constraint $C^{ij}k_j=f^i-16 S^{i}$. On the
		other hand the 4d CFT condition in equation \eqref{beta-function-bif}
		sets $C^{ij}k_j=f^i - 4 S^{i}$. For unitary quivers, they are therefore
		the same, but for quivers involving $\mathfrak{so}(k)$ or
		$\mathfrak{usp}(k)$ gauge algebras, they are slightly different. Upon
compactification on a torus, for the former we directly reach an SCFT, while
for the latter one obtains an $\mathcal{N}=2$ quiver SQFT that then flows to a
Lagrangian SCFT.}

In the previous section, we have seen that finite quivers can be obtained by
decoupling a node of an affine quiver. Similarly, six-dimensional SCFTs can be
obtained by going to an infinite-distance point of the K\"ahler moduli space of
the Calabi--Yau $Y_3$ associated with an LST that is equivalent to removing one
of the curves $\Sigma_i$ \cite{Bhardwaj:2015oru}. Doing so, one is then unable
to find a combination of the remaining curves $\Sigma_\text{LST}$ with
self-intersection zero. The resulting six-dimensional theory is associated with
an SCFT rather than an LST.

As an example, consider three curves $\Sigma^0, \Sigma^1,\Sigma^2\subset B_2$
whose intersection pairing is that of (minus) the Cartan matrix of $A_2^{(1)}$,
$\eta^{ij}=-C^{ij}$, see below for a pictorial description. It is straightforward to see
that the curve $\Sigma_\text{LST} = \Sigma^0 + \Sigma^1 + \Sigma^2$ has
self-intersection zero, and the F-theory construction leads to a
six-dimensional LST. Performing a deformation where
$\text{Vol}(\Sigma^0)\to\infty$, we are removing this curve from the base
$B_2$:
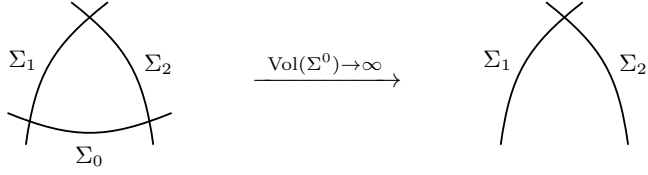
\begin{equation*}
	\begin{matrix}\begin{tikzpicture}[
	  component/.style={
	    draw=black,
		line width=0.7pt,
	    line cap=round,
	    shorten <= -3mm,
	    shorten >= -3mm
	  },
	  curve label/.style={
	    font=\small,
	    fill=white,
	    inner sep=1.2pt
	  }
	]
	\path[use as bounding box] (-1.35,-1.90) rectangle (1.35,0.35);
	  \node[coordinate] (A1) at (0,0) {};
	  \node[coordinate, below left=13.9mm and 8mm of A1] (A2) {};
	  \node[coordinate, right=16mm of A2] (A3) {};
	  \draw[component]
	    (A1) to[bend right=22]
	    node[curve label, pos=0.48, left=1mm] {$\Sigma_1$}
	    (A2);
	  \draw[component]
	    (A2) to[bend right=22]
	    node[curve label, pos=0.50, below=1mm] {$\Sigma_0$}
	    (A3);
	  \draw[component]
	    (A3) to[bend right=22]
	    node[curve label, pos=0.48, right=1mm] {$\Sigma_2$}
	    (A1);
	\end{tikzpicture}\end{matrix}
	%
	\qquad\xrightarrow{~\text{Vol}(\Sigma^0)\to \infty~}\qquad
	\begin{matrix}\begin{tikzpicture}[
	  component/.style={
	    draw=black,
		line width=0.7pt,
	    line cap=round,
	    shorten <= -3mm,
	    shorten >= -3mm
	  },
	  curve label/.style={
	    font=\small,
	    fill=white,
	    inner sep=1.2pt
	  }
	]
	\path[use as bounding box] (-1.35,-1.90) rectangle (1.35,0.35);
	  \node[coordinate] (A1) at (0,0) {};
	  \node[coordinate, below left=13.9mm and 8mm of A1] (A2) {};
	  \node[coordinate, right=16mm of A2] (A3) {};
	  \draw[component]
	    (A1) to[bend right=22]
	    node[curve label, pos=0.48, left=1mm] {$\Sigma_1$}
	    (A2);
	  \draw[component]
	    (A3) to[bend right=22]
	    node[curve label, pos=0.48, right=1mm] {$\Sigma_2$}
	    (A1);
	\end{tikzpicture}\end{matrix}
\end{equation*}
The resulting two curves then intersect like the finite Dynkin diagram
associated with $A_2^{(0)}$, and it is easy to show that there are
no combinations of curves $a_1 \Sigma^1+a_2 \Sigma^2$ with self-intersection
zero. At the singular point where $\text{Vol}(\Sigma_{1,2})\to0$, we then
obtain a six-dimensional SCFT called type-A conformal matter
\cite{Heckman:2018jxk}, as advertised above.  The same procedure holds away
from the singular point of the moduli space, where all curves have finite
volume, and which admits a weakly-coupled description in terms of a
six-dimensional gauge theory. 

Further reducing on a torus $T^2$, we obtain a four-dimensional quiver
with gauge coupling 
\begin{equation}\label{coupling-to-volume}
		\frac{1}{g_{\text{4d},i}^{2}} \propto \text{Vol}(T^2)\,\text{Vol}(\Sigma^i)\,,
\end{equation}
where the type of gauge symmetry is determined by the type of elliptic fiber
over each curve $\Sigma^i$. If they are all taken to be associated with the
gauge algebra $\mathfrak{su}(N)$, before decoupling we have the unitary affine
quiver of type $\mathfrak{b}=A_2^{(1)}$. We therefore see that the affine-node
decoupling has a geometric avatar in terms of the volumes of the curves in the
base. After reducing on the torus, this is equivalent to the operation we have
discussed in Section \ref{sec:affine-decoupling}: 
\begin{equation}
	\begin{matrix}\begin{tikzpicture}
		\node[node, label=above:{\footnotesize $N$}, fill=black] (A1) at (0mm,0mm) {};
	    \node[node, label=above:{\footnotesize $N$}, fill=black] (A2) at (8mm,0mm) {};
	    \node[node, label=below:{\footnotesize $N$}, fill=black] (A0) at (4mm,-5.93mm) {};
	    \draw (A1.east) -- (A2.west);
	    \draw (A1.south east) -- (A0.north west);
	    \draw (A2.south west) -- (A0.north east);
	\end{tikzpicture}\end{matrix}
	\qquad\xrightarrow{~g_{\text{4d},0}\to 0~}\qquad
	\begin{matrix}\begin{tikzpicture}
		\node[node, label=above:{\footnotesize $N$}, fill=black] (A1) at (0mm,0mm) {};
	    \node[node, label=above:{\footnotesize $N$}, fill=black] (A2) at (8mm,0mm) {};
	    \node[rectangle, label=below:{\footnotesize $N$}, fill=black] (F1) at (0mm,-5.93mm) {};
	    \node[rectangle, label=below:{\footnotesize $N$}, fill=black] (F2) at (8mm,-5.93mm) {};
	    \draw (A1.east) -- (A2.west);
	    \draw (A1.south) -- (F1.north);
	    \draw (A2.south) -- (F2.north);
		\end{tikzpicture}\end{matrix}
\end{equation}

In the type-II string realisation, the gauge couplings are also related to the
string coupling $g_s^{-1}\propto g^{-2}_0 + g^{-2}_1 + g^{-2}_2$. In the
F-theory picture, this constraint arises from the fact that the volumes of the
curves control these couplings, and that $\Sigma_\text{LST}$ is a combination of
the other curves:
\begin{equation}\label{vol-lst-constraint}
		\text{Vol}(\Sigma_\text{LST}) = \ell_i\,\text{Vol}(\Sigma^i)\,.
\end{equation}
This simple example can be generalised straightforwardly to more involved
cases, in particular all theories with classical gauge algebras. 

Furthermore,
the F-theory framework incorporates (anti-)symmetric representations as well,
as the matter spectrum is controlled by singularity enhancements at the
intersection of the curves. Choosing the type of elliptic fibers appropriately,
all possible gauge representations for hypermultiplets that we have discussed in
the previous sections can be obtained easily. In general, different matter and
gauge content associated with a particular curve will change its
self-intersection, and therefore the null vector $\ell_i$, meaning that quivers
with the same shape but different gauge data will have generically different
constraints on $\Sigma_\text{LST}$ via equation \eqref{vol-lst-constraint}.

Intriguingly, in the previous section we have seen that all but three finite
quivers can be obtained by decoupling a node of an affine quiver. Their 6d
lifts are exactly the quivers associated with SCFTs discussed in reference
\cite{Bhardwaj:2019hhd}, and that cannot be obtained through a deformation of
an LST. In all other cases, a finite six-dimensional quiver associated with an
SCFT can be reached by decoupling the zero curve of an LST. However, to our
knowledge there is no known explicit geometric realisation for these theories,
which are expected to be part of the frozen phase of F-theory. If these outliers
could be shown to be inconsistent, this would mean that all six-dimensional
gauge theories and by extension all four-dimensional SCFTs discussed in this
work can be obtained by a deformation of an LST, or a compactification thereof.

For LSTs the little string itself, obtained by wrapping a D3-brane on the
zero-curve $\Sigma_\text{LST}$, can be related to a solitonic fundamental
type-II or heterotic string in a dual frame \cite{Bhardwaj:2015xxa,
		Bhardwaj:2015oru, DelZotto:2023ahf, DelZotto:2022ohj, Baume:2024oqn,
Ahmed:2025fmq, Bhardwaj:2019hhd}. After compactification, we therefore
naturally expect a critical fundamental string in the $\text{AdS}_5\times
S^5$ bulk. After partial decoupling however, the absence of one of the curves
prohibits the existence of $\Sigma_\text{LST}$, and we instead have a
collection of six-dimensional field-theory BPS strings, and identifying the
bulk string is more arduous. 

\paragraph{The Double-Scaling Limit and Decoupling:} As mentioned above,
finite quivers do not have a known bulk dual realisation. However, it has
been proposed that the flavour-singlet sector of some of these theories can be
obtained via ``little-string holography'' using a particular double-scaling
limit. For $\mathcal{N}=2$ superconformal QCD in the Veneziano
limit---corresponding to $\mathfrak{b} = A_{1}^{(0)}$ with $N\to\infty$, and
$g^2 N$ fixed in our notation---this is achieved by starting with the
$\mathbb{Z}_2$ orbifold theory, and considering the following limit
\cite{Gadde:2009dj}:
\begin{equation}
		\mathbb{Z}_2~\text{orbifold}:\qquad	g_0\to0\,,\qquad g_1~\text{fixed}\,,\qquad \frac{1}{g_0^2} + \frac{1}{g_1^2} = \frac{1}{g_s}\to \infty\,,
\end{equation}
with $g_0$ the gauge coupling associated with the affine node.

In the Hanany--Witten picture, this limit corresponds to collapsing two
NS5-branes on top of each other, leading to a worldsheet description in terms
of a linear dilaton model on the background
\begin{equation}\label{chs-background}
		\mathbb{R}^{1,5}\times \frac{SL(2,\mathbb{R})_2}{U(1)}/\mathbb{Z}_2\,.
\end{equation}
It has been proposed that when taking the backreaction of D3- and D5-branes
into account, this leads to the holographic dual of conformal SQCD
\cite{Gadde:2009dj}.

While reproducing this result and generalising it to more involved quivers is
certainly beyond the scope of this work, see e.g. references
\cite{Dei:2024frl, Dei:2025ilx, Calderon-Infante:2026zmd} for recent advances on the
topic, we would like to point out that this double-scaling limit is easily
implemented for all possible affine-decoupling limits via F-theory. 

Starting with an LST and keeping the extra torus as a spectator, we can obtain
the finite quiver as follows. First, we need to decouple the string modes.
These are controlled by the little-string scale $M^2_{s}\propto
\text{Vol}(\Sigma_\text{LST})$ which must be taken to be large. Decoupling
the affine node, $g_0\to 0$, then requires us to take the large-volume limit of
the corresponding curve, $\text{Vol}(\Sigma^0)\to\infty$, see equation
\eqref{coupling-to-volume}. As the LST curve is obtained as a combination of
the $\Sigma^i$, we must also ensure that the remaining gauge couplings of the
four-dimensional quiver remain finite. All in all, this leads us to the
following limit:
\begin{equation}\label{f-theory-double-scaling}
		\frac{\text{Vol}(\Sigma^{i\neq0})}{\text{Vol}(\Sigma_\text{LST})}\to 0\,,\qquad
		\frac{\text{Vol}(\Sigma^{0})}{\text{Vol}(\Sigma_\text{LST})}\quad\text{fixed}\,,\qquad
		\Sigma_\text{LST} = \ell_i\,\Sigma^i\,,
\end{equation}
where $i=0$ corresponds to the index of the curve / 4d Yang--Mills coupling of
the node that is decoupled.

If a Hanany--Witten setup is available, this will correspond to collapsing
multiple NS5-branes on top of each other, the simplest case being the necklace
quivers with $\mathfrak{b}=A^{(1)}_r$. This generalises the $\mathbb{Z}_2$ and $\mathbb{Z}_3$
orbifolds discussed above, where $r+1$ NS5-branes are put on top of each other in
that limit \cite{Calderon-Infante:2026rkj}.

However, an additional advantage of the F-theory picture is that this limit can
be performed even when a brane construction is not available, and covers
quivers with exceptional shapes, and with all types of classical algebras.
Different gauge algebras will nonetheless lead to different intersection
matrices $\eta^{ij} = \Sigma^i\cdot \Sigma^j$ and therefore different
null vectors $\ell^i$ setting the zero-curve $\Sigma_\text{LST}$. 

Given a 4d quiver, its six-dimensional lift and the corresponding intersection
matrix are easily obtained from the triplet $(\mathfrak{b}, S^i, f^i)$, see e.g.
reference \cite{Ahmed:2025fmq}, and ultimately the double-scaling limit given
in equation \eqref{f-theory-double-scaling} only depends on its null vector
$\ell^i$. This therefore gives us a handle on the decoupling limit in terms of
simple Lie-algebraic data.

Finding a map between this data and the possible worldsheet model describing
part of the decoupled theory, as in equation \eqref{chs-background} for SQCD, is
on the other hand more involved. At least in principle however, one can obtain part
of the worldsheet data of a D3-brane wrapping a curve from the geometric data
in the F-theory picture, see e.g. \cite{Lawrie:2016axq} and references therein.
It would therefore be interesting to further study these decoupling limits
geometrically, and whether one can obtain a map between the triplet
$(\mathfrak{b}, S^i, f^i)$ and the corresponding background. Such a map would
certainly be an important step towards a better understanding of the
holographic picture, and the nature of the tensionless bulk strings at infinite
distance in moduli space.

\section{Bounds on Distance Conjecture Rates}\label{sec:bound-rate}

In Sections \ref{sec:classification-quivers} and \ref{sec:thermal} we have
focussed on a very particular point of the conformal manifold, namely the one
where the entire SCFT becomes free. We now make use of the Lie-algebraic
properties of $\mathcal{N}=2$ quiver theories to explore the properties of the
conformal manifold along trajectories where only a subsector of the theory
becomes free.

The standard way to define a notion of distance in the conformal manifold is
through the Zamolodchikov metric, defined in terms of the two-point function of
exactly-marginal operators $\mathcal{O}^i(x)$ associated with marginal
couplings $g^i$:
\begin{equation}
		\langle \mathcal{O}_i(x)\mathcal{O}_j(y)\rangle = \frac{\chi_{ij}(g)}{|x-y|^{8}}\,.
\end{equation}
For $\mathcal{N}=2$ four-dimensional SCFTs, the operator $\mathcal{O}_i$ must
be in the same superconformal multiplet as the operator corresponding to the
kinetic term $\text{Tr}F^2$ and therefore $(\mathcal{N}=2)$-preserving marginal
deformations are associated with complexified Yang--Mills couplings $\tau^i$.

Due to the presence of supersymmetry, the conformal manifold is endowed with
additional geometric structure and must be Hodge--K\"ahler \cite{Gomis:2015yaa,
Niarchos:2021iax}. For $\mathcal{N}=2$ SCFTs, whether Lagrangian or not, it was
furthermore shown that the Zamolodchikov metric can be obtained from the
partition function on $S^4$ \cite{Gomis:2014woa, Gerchkovitz:2014gta,
Niarchos:2021iax}, and it follows that the metric is always approximately
hyperbolic at weak coupling, see e.g. reference \cite{Baume:2020dqd}:
\begin{equation}
		ds^2 = \chi_{i\overline{\jmath}}\,d\tau^i\,d\overline{\tau}^{\bar{\jmath}} 
		\simeq \sum_{i=1}^{\tilde{r}}\frac{\text{dim}(\mathfrak{g}_i)}{8c}\frac{d\tau^i\,d\overline{\tau}^{\bar{\imath}}}{\text{Im}(\tau^i)^2} 
		= \sum_{i=1}^{\tilde{r}}\frac{\text{dim}(\mathfrak{g}_i)}{2c}\frac{(d g_i)^2}{(g_i)^2} \,,
\end{equation}
where we use the same conventions as in reference \cite{Perlmutter:2020buo} in
order to have Planck units in the bulk dual. Following reference
\cite{Calderon-Infante:2026rkj}, we introduce flat coordinates on the conformal
manifold:
\begin{equation}
		g_i\sim e^{-\kappa_i\, \hat{g}_i}\,,\qquad \kappa_i = \sqrt{\frac{2\,c}{\text{dim}(\mathfrak{g}_i)}}\,.
\end{equation}
These coordinates are useful to study the leading tensionless tower in the
bulk, which is the one associated with the theory becoming free, and the
infinite tower of massless states corresponds in the CFT to the operators
becoming higher-spin conserved currents. 

In the weakly-coupled regime, the anomalous dimension of higher-spin currents
receives quantum corrections at order $g_i^2$ in perturbation theory, as can be
seen from a diagrammatic approach. In the bulk, the mass of the higher-spin
tower associated with the $i$-th node then behaves as
\begin{equation}
		\frac{M_\text{HS}^i}{M_\text{Pl}}\sim \sqrt{\gamma_i}\,,\qquad \gamma_i\sim g_i^2 \sim \exp(-2\,\kappa_i \,\hat{g}_i)\,.
\end{equation}
If the rates are chosen such that $\gamma_i\sim\gamma_j$, they will be part of
the same leading tower, and therefore follow simple ``taxonomy rules''
\cite{Calderon-Infante:2020dhm, Castellano:2023stg, Castellano:2023jjt,
Etheredge:2024tok}, see reference \cite{Calderon-Infante:2026rkj} for a
discussion of unitary quivers. 

The root and coweight lattices of the algebra $\mathfrak{b}$ defining the shape
of the quiver can then be used to classify the possible infinite-distance
weak-coupling limits. Since each gauge coupling is associated with a node, a
decoupling limit associated with a particular leading tower corresponds to
selecting a subset of the nodes, or equivalently a choice of subalgebra
$\mathfrak{b}_\text{dec.}\subset \mathfrak{b}$. At the decoupling point, we
then have a free sub-quiver. As it is associated with the algebra
$\mathfrak{b}_\text{dec.}$, we have already computed its Hagedorn temperature
in Section \ref{sec:thermal}. The only subtlety is when
$\mathfrak{b}_\text{dec.}$ is semi-simple, i.e. when the free subquiver is the
disjoint union of two diagrams. In those cases, the Hagedorn temperature is set
by the largest $\lambda_\text{max}$ of each component.

We therefore see that in coordinates appropriate for comparison with the bulk,
the rate at which the higher-spin would-be currents become conserved is once
again set by group theory. We can use the results obtained in Section
\ref{sec:holography} to obtain bounds on the rate $\kappa_i$. Recall that at
large $N$, the central charge $c$ is given by
\begin{equation}
		24c \sim 6n_\text{V} + \frac{1}{2} k\cdot D\cdot f\,,\qquad n_\text{V}\sim \frac{1}{2}k \cdot D\cdot k\,.
\end{equation}

The exponential rates can then be written as 
\begin{equation}
		\kappa_i = \sqrt{\frac{2c}{\text{dim}(\mathfrak{g}_i)}} = \kappa_\text{min}\,\sqrt{\frac{n_\text{V}}{\text{dim}(\mathfrak{g}_i)}}\,,\qquad \kappa_\text{min} = \sqrt{\frac{2c}{n_\text{V}}} = \sqrt{\frac{1}{6} \bigg(2+\frac{n_\text{H}}{n_\text{V}}\bigg)} = \frac{1}{\sqrt{2}}\frac{1}{\sqrt{2\,\frac{a}{c} - 1}}\,.
\end{equation}
The coefficient $\kappa_\text{min}$ corresponds to the decay rate in the
overall-free limit and depends on the ratio $n_\text{H}/n_\text{V}$, or
equivalently $a/c$.
We see that whenever $a\sim c$
and the theory has a holographic dual we obtain
\begin{equation}
		a\sim c:\qquad \kappa_\text{min}=\frac{1}{\sqrt{2}}\,,
\end{equation}
which is the proposed lower bound for the exponential rate in the CFT Distance
Conjecture \cite{Perlmutter:2020buo}.

We investigated these ratios defining $\kappa$ in Section \ref{sec:holography} when discussing
the possibility of holographic duals. There, we have found that they can
be related to the Rayleigh quotient of the Cartan matrix $C$ of $\mathfrak{b}$,
which in turn obeys a lower bound given in terms of the largest eigenvalue
$\lambda_\text{max}$ of the adjacency matrix $A=2\,\mathbf{1}-C$, see equation
\eqref{bound-rayleigh}. Although it was found for finite algebras, it remains
valid when $\mathfrak{b}$ is affine. We reproduce it here for ease of reading: 
\begin{equation}
		\frac{n_\text{H}}{n_\text{V}} \geq 1 + \frac{1}{2}(2-\lambda_\text{max})\,,\qquad \lambda_\text{max}\leq 2\,.
\end{equation}
It is saturated for $\mathfrak{b}$ affine i.e. by $\mathcal{N}=4$
super-Yang--Mills and its orbifolds as $n_\text{V}\sim n_\text{H}$, and for
finite type we have $\lambda_\text{max}=2\,\text{cos}(\pi/h_{\mathfrak{b}})$.
Furthermore, as we have seen in Section \ref{sec:holography}, it is also valid
for theories that have (anti-)symmetric representations when considering the
appropriate doubled algebra. This leads to the universal bound:
\begin{equation}\label{alpha-min}
		\kappa_\text{min} \geq \sqrt{\frac{1}{2} + \frac{1}{12}(2-\lambda_\text{max})}\geq \frac{1}{\sqrt{2}}\,,
\end{equation}
where we have used that for finite quivers $\lambda_\text{max}<2$. The last
bound $\kappa_\text{min}\geq \frac{1}{\sqrt{2}}$ is saturated by
$\mathcal{N}=4$ super-Yang--Mills and its orientifolds at all values of $N$,
and is strictly larger for finite quivers
$\kappa_\text{min}>\frac{1}{\sqrt{2}}$, as expected \cite{Perlmutter:2020buo}.

We therefore see that the very same quantity $\lambda_\text{max}$ setting the
Hagedorn temperature also gives a lower bound on the rate at which the string
becomes tensionless in the bulk, and both are controlled by the algebra
$\mathfrak{b}$ giving the shape of the quiver.

On the other hand, $\kappa_\text{min}$ can be a very crude estimate of the
actual rate at which the string becomes tensionless. For instance, in the case
of affine unitary quivers, we have seen that the large-$N$ scaling of the gauge
symmetries is of the form $k_i\sim K_i\, N$ where the integers $K_i$ are the
``Dynkin comarks'' of the algebra $\mathfrak{b}$. It immediately follows that,
decoupling the $i$-th node of the quiver, the exponential rate is given by:
\begin{equation}
		\text{unitary affine}:\qquad \kappa_i = \sqrt{\frac{K\cdot K}{2\,K_i^2}}\,.
\end{equation}
As an example, for $\mathfrak{b}=E_8^{(1)}$, the largest possible comark
is $\text{max}(K_i)=6$. We then have $K\cdot K= |\Gamma_{\widehat{E}_8}|=120$ so that $\kappa_i \sim 1.3$. For
exceptional shapes, $\kappa_i$ will always be roughly of order one, but strictly
larger than $\frac{1}{\sqrt{2}}$. For the classical cases on the other hand it
can be arbitrarily large. For instance with $\mathfrak{b}=A_r^{(1)}$, we find $\kappa_i
\sim \sqrt{r}$ and long quivers will generically have large exponential rates.

On the other hand, for finite quivers we also have an upper bound on the rate,
see equation \eqref{bound-rayleigh}: $\frac{n_\text{H}}{n_\text{V}} \leq 1 +
\frac{1}{2}(2+\lambda_\text{max})$. The resulting bound is however weaker than
$\kappa_{\text{min}}\leq \sqrt{2/3}$, which follows from the Hofman--Maldacena
\cite{Hofman:2008ar} conformal-collider bound on the central charges
\cite{Calderon-Infante:2026rkj}. Instead we can use the stronger bound on the
Rayleigh quotient discussed around equation \eqref{Rmax}. Substituting it in
the definition of $\kappa_i$, we find that 
\begin{equation}\label{upper-bound-finite}
		\text{finite:}\qquad \kappa_i = \kappa_\text{min}\sqrt{\frac{n_\text{V}}{\text{dim}(\mathfrak{g}_i)}}\,,\qquad \kappa_\text{min} \leq \sqrt{\frac{1}{2} + \frac{1}{12}R_{\mathfrak{b}}^\text{max}}\leq\sqrt{\frac{1}{2} + \frac{1}{3h^\vee_{\mathfrak{b}}}}\,,
\end{equation}
where $R^\text{max}_{\mathfrak{b}}$ is given in Table \ref{tab:exponents}. The
bound given in terms of $h^\vee_{\mathfrak{b}}$ is not sharp, as
$R^\text{max}_{\mathfrak{b}}\leq4/h^\vee_{\mathfrak{b}}$ is a correct but weak
estimator for $R^\text{max}_{\mathfrak{b}}$. However, the bounds given in
equation \eqref{upper-bound-finite} are an improvement over
$\kappa_\text{min}\leq \sqrt{2/3}$ in all cases, although the latter is valid
for $\mathcal{N}=1$ theories as well. 

As for the affine case, this does not prevent us from obtaining a large decay rate:
if the number of nodes is large, the ratio between $n_\text{V}$ and the
dimension of the decoupling gauge symmetry can be made arbitrarily large.

\paragraph{The One-Node Case:} Equation \eqref{alpha-min} also gives an
explanation as to the observation in reference \cite{Calderon-Infante:2024oed}
that for rank-one theories the bound is set by the quantity $a/c$, which also
sets the temperature. This ratio is equivalent to $n_\text{H}/n_\text{V}$, and if the quiver has only one node, the Rayleigh quotient
$R_\mathfrak{b}$ is that of a $1\times1$ matrix and therefore has a single
eigenvalue. The bound $R_{\mathfrak{b}}\geq 2 - \lambda_\text{max}$ is then
trivially saturated, and equation \eqref{alpha-min} is an equality. Since
$\lambda_\text{max}$ is the very quantity setting both the Hagedorn temperature
and the bound on $\kappa_\text{min}$, we can only have the three classes
obtained in that work: the universality class of $\mathcal{N}=4$
super-Yang--Mills has $\lambda_\text{max}=2$, the one of SQCD ($\mathfrak{b}=A_1^{(0)}$) has $\lambda_\text{max}=0$. The subtlety lies in the
third class, which has $\lambda_\text{max}=1$ as the quiver has an
(anti-)symmetric representation and mimics the case with
$\mathfrak{b}=A_2^{(0)}$.

\paragraph{$\mathcal{N}=1$ Quivers:} We expect a similar behaviour at large $N$
for $\mathcal{N}=1$ quiver theories. The central charges can be expressed in
terms of the number of $\mathcal{N}=1$ chiral and vector multiplets similarly
to equation \eqref{central-charges-def}. As we have seen in Section
\ref{sec:N=1}, the possible adjacency matrices $A$ are not as constrained as
for $\mathcal{N}=2$ quivers, but the number of chiral multiplets can also be
written in terms of a Rayleigh quotient related to $A$, and therefore
$\kappa_\text{min}$ will ultimately be bounded by the largest eigenvalue
$\lambda_\text{max}$ of $A$ similarly as in equation \eqref{alpha-min}. For the
one-node cases, we have an equality instead of a bound, and we again have that
both the Hagedorn temperature and $\kappa_\text{min}$ are set by the same
quantity $\lambda_\text{max}$, completing the explanation of the mini-landscape
of reference \cite{Calderon-Infante:2024oed} we started in Section
\ref{sec:mini-landscape}. However, in the absence of a full classification of
$\mathcal{N}=1$ quiver SCFTs and their adjacencies, it is difficult to state
whether a stronger lower bound on $\kappa_\text{min}$ can be found, and
therefore different classes of theories with the same
$\lambda_\text{max}$---i.e. the same Hagedorn temperature---can have different
bounds on the exponential rate.

\section{Conclusions}\label{sec:conclusions}

In this work, we have performed a systematic analysis of the Hagedorn behaviour
of $\mathcal{N}=2$ Lagrangian SCFTs admitting both large-$N$ and overall
free-field limits, and its relation to the CFT Distance Conjecture. This
continues the line of research initiated in references \cite{Calderon-Infante:2024oed,
Calderon-Infante:2026rkj} to larger families of theories with multi-dimensional
conformal manifolds parametrised by Yang--Mills couplings. We have focussed on
these theories due to their close connection to group theory: as we have
reviewed in detail in Section \ref{sec:classification-quivers}, Classification
\ref{classification-scft} associates a finite or affine Lie algebra
$\mathfrak{b}$ to all quivers, describing their shape, with the rest of the
gauge and flavour data controlled by its root system. 

Our main result is that while the thermal partition function does depend on the
gauge data, the Hagedorn behaviour is set solely by the shape of the quiver.
More precisely, in all cases---including quivers that are not made out of only
bifundamental matter---it depends on the largest eigenvalue
$\lambda_\text{max}\leq2$ of the Cartan adjacency $A$ of the
algebra $\mathfrak{b}$.

The Hagedorn growth of states then defines universality classes of SCFTs with
the same string-like spectrum at high energy. We find two types of such
universality classes. The first is associated with all affine Lie algebras as
they all have $\lambda_\text{max}=2$, and correspond to orientifold projections
of $\mathcal{N}=4$ super-Yang--Mills. The others are related to finite
algebras $\mathfrak{b}$, which have their Hagedorn temperature set by their
Coxeter number $h$: $\lambda_\text{max}=2\cos(\pi/h)$. After unfolding the
quiver, the corresponding universality classes follow an ADE classification.
Since non-simply-laced quivers are related to O-planes, we conclude that the
Hagedorn behaviour of the thermal partition function is unaffected by their
presence. 

Furthermore, up to three outliers, all SCFTs associated with a
finite algebra can be obtained by decoupling a single node from their affine
version. In the F-theory picture, this can be understood through
compactifications of six-dimensional Little String Theories (LSTs) on a torus,
where this operation corresponds geometrically to removing the curve
associated with this non-critical string. 

In the holographic bulk dual, the interpretation of this Hagedorn behaviour is a
tensionless string. For the universality class encompassing all affine quivers,
the bulk string corresponds to a type-II fundamental string propagating in a
(backreacted) $\text{AdS}_5\times S^5$ background and its orientifolds. For
finite quivers, an interpretation is more difficult, as these theories do not
have a weakly-coupled gravity dual. However, all these theories except the three outliers can be
understood as torus compactifications of (deformations of) LSTs, and it would be
interesting to see if the nature of these tensionless strings can be understood
in terms of ``little-string holography'' \cite{Giveon:1998sr, Dei:2024frl,
Dei:2025ilx, Calderon-Infante:2026zmd}.

We have focussed on the thermal partition function at the overall-free point,
and relied on the fact that it can be expressed in terms of single-letter
partition functions. This is not possible when interactions are turned on. A
natural perturbative extension would be to incorporate the one-loop dilatation
operator into the counting problem, as was done for planar $\mathcal N=4$
super-Yang--Mills using a spin-chain generalisation of Pólya counting
\cite{Spradlin:2004pp}. The question is then whether an analogous
construction exists for the large-$N$ $\mathcal N=2$ quiver SCFTs considered
here, and if the Lie-algebraic data controlling the free Hagedorn temperature
also organises its first perturbative correction, and can be used to refine the
universality classes. Beyond perturbation theory, recent integrability
techniques have made it possible to compute the Hagedorn temperature of planar
$\mathcal N=4$ super-Yang--Mills at finite coupling
\cite{Harmark:2017yrv,Ekhammar:2023cuj,Ekhammar:2023glu}. For the special
subclass of affine quivers arising as orbifold or orientifold projections of
$\mathcal N=4$ super-Yang--Mills, such a perturbative analysis could provide a
bridge to finite-coupling integrability methods. More generally, developing
tools to follow the Hagedorn temperature away from the free point, whether
perturbatively or through integrability in special cases, could shed additional
light on the CFT Distance Conjecture and on how spectral data vary over the
conformal manifold of both integrable and non-integrable SCFTs.

The Lie-algebraic point of view is also very useful to study
partial-decoupling limits. Through similar techniques used to relate the
Hagedorn temperature to the largest adjacency eigenvalue, we have found that it
also bounds the exponential rate $\kappa$ dictated by the CFT Distance Conjecture from
below. In the one-node case, this bound is saturated, and $\lambda_\text{max}$
sets both the Hagedorn temperature and the exponential rates. Our results
therefore give an elegant explanation of the mini-landscape discovered in
reference \cite{Calderon-Infante:2024oed}: the three universality classes of
SCFTs with a simple gauge algebra consist of orientifolds of $\mathcal{N}=4$
super-Yang--Mills ($\lambda_\text{max}=2$), or partial decouplings of affine
quivers associated with $\mathbb{Z}_2$ or $\mathbb{Z}_3$ orientifolds
($\lambda_{\text{max}}=0,1$ respectively).

Our results can also be generalised to $\mathcal{N}=1$ quivers, as the
computation of the partition function at large $N$ is similar. The matter
content is however encoded in an adjacency matrix that is much less constrained
than for $\mathcal{N}=2$ theories, although for $\mathcal{N}=1$ orbifolds we
find that the Hagedorn temperature is also the same as that of $\mathcal{N}=4$
super-Yang--Mills. Our interpretation is that orbifold and/or orientifold
projections cannot significantly affect the large-energy growth of states and
we have a tensionless fundamental type-II string propagating on a backreacted
$\text{AdS}_5\times S^5$ background. 

These theories are a very small part of the landscape of $\mathcal{N}=1$ quiver
SCFTs. A natural extension of our work would be to study larger classes of
SCFTs with fewer supercharges. Despite the lack of a full classification for
$\mathcal{N}=1$ theories, recent progress has been made for (possibly
non-Lagrangian) theories with a simple gauge group \cite{Cho:2025xod}. Another
class of theories where our techniques can be applied readily consists of those
associated with D3-branes probing a Calabi--Yau singularity. In the toric case,
an algorithm exists to find the quiver, and it would be interesting to check if
their Hagedorn behaviour can teach us more about their universality classes,
even if the overall-free point is not part of the conformal manifold. Along
these lines, another natural question is how much information the Hagedorn
temperature and the associated universality classes encode about how the bulk
strings transform along Renormalisation-Group flows on the CFT side.

The techniques we have used in Section \ref{sec:thermal} to compute the thermal
partition function can also be applied to the superconformal index. The
single-letter contributions are in those cases more structured and satisfy
certain relations which can be used to simplify the infinite product and write
it in terms of single-trace contributions. It has recently been used to
characterise the infinite tower of BPS states appearing in the partial
decoupling of the $\mathbb{Z}_2$ orbifold \cite{Mantegazza:2026spd}.
Furthermore, the superconformal index is an invariant on the conformal manifold
and therefore can probe its structure in the non-perturbative regime. In
$\mathcal{N}=1$ theories where more types of marginal deformations exist, this
could also be used to probe the structure of the conformal manifold further,
such as the neighbourhood of finite-distance singularities.

Contrary to the thermal partition function, the index can also be computed for
theories that do not have a free-field regime, such as class-S constructions
\cite{Gadde:2011uv}. Our results therefore open the way to a study
of similar towers in more general SCFTs, and could be used to
shed some light on the nature of these towers in more involved cases, and could
in turn teach us more about the bulk tensionless strings, and how much is
covered in the ADE classification we have found.

\subsection*{Acknowledgements}
We thank A. Antunes, G. Bonori, V. Chakrabhavi, J. J. Heckman, C. Lawrie, F.
Mangialardi, J. Monnee, A. \c{C}avu\c{s}o\u{g}lu, T. Skrzypek, and M. Sperling
for helpful discussions. We are also particularly grateful to J. Calder\'on
Infante, E. Pomoni, and T. Weigand for discussions and comments on an early version of the
manuscript. FB is particularly indebted to H. Ahmed, G. Bonori, J.
Calder\'on Infante, C. Lawrie, P.K. Oehlmann, and F. R\"uhle for collaborations on
related topics and for enabling his obsession with Lie algebras.

The authors are supported by the Deutsche Forschungsgemeinschaft under
Germany’s Excellence Strategy -- EXC 2121 ``Quantum Universe'' -- 390833306,
and the Collaborative Research Center --- SFB 1624 ``Higher Structures, Moduli
Spaces, and Integrability'' --- 506632645. FB is also partially funded by the
German Research Foundation through a German-Israeli Project Cooperation (DIP)
grant ``Holography and the Swampland''. 

\appendix

\section{Lie Algebras and Their Symmetric Functions}\label{app:lie-algebras}

We present here a short technical review of quantities related to Lie algebras that are
relevant to this work, and how to compute integrals involving the Haar measure
in the large-$N$ limit.

Consider a semi-simple algebra $\mathfrak{g}$. We work with the root system
$\Phi$ of $\mathfrak{g}$ defined by a choice of simple roots $\alpha^1,\dots,
\alpha^{\text{rk}(\mathfrak{g})}$. The roots can be used to decompose
$\mathfrak{g}$ into its Cartan subalgebra
$\mathfrak{h}\subset\mathfrak{g}$---the set of maximally commuting generators
$[H_i,H_j]=0$---and its one-dimensional root spaces $\mathfrak{g}_\alpha$
through the triangular decomposition:
\begin{equation}\label{triangular-decomp-algebra}
		\mathfrak{g} = \mathfrak{n}^-\oplus\mathfrak{h}\oplus \mathfrak{n}^+\,,\qquad \mathfrak{n}^{\pm} = \bigoplus_{\alpha\in\Phi^\pm}\mathfrak{g}_{\alpha}\,,
\end{equation}
where $\Phi^\pm$ indicates the set of positive and negative roots, defined as
positive (respectively negative) integer combinations of the simple roots
$\alpha^i$. We also define the root lattice whose dual, the coweight lattice,
is generated by fundamental coweights $\omega_i^\vee$:
\begin{equation}\label{lattices}
		Q = \text{span}_{\mathbb{Z}}(\alpha^i)\,,\qquad P^\vee=\text{span}_{\mathbb{Z}}(\omega_i^\vee)\,,\qquad \langle\alpha^i,\omega^\vee_j\rangle = \delta^i_j\,,
\end{equation}
where $\langle\cdot,\cdot\rangle$ is the natural lattice pairing. We further
work with coroots $\alpha^{i\,\vee}$ and fundamental weights $\omega_i$
satisfying
\begin{equation}\label{lattice-pairings}
		\langle\alpha^i, \alpha^{j\vee}\rangle = C^{ij}\,,\qquad \langle \omega_j,\alpha^{i\vee}\rangle = \delta^i_j\,,
\end{equation}
where $C^{ij}$ is the Cartan matrix. Our conventions for the numbering of the
roots and the Cartan matrix are the same as Kac \cite{Kac:1990gs}. The coroot
and weight lattices $Q^\vee$ and $P$, respectively, are defined similarly as in
equation \eqref{lattices}. We have the inclusions $\Phi\subset
Q\subset P$, which means that all roots are weights, but not vice versa.

If a coroot $\gamma = k_i\alpha^{i\,\vee}$ has only positive coefficients
$k_i\geq0$, it is called a positive coroot $\gamma\in Q_+^\vee$, while a coweight
$\mu = \mu^i\omega_i^\vee$ with $\mu^i\geq0$ is called dominant,
and the set of dominant coweights is denoted $P_+^\vee$. The same applies to
roots and weights in the obvious notation.

This language allows us to give a Lie-algebraic meaning for the SCFT condition
given in equation \eqref{beta-function-bif}. Indeed, we have seen in Section
\ref{sec:classification-quivers} that the shape of a large-$N$ quiver SCFT is
encoded in a finite or affine algebra $\mathfrak{b}$. The rest of the data is
given by dimensions of the fundamental representations of the gauge and flavour
symmetries $k_i$ and $f^i$, respectively.

These two quantities can be understood as the coefficients of elements of the
root system. Indeed, using the pairings \eqref{lattice-pairings}, the SCFT
condition can be rewritten as
\begin{equation}
		C^{ij}k_j = \langle \alpha^i, \alpha^{j\,\vee}\rangle k_j\,,\qquad f^i-4S^i = (f^j-4S^j) \langle \alpha^i, \omega_j^\vee\rangle\,.
\end{equation}
We therefore see that both the flavour and gauge data can be understood as the
non-negative coefficients in the fundamental-coweight and simple-coroot basis,
respectively:
\begin{equation}
		\mu^\vee = k_j \,\alpha^{j\,\vee} = (f^i - 4S^i)\,\omega_i^\vee\,,\qquad \mu^\vee \in Q^\vee_+ \cap P^\vee_+\,.
\end{equation}
The element $\mu^\vee$ must therefore be both a positive coroot and a
dominant coweight to satisfy the SCFT condition, and we therefore see that the
full data of the quiver is then described by the algebra $\mathfrak{b}$ and its
root system. 

If the base algebra $\mathfrak{b}$ is of affine type, the equations above
remain correct but the full root and weight lattices must be supplemented by
additional generators to be consistent: the simple root $\alpha^0$ associated
with the affine node is found to be a combination of the other simple roots, and
therefore the $\alpha^i$ do not form an independent basis of the root lattice.
To ensure consistency of the full affine root system, one must then introduce a
``scaling element''. These subtleties will not be important in this work, the
salient point being that as the Cartan matrix has a one-dimensional null space
$C^{ij}K_j = 0 $, it is not invertible. At the level of the SCFT condition,
this gives rise to a one-parameter family of solutions for fixed $f^i$:
$k_i\propto K_i\, N +\dots$. The integers $K_i$ are the Dynkin comarks of the algebra, and the parameter $N$ is the coefficient of the
scaling element, and is part of the definition of affine coweights. This means
that $\mu^\vee$ encodes both the possible flavour \emph{and} the value of $N$
in the affine case. The large-$N$ limit therefore does not introduce a new
parameter, but rather a restriction of the allowed affine coweights. We defer
to e.g. references \cite{Fazzi:2023ulb, Ahmed:2025fmq} for constructions of
affine root systems in the context of SCFTs similar to those encountered in this work. In
particular, we follow the same conventions as reference \cite{Ahmed:2025fmq}.

\subsection{Representations and Weyl Characters}\label{app:weyl-characters}

Irreducible representations of a semi-simple algebra $\mathfrak{g}$ are in
one-to-one correspondence with dominant weights, i.e. those weights $\mu =
\mu^i\omega_i\in P$ for which $\mu^i\geq0$ for all components. For the adjoint
representation, this is equivalent to the decomposition in equation
\eqref{triangular-decomp-algebra}. In the same way the algebra $\mathfrak{g}$
can be decomposed into root components, every irreducible representation
$\bm{R}$ of $\mathfrak{g}$ can be decomposed in terms of its weight system
$\text{Wt}(\mu)$:
\begin{equation}
		\bm{R} = \bigoplus_{\nu\in \text{Wt}(\mu)} m_{\nu} V_{\nu}\,,
\end{equation}
where $V_\nu$ are one-dimensional vector spaces defined by the weight $\nu$,
and $m_\nu$ their multiplicity. The usefulness of this language comes about
when discussing the thermal partition function. It is often useful to refine
it by turning on fugacities $u^i$ for the flavour and gauge
symmetries, one for each basis generator $H_i$ of the Cartan subalgebra
$\text{span}(H_i) = \mathfrak{h}$. If $\mathfrak{g}$ is a symmetry of the
system, the thermal partition function can be written as a sum of irreducible
representations labelled by their highest weight $\mu$:
\begin{equation}
		Z[x;u] = \text{Tr}_{\mathcal{H}}\, \big[ x^{E}\,\prod_{i=1}^{\text{rk}(\mathfrak{g})}u_i^{H_i}\big] = 
		\sum_{(E,\mu)}\sum_{\nu\in\text{Wt}(\mu)} m_\nu \,\langle E,\nu|x^E u^\nu|E, \nu\rangle\,,
\end{equation}
where we have used the multi-index notation
\begin{equation}\label{multi-index}
		u^\nu = \prod_{i=1}^{\text{rk}(\mathfrak{g})}\, u_i^{\langle \nu,\alpha^{i\vee}\rangle}\,,\qquad H_i\, |\nu\rangle =\, \langle\nu, \alpha^{i\,\vee} \rangle\, |\nu \rangle\,,
\end{equation}
for all $|\nu\rangle\in V_\nu$. If we use the fundamental-weight basis
$\nu=\nu^i\omega_i$, then $u^\nu$ is the product of its components
$u_i^{\nu^i}$, and the partition function reorganises into a sum of Weyl
characters:
\begin{equation}
		Z[x;u] = \sum_{(E, \mu)} \chi_{\mu}(u) \, x^E\,,\qquad 
		\chi_{\mu}(u) = \sum_{\nu\in \text{Wt}(\mu)}\, m_\nu\, u^{\nu}\,.
\end{equation}
In the free-field limit, one can then use an oscillator construction to
obtain the multi-particle Fock space, and its contribution to the thermal
partition function, given in terms of a single-letter partition function
\cite{Aharony:2003sx}. If $\mathfrak{g}$ is a gauge symmetry, we need to
project to the singlet sector, which is achieved through the Haar measure,
leading to equation \eqref{partition-function-as-haar-integral}.

\paragraph{The Stable Limit:} 
Weyl characters are Laurent polynomials in terms of the fugacities $u_i$ with
integer coefficients, and their exponents are associated with weights $\mu\in
P$. In addition, they have the property that they are invariant under the
Weyl group $W$. In the theory of symmetric functions, the set of all Weyl
characters is often denoted as 
\begin{equation}
		\chi_\mu(u) \in \mathbb{Z}[P]^W = \{f(u) \in \mathbb{Z}[u_1^{\pm1}, \dots, u_{\text{rk}(\mathfrak{g})}^{\pm1}]\,|\, f(w\cdot u) = f(u)\,,\forall\, w\in W\} \,.
\end{equation}

Let $\mathfrak{g}$ be a classical algebra $\mathfrak{su}(N), \mathfrak{so}(2N)$
or $\mathfrak{usp}(2N)$, with weight lattice $P_N$ and Weyl group $W_N$. For
these classical algebras, it is sometimes useful to introduce new variables
$U_a$, $a=1,\dots,N$, and define the following power sums
\cite{macdonald1998symmetric}:
\begin{equation}\label{power-sums}
		p_n(U) = 
		\begin{cases}
				~\sum_{a=1}^N U_a^n\,,&\qquad \text{for }\mathfrak{g}=\mathfrak{su}(N)\,;\\
				~\sum_{a=1}^N (U_a^n + U_a^{-n})\,,&\qquad \text{for }\mathfrak{g}=\mathfrak{so}(2N),\mathfrak{usp}(2N)\,.
		\end{cases}
\end{equation}
These polynomials satisfy $p_n(U)=p_1(U^n)$, and it can be shown that, except
for spinor representations of $\mathfrak{so}(N)$ irrelevant here, any Weyl
character can be written in terms of power sums:
\begin{equation}
		\chi_\mu(U)\in \mathbb{Z}[P_N]^{W_N} \subset \mathbb{Q}[p_1(U), p_2(U),\dots, p_N(U)]\,.\qquad 
\end{equation}
While Weyl characters are integer-valued Laurent polynomials with respect to the
variables $u_i$ associated with the fundamental-weight basis, they are
rational-valued combinations of the $p_n(U)$. 

When $n\leq N$, the power sums are independent. In the theory of symmetric
functions, this is called the stable regime. On the other hand, they satisfy
certain relations when $n> N$. For the case where
$\mathfrak{g}=\mathfrak{su}(N)$, these are Newton's identities. There are a
few more subtleties, which we will not discuss here as we are ultimately
interested in the large-$N$ limit, see Footnote \ref{fn:weyl}.

As $N\to\infty$, the so-called \emph{stable limit}, all $p_n$ become
independent and freely generate the ring of Weyl characters over rational
numbers \cite{macdonald1998symmetric}:
\begin{equation}\label{stable-limit-ring}
		\mathbb{Q}[P_\infty]^{W_\infty} \cong \mathbb{Q}[p_1, p_2,\dots]\,,\qquad 
\end{equation}
and every character $\chi_{\bm{R}}(U^m)$ can be expressed in terms of the power
sums $p_n$. They are therefore particularly amenable to the computation of the
thermal partition function, as we can treat them as independent integration
variables.

\subsection{The Haar Measure}\label{app:Haar}

The result above can be used to show that the Haar measure becomes Gaussian in
the large-$N$ limit. Let us consider a compact connected Lie group $G$ with $\mathfrak{g}=\text{Lie}(G)$ and make
a choice of maximal torus $T$---the largest compact, connected, and Abelian Lie subgroup of
$G$. By the torus theorem, every element $g\in G$ is conjugate to an element of
$T$: $\forall g \in G\,,\exists h\in G $ such that $h^{-1}gh \in T$.

Given a class function of $G$, namely a function invariant under conjugation,
$f(h^{-1}g h) = f(g) $ $\forall g,h\in G$, the Haar measure can be written as an
integral over torus coordinates $u\in T$ via Weyl's integration formula
\cite{Fulton:2004uyc}
\begin{equation}
		\int_G d\mu\, f(g) = \frac{1}{|W|} \int_T d\mu(u) |\Delta(u)|^2\,f(u)\,,
\end{equation}
where $|W|$ is the order of the Weyl group of $G$. We have also defined the Weyl
determinant
\begin{equation}
		\Delta(u) = \prod_{\alpha\in\Phi^+} (u^{\alpha/2}-u^{-\alpha/2})\,,\qquad |\Delta(u)|^2 = \prod_{\alpha\in\Phi^+}(1-u^{\alpha})(1-u^{-\alpha})\,,
\end{equation}
with the product taken over all positive roots, and we used the multi-index
notation $u^\alpha$ as in equation \eqref{multi-index}. For $G=SU(N)$,
$\Delta(U)$ is the Vandermonde determinant. Expanding the Weyl determinant as a
formal power series,
\begin{equation}\label{determinant-expansion}
		\log(|\Delta(u)|^2) 
		= -\sum_{n>0}\frac{1}{n}\sum_{\alpha\in\Phi^+}(u^{n\alpha}+u^{-n\alpha}) = - \sum_{n>0}\frac{1}{n}\sum_{\alpha\in\Phi}u^{n\alpha}\,,
\end{equation}
it can be rewritten in terms of the Weyl character of the adjoint representation. Following the triangular decomposition of the algebra $\mathfrak{g}$ given in
\eqref{triangular-decomp-algebra}, the weight system of the adjoint is given by
the roots, supplemented by the Cartan elements. The last term in equation
\eqref{determinant-expansion} is therefore the Weyl character of the adjoint
representation up to the rank of the algebra, and we obtain
\begin{equation}
		|\Delta(u)|^2 = \text{PE}[\text{rk}(\mathfrak{g})-\chi_{\bm{adj}}(u)] = \text{exp}\big[\sum_{n>0}\frac{1}{n}\big(\text{rk}(\mathfrak{g})-\chi_{\bm{adj}}(u^n)\big)\big]\,.
\end{equation}
In practice, $f(u)$ will be a function of the Weyl characters, and therefore a
class function of $G$ as they are invariant under the Weyl group, the action of
Weyl-group elements simply reshuffling the sum. As we are interested in the large-$N$
limit, we can use the power sums for the Weyl characters:
\begin{equation}
		\chi_{\bm{F}}(U^n) = p_n(U)\,.
\end{equation}
We have seen around equation \eqref{stable-limit-ring} that they form an
independent basis for all Weyl characters as $N\to\infty$. We then follow the
eigenvalue method \cite{Aharony:2003sx, Imamura:2016abe,
Calderon-Infante:2024oed} and introduce a density for the angles
$U_a=e^{i\theta_a}$:
\begin{equation}
	\begin{aligned}
			SU(N):&\qquad \rho(\theta) = \sum_{a=1}^N\delta(\theta-\theta_a)\,, & \qquad \int_{-\pi}^\pi d\theta\, \rho(\theta) = N\,;\\
			SO(2N)/USp(2N):&\qquad \rho(\theta) = \sum_{a=1}^N\big(\delta(\theta-\theta_a)+\delta(\theta+\theta_a)\big)\,, & \qquad \int_{-\pi}^\pi d\theta\, \rho(\theta) = 2N\,,
\end{aligned}
\end{equation}
where we have used that for $SO(2N)$ and $USp(2N)$, the torus coordinates come
in pairs $\pm\theta_a$ in the orthogonal basis, see equations \eqref{def:p_n}
or \eqref{power-sums}. As the torus angles are periodic, so is the density
$\rho(\theta)$ and we can perform a Fourier expansion:
\begin{equation}
		\rho(\theta) = \frac{1}{2\pi}\sum_{n} \rho_n e^{-in\theta}\,,\qquad
		\rho_n = \int_{-\pi}^\pi d\theta\, \rho(\theta) e^{in\theta} = p_n\,.
\end{equation}
The Fourier modes are therefore simply the Weyl characters themselves,
$p_n=\chi_{\bm{F}}(U^n)$, and become integration variables. The normalisation
of the modes is fixed by using that the Haar measure over the trivial character
is one:
\begin{equation}
		1\overset{!}{=}
		\int_G d\mu \, \chi_{\bm{1}}(U) =
		\int_G d\mu \cdot 1 = 
		\frac{1}{|W|}\int_T\prod_{a=1}^{N}\frac{d\theta_a}{2\pi} \text{PE}[\text{rk}(\mathfrak{g}) -\chi_{\bm{adj}}(U)] \,.
\end{equation}
The order of the Weyl group and other formally infinite constants are then
absorbed in the normalisation, and we ultimately can make the replacement
\cite{Aharony:2003sx, Imamura:2016abe, Calderon-Infante:2026rkj}:
\begin{equation}
		\int_G d\mu\quad \longrightarrow \quad \int \prod_{n>0}d^{d}p_n\, \left( \frac{d}{2\pi n} \right)^{d/2} \text{exp}\left(-\sum_{n>0}\frac{d}{2n}(|p_n|^2 +S\, p_{2n} )\right)\,,
\end{equation}
where $d, S$ are defined in Table \ref{tab:classical-quantities}. The
integration is complex for $G=SU(N)$, $\overline{p}_n=p_{-n}$, and real for $G =
SO(2N), USp(2N)$, $\overline{p}_n=p_{n}$. If the group is semi-simple, $G =
G_1\times G_2\times \dots$, that is the quiver has more than a single gauge
node, the measure factorises into products associated with each simple factor,
and we obtain the result quoted in Section \ref{sec:thermal}.

\bibliography{references}
\bibliographystyle{JHEP}

\end{document}

%% file: tikz_data.tex
\usetikzlibrary{arrows}
\usetikzlibrary{calc}
\tikzstyle{every picture}+=[remember picture]
\tikzstyle{na} = [baseline=-.5ex]
\tikzstyle{mine}= [arrows={angle 90}-{angle 90},thick]

\def\Llleftarrow{%
\lower2pt\hbox{\begingroup
\tikz
\draw[shorten >=0pt,shorten <=0pt] (0,3pt) -- ++(-1em,0) (0,1pt) -- ++(-1em-1pt,0) (0,-1pt) -- ++(-1em-1pt,0) (0,-3pt) -- ++(-1em,0) (-1em+1pt,5pt) to[out=-105,in=45] (-1em-2pt,0) to[out=-45,in=105] (-1em+1pt,-5pt);
\endgroup}
}

\tikzset{
  doublearrow/.style={
    double distance=2pt,
    shorten <=0.2pt,
    shorten >=0.2pt,
    postaction={decorate},
    decoration={markings,
      mark=at position 0.1 with {
			  \draw[line cap=round] (0.06,0.14) -- (0.16,0) -- (0.06,-0.14);
      }
    }
  }
}

\tikzset{
  doublearrowright/.style={
    preaction={draw,transform canvas={yshift=1pt}},
    draw,transform canvas={yshift=-1pt},
    postaction={decorate},
    decoration={markings,
      mark=at position 0.1 with {
			  \draw[line cap=round] (0.08,0.14) -- (0.19,0.035) -- (0.08,-0.075);
      }
    }
  }
}

\tikzset{
  doublearrowleft/.style={
    preaction={draw,transform canvas={yshift=1pt}},
    draw,transform canvas={yshift=-1pt},
    postaction={decorate},
    decoration={markings,
      mark=at position 0.16 with {
			  \draw[line cap=round] (0.19,0.15) -- (0.08,0.035) -- (0.19,-0.075);
      }
    }
  }
}

\usetikzlibrary{arrows,shapes.misc,positioning,decorations.pathmorphing,decorations.markings,decorations.pathreplacing,matrix,patterns,backgrounds,calligraphy}

\tikzset{
scale cd/.style={every label/.append style={scale=#1}, cells={nodes={scale=#1}}}}

\tikzset{gauge/.style={rounded rectangle, draw=black!100, thick, minimum size=5mm},  gaugeD/.style={rounded rectangle, draw=black!100,double,thick,minimum size=5mm},  empty/.style={rounded rectangle, draw=white!100, thick, minimum size=5mm}, flavor/.style={rectangle, draw=black!100, thick, minimum size=5mm},flavorD/.style={rectangle, draw=black!100, double,thick, minimum size=5mm}}

\tikzset{
node/.style={circle, thick, draw=black!100,fill=white!100,  minimum size=2mm, inner sep=0pt},
sqnode/.style={rectangle
, thick, draw=black!100,fill=white!100,  minimum size=2mm, inner sep=0pt
},
sonode/.style={circle, thick, draw=black!100,fill=red!100,  minimum size=3mm, inner sep=0pt},
spnode/.style={circle, thick, draw=black!100,fill=blue!100,  minimum size=3mm, inner sep=0pt},
fnode/.style={rectangle, thick, draw=black!100,fill=white!100,  minimum size=3mm, inner sep=0pt},
tnode/.style={rounded rectangle, outer sep=0pt, thick, minimum size=5mm},
Rightarrow/.style={double equal sign distance,>={Implies},->},
triplearrow/.style={-,preaction={draw,Rightarrow}},
 triple/.style={
        double distance=2pt,
        line width=0.4pt,
    }
}
\tikzset{
    node/.style={circle, draw, minimum size=5pt, inner sep=0pt},
    doublemidarrow/.style={
        postaction={
            decorate,
            decoration={
                markings,
                mark=at position 0.50 with {
                    \arrow{>}
                    \arrow[xshift=3pt]{>}
                }
            }
        }
    }
}

%% file: figures/Ak_flavor.tex
\begin{tikzpicture}
    \node[node, fill=black] (A1)  {};
    \node[node, fill=black] (A2) [right=6mm of A1] {};
    \node (A3) [right=6mm of A2] {\dots};
	\node[node, fill=black] (A4) [right=6mm of A3] {};
    \node[node, fill=black] (A5) [right=6mm of A4] {};
    \node[rectangle, fill=black] (F1) [below=4mm of A1] {};
    \node[rectangle, fill=black] (F2) [below=4mm of A2] {};
    \node[rectangle, fill=black] (F4) [below=4mm of A4] {};
    \node[rectangle, fill=black] (F5) [below=4mm of A5] {};
    \draw (A1.east) -- (A2.west);
    \draw (A2.east) -- (A3.west);
    \draw (A3.east) -- (A4.west);
    \draw (A4.east) -- (A5.west);
    \draw (A1.south) -- (F1.north);
    \draw (A2.south) -- (F2.north);
    \draw (A4.south) -- (F4.north);
    \draw (A5.south) -- (F5.north);
\end{tikzpicture}

%% file: figures/Ah.tex
\begin{tikzpicture}
    \node[node, fill=black] (A1)  {};
    \node[node, fill=black] (A2) [right=6mm of A1] {};
    \node (A3) [right=6mm of A2] {\dots};
	\node[node, fill=black] (A4) [right=6mm of A3] {};
    \node[node, fill=black] (A5) [right=6mm of A4] {};
    \node[node, fill=black] (A0) [above=6mm of A3] {};
    \draw (A1.east) -- (A2.west);
    \draw (A2.east) -- (A3.west);
    \draw (A3.east) -- (A4.west);
    \draw (A4.east) -- (A5.west);
	\draw (A1.north east) -- (A0.south west);
	\draw (A5.north west) -- (A0.south east);
\end{tikzpicture}

%% file: figures/Dk_flavor.tex
\begin{tikzpicture}
    \node[node, fill=black]   (A1) [right=6mm of A0] {};
    \node[node, fill=black] (A2) [right=6mm of A1] {};
    \node[node, fill=black]   (A3) [right=6mm of A2] {};
    \node (A4) [right=6mm of A3] {\dots};
    \node[node, fill=black]   (A5) [right=6mm of A4] {};
    \node[node, fill=black] (A6) [right=6mm of A5] {};
    \node[node, fill=black] (A7) [right=6mm of A6] {};
    \node[node, fill=black] (A8) [above=6mm of A6] {};
    \node[rectangle, fill=black] (F1) [below=4mm of A1] {};
    \node[rectangle, fill=black] (F2) [below=4mm of A2] {};
    \node[rectangle, fill=black] (F3) [below=4mm of A3] {};
    \node[rectangle, fill=black] (F5) [below=4mm of A5] {};
    \node[rectangle, fill=black] (F6) [below=4mm of A6] {};
    \node[rectangle, fill=black] (F7) [below=4mm of A7] {};
    \node[rectangle, fill=black] (F8) [above=4mm of A8] {};
    \draw (A1.east) -- (A2.west);
    \draw (A2.east) -- (A3.west);
    \draw (A3.east) -- (A4.west);
    \draw (A4.east) -- (A5.west);
    \draw (A5.east) -- (A6.west);
    \draw (A6.east) -- (A7.west);
    \draw (A6.north) -- (A8.south);
    \draw (A1.south) -- (F1.north);
    \draw (A2.south) -- (F2.north);
    \draw (A3.south) -- (F3.north);
    \draw (A5.south) -- (F5.north);
    \draw (A6.south) -- (F6.north);
    \draw (A7.south) -- (F7.north);
    \draw (A8.north) -- (F8.south);
\end{tikzpicture}

%% file: figures/Dh.tex
\begin{tikzpicture}
    \node[node, fill=black]   (A1) [right=6mm of A0] {};
    \node[node, fill=black] (A2) [right=6mm of A1] {};
    \node[node, fill=black]   (A3) [right=6mm of A2] {};
    \node (A4) [right=6mm of A3] {\dots};
	\node[node, fill=black]   (A5) [right=6mm of A4] {};
    \node[node, fill=black] (A6) [right=6mm of A5] {};
    \node[node, fill=black] (A7) [right=6mm of A6] {};
    \node[node, fill=black] (A8) [above=6mm of A6] {};
	\node[node, fill=black] (A0) [above=6mm of A2] {};
    \draw (A0.south) -- (A2.north);
    \draw (A1.east) -- (A2.west);
    \draw (A2.east) -- (A3.west);
    \draw (A3.east) -- (A4.west);
    \draw (A4.east) -- (A5.west);
    \draw (A5.east) -- (A6.west);
    \draw (A6.east) -- (A7.west);
    \draw (A6.north) -- (A8.south);
\end{tikzpicture}

%% file: figures/E6_flavor.tex
\begin{tikzpicture}
    \node[node, fill=black] (A1) [right=6mm of A0] {};
    \node[node, fill=black]   (A2) [right=6mm of A1] {};
    \node[node, fill=black] (A3) [right=6mm of A2] {};
    \node[node, fill=black]   (A4) [right=6mm of A3] {};
    \node[node, fill=black] (A5) [right=6mm of A4] {};
    \node[node, fill=black]   (A6) [above=6mm of A3] {};
    \node[rectangle, fill=black] (F1) [below=4mm of A1] {};
    \node[rectangle, fill=black] (F2) [below=4mm of A2] {};
    \node[rectangle, fill=black] (F3) [below=4mm of A3] {};
    \node[rectangle, fill=black] (F4) [below=4mm of A4] {};
    \node[rectangle, fill=black] (F5) [below=4mm of A5] {};
    \node[rectangle, fill=black] (F6) [above=4mm of A6] {};
    \draw (A1.east) -- (A2.west);
    \draw (A2.east) -- (A3.west);
    \draw (A3.east) -- (A4.west);
    \draw (A4.east) -- (A5.west);
    \draw (A3.north) -- (A6.south);
    \draw (A1.south) -- (F1.north);
    \draw (A2.south) -- (F2.north);
    \draw (A3.south) -- (F3.north);
    \draw (A4.south) -- (F4.north);
    \draw (A5.south) -- (F5.north);
    \draw (A6.north) -- (F6.south);
\end{tikzpicture}

%% file: figures/E6h.tex
\begin{tikzpicture}
    \node[node, fill=black] (A1) [right=6mm of A0] {};
    \node[node, fill=black]   (A2) [right=6mm of A1] {};
    \node[node, fill=black] (A3) [right=6mm of A2] {};
    \node[node, fill=black]   (A4) [right=6mm of A3] {};
    \node[node, fill=black] (A5) [right=6mm of A4] {};
    \node[node, fill=black]   (A6) [above=6mm of A3] {};
    \node[node, fill=black]   (A0) [above=6mm of A6] {};
    \draw (A0.south) -- (A6.north);
    \draw (A1.east) -- (A2.west);
    \draw (A2.east) -- (A3.west);
    \draw (A3.east) -- (A4.west);
    \draw (A4.east) -- (A5.west);
    \draw (A3.north) -- (A6.south);
\end{tikzpicture}

%% file: figures/E7_flavor.tex
\begin{tikzpicture}
    \node[node, fill=black]   (A1) [right=6mm of A0] {};
    \node[node, fill=black] (A2) [right=6mm of A1] {};
    \node[node, fill=black]   (A3) [right=6mm of A2] {};
    \node[node, fill=black] (A4) [right=6mm of A3] {};
    \node[node, fill=black]   (A5) [right=6mm of A4] {};
    \node[node, fill=black] (A6) [right=6mm of A5] {};
    \node[node, fill=black]   (A7) [above=6mm of A3] {};
    \node[rectangle, fill=black] (F1) [below=4mm of A1] {};
    \node[rectangle, fill=black] (F2) [below=4mm of A2] {};
    \node[rectangle, fill=black] (F3) [below=4mm of A3] {};
    \node[rectangle, fill=black] (F4) [below=4mm of A4] {};
    \node[rectangle, fill=black] (F5) [below=4mm of A5] {};
    \node[rectangle, fill=black] (F6) [below=4mm of A6] {};
    \node[rectangle, fill=black] (F7) [above=4mm of A7] {};
    \draw (A1.east) -- (A2.west);
    \draw (A2.east) -- (A3.west);
    \draw (A3.east) -- (A4.west);
    \draw (A4.east) -- (A5.west);
    \draw (A5.east) -- (A6.west);
    \draw (A3.north) -- (A7.south);
    \draw (A1.south) -- (F1.north);
    \draw (A2.south) -- (F2.north);
    \draw (A3.south) -- (F3.north);
    \draw (A4.south) -- (F4.north);
    \draw (A5.south) -- (F5.north);
    \draw (A6.south) -- (F6.north);
    \draw (A7.north) -- (F7.south);
\end{tikzpicture}

%% file: figures/E7h.tex
\begin{tikzpicture}
    \node[node, fill=black] (A0)  {};
    \node[node, fill=black]   (A1) [right=6mm of A0] {};
    \node[node, fill=black] (A2) [right=6mm of A1] {};
    \node[node, fill=black]   (A3) [right=6mm of A2] {};
    \node[node, fill=black] (A4) [right=6mm of A3] {};
    \node[node, fill=black]   (A5) [right=6mm of A4] {};
    \node[node, fill=black] (A6) [right=6mm of A5] {};
    \node[node, fill=black]   (A7) [above=6mm of A3] {};
    \draw (A0.east) -- (A1.west);
    \draw (A1.east) -- (A2.west);
    \draw (A2.east) -- (A3.west);
    \draw (A3.east) -- (A4.west);
    \draw (A4.east) -- (A5.west);
    \draw (A5.east) -- (A6.west);
    \draw (A3.north) -- (A7.south);
\end{tikzpicture}

%% file: figures/E8_flavor.tex
\begin{tikzpicture}
    \node[node, fill=black]   (A1) [right=6mm of A0] {};
    \node[node, fill=black] (A2) [right=6mm of A1] {};
    \node[node, fill=black]   (A3) [right=6mm of A2] {};
    \node[node, fill=black] (A4) [right=6mm of A3] {};
    \node[node, fill=black]   (A5) [right=6mm of A4] {};
    \node[node, fill=black] (A6) [right=6mm of A5] {};
    \node[node, fill=black] (A7) [right=6mm of A6] {};
    \node[node, fill=black]   (A8) [above=6mm of A5] {};
    \node[rectangle, fill=black] (F1) [below=4mm of A1] {};
    \node[rectangle, fill=black] (F2) [below=4mm of A2] {};
    \node[rectangle, fill=black] (F3) [below=4mm of A3] {};
    \node[rectangle, fill=black] (F4) [below=4mm of A4] {};
    \node[rectangle, fill=black] (F5) [below=4mm of A5] {};
    \node[rectangle, fill=black] (F6) [below=4mm of A6] {};
    \node[rectangle, fill=black] (F7) [below=4mm of A7] {};
    \node[rectangle, fill=black] (F8) [above=4mm of A8] {};
    \draw (A1.east) -- (A2.west);
    \draw (A2.east) -- (A3.west);
    \draw (A3.east) -- (A4.west);
    \draw (A4.east) -- (A5.west);
    \draw (A5.east) -- (A6.west);
    \draw (A6.east) -- (A7.west);
    \draw (A5.north) -- (A8.south);
    \draw (A1.south) -- (F1.north);
    \draw (A2.south) -- (F2.north);
    \draw (A3.south) -- (F3.north);
    \draw (A4.south) -- (F4.north);
    \draw (A5.south) -- (F5.north);
    \draw (A6.south) -- (F6.north);
    \draw (A7.south) -- (F7.north);
    \draw (A8.north) -- (F8.south);
\end{tikzpicture}

%% file: figures/E8h.tex
\begin{tikzpicture}
    \node[node, fill=black] (A0)  {};
    \node[node, fill=black]   (A1) [right=6mm of A0] {};
    \node[node, fill=black] (A2) [right=6mm of A1] {};
    \node[node, fill=black]   (A3) [right=6mm of A2] {};
    \node[node, fill=black] (A4) [right=6mm of A3] {};
    \node[node, fill=black]   (A5) [right=6mm of A4] {};
    \node[node, fill=black] (A6) [right=6mm of A5] {};
    \node[node, fill=black] (A7) [right=6mm of A6] {};
    \node[node, fill=black]   (A8) [above=6mm of A5] {};
    \draw (A0.east) -- (A1.west);
    \draw (A1.east) -- (A2.west);
    \draw (A2.east) -- (A3.west);
    \draw (A3.east) -- (A4.west);
    \draw (A4.east) -- (A5.west);
    \draw (A5.east) -- (A6.west);
    \draw (A6.east) -- (A7.west);
    \draw (A5.north) -- (A8.south);
\end{tikzpicture}

%% file: figures/Ak_os_flavor.tex
\begin{tikzpicture}
    \node[node, fill=red] (A1)  {};
    \node[node, fill=blue] (A2) [right=6mm of A1] {};
    \node (A3) [right=6mm of A2] {\dots};
	\node[node, fill=red] (A4) [right=6mm of A3] {};
    \node[node, fill=blue] (A5) [right=6mm of A4] {};
    \node[rectangle, fill=blue] (F1) [below=4mm of A1] {};
    \node[rectangle, fill=red] (F2) [below=4mm of A2] {};
    \node[rectangle, fill=blue] (F4) [below=4mm of A4] {};
    \node[rectangle, fill=red] (F5) [below=4mm of A5] {};
    \draw (A1.east) -- (A2.west);
    \draw (A2.east) -- (A3.west);
    \draw (A3.east) -- (A4.west);
    \draw (A4.east) -- (A5.west);
    \draw (A1.south) -- (F1.north);
    \draw (A2.south) -- (F2.north);
    \draw (A4.south) -- (F4.north);
    \draw (A5.south) -- (F5.north);
\end{tikzpicture}

%% file: figures/Ak_os_flavor_3.tex
\begin{tikzpicture}
    \node[node, fill=blue] (A1)  {};
    \node[node, fill=red] (A2) [right=6mm of A1] {};
    \node (A3) [right=6mm of A2] {\dots};
	\node[node, fill=red] (A4) [right=6mm of A3] {};
    \node[node, fill=blue] (A5) [right=6mm of A4] {};
    \node[rectangle, fill=red] (F1) [below=4mm of A1] {};
    \node[rectangle, fill=blue] (F2) [below=4mm of A2] {};
    \node[rectangle, fill=blue] (F4) [below=4mm of A4] {};
    \node[rectangle, fill=red] (F5) [below=4mm of A5] {};
    \draw (A1.east) -- (A2.west);
    \draw (A2.east) -- (A3.west);
    \draw (A3.east) -- (A4.west);
    \draw (A4.east) -- (A5.west);
    \draw (A1.south) -- (F1.north);
    \draw (A2.south) -- (F2.north);
    \draw (A4.south) -- (F4.north);
    \draw (A5.south) -- (F5.north);
\end{tikzpicture}

%% file: figures/Ah_os.tex
\begin{tikzpicture}
    \node[node, fill=red] (A1)  {};
    \node[node, fill=blue] (A2) [right=6mm of A1] {};
    \node (A3) [right=6mm of A2] {\dots};
	\node[node, fill=blue] (A4) [right=6mm of A3] {};
    \node[node, fill=red] (A5) [right=6mm of A4] {};
    \node[node, fill=blue] (A0) [above=6mm of A3] {};
    \draw (A1.east) -- (A2.west);
    \draw (A2.east) -- (A3.west);
    \draw (A3.east) -- (A4.west);
    \draw (A4.east) -- (A5.west);
	\draw (A1.north east) -- (A0.south west);
	\draw (A5.north west) -- (A0.south east);
\end{tikzpicture}

%% file: figures/Dk_os_flavor.tex
\begin{tikzpicture}
    \node[node, fill=blue]   (A1) [right=6mm of A0] {};
    \node[node, fill=red] (A2) [right=6mm of A1] {};
    \node[node, fill=blue]   (A3) [right=6mm of A2] {};
    \node (A4) [right=6mm of A3] {\dots};
    \node[node, fill=blue]   (A5) [right=6mm of A4] {};
    \node[node, fill=red] (A6) [right=6mm of A5] {};
    \node[node, fill=blue] (A7) [right=6mm of A6] {};
    \node[node, fill=blue] (A8) [above=6mm of A6] {};
    \node[rectangle, fill=blue] (F1) [below=4mm of A1] {};
    \node[rectangle, fill=red] (F2) [below=4mm of A2] {};
    \node[rectangle, fill=blue] (F3) [below=4mm of A3] {};
    \node[rectangle, fill=red] (F5) [below=4mm of A5] {};
    \node[rectangle, fill=blue] (F6) [below=4mm of A6] {};
    \node[rectangle, fill=red] (F7) [below=4mm of A7] {};
    \node[rectangle, fill=red] (F8) [above=4mm of A8] {};
    \draw (A1.east) -- (A2.west);
    \draw (A2.east) -- (A3.west);
    \draw (A3.east) -- (A4.west);
    \draw (A4.east) -- (A5.west);
    \draw (A5.east) -- (A6.west);
    \draw (A6.east) -- (A7.west);
    \draw (A6.north) -- (A8.south);
    \draw (A1.south) -- (F1.north);
    \draw (A2.south) -- (F2.north);
    \draw (A3.south) -- (F3.north);
    \draw (A5.south) -- (F5.north);
    \draw (A6.south) -- (F6.north);
    \draw (A7.south) -- (F7.north);
    \draw (A8.north) -- (F8.south);
\end{tikzpicture}

%% file: figures/Dk_os_flavor_2.tex
\begin{tikzpicture}
    \node[node, fill=red]   (A1) [right=6mm of A0] {};
    \node[node, fill=blue] (A2) [right=6mm of A1] {};
    \node[node, fill=red]   (A3) [right=6mm of A2] {};
    \node (A4) [right=6mm of A3] {\dots};
    \node[node, fill=red]   (A5) [right=6mm of A4] {};
    \node[node, fill=blue] (A6) [right=6mm of A5] {};
    \node[node, fill=red] (A7) [right=6mm of A6] {};
    \node[node, fill=red] (A8) [above=6mm of A6] {};
    \node[rectangle, fill=blue] (F1) [below=4mm of A1] {};
    \node[rectangle, fill=red] (F2) [below=4mm of A2] {};
    \node[rectangle, fill=blue] (F3) [below=4mm of A3] {};
    \node[rectangle, fill=blue] (F5) [below=4mm of A5] {};
    \node[rectangle, fill=red] (F6) [below=4mm of A6] {};
    \node[rectangle, fill=blue] (F7) [below=4mm of A7] {};
	\node[rectangle, fill=blue] (F8) [above=4mm of A8] {};
    \draw (A1.east) -- (A2.west);
    \draw (A2.east) -- (A3.west);
    \draw (A3.east) -- (A4.west);
    \draw (A4.east) -- (A5.west);
    \draw (A5.east) -- (A6.west);
    \draw (A6.east) -- (A7.west);
    \draw (A6.north) -- (A8.south);
    \draw (A1.south) -- (F1.north);
    \draw (A2.south) -- (F2.north);
    \draw (A3.south) -- (F3.north);
    \draw (A5.south) -- (F5.north);
    \draw (A6.south) -- (F6.north);
    \draw (A7.south) -- (F7.north);
    \draw (A8.north) -- (F8.south);
\end{tikzpicture}

%% file: figures/Dh_os_flavor_2.tex
\begin{tikzpicture}
    \node[node, fill=red]   (A1) [right=6mm of A0] {};
    \node[node, fill=blue] (A2) [right=6mm of A1] {};
    \node[node, fill=red]   (A3) [right=6mm of A2] {};
    \node (A4) [right=6mm of A3] {\dots};
    \node[node, fill=blue]   (A5) [right=6mm of A4] {};
    \node[node, fill=red] (A6) [right=6mm of A5] {};
    \node[node, fill=blue] (A7) [right=6mm of A6] {};
    \node[node, fill=blue] (A8) [above=6mm of A6] {};
    \node[node, fill=red] (A0) [above=6mm of A2] {};
    \draw (A0.south) -- (A2.north);
    \draw (A1.east) -- (A2.west);
    \draw (A2.east) -- (A3.west);
    \draw (A3.east) -- (A4.west);
    \draw (A4.east) -- (A5.west);
    \draw (A5.east) -- (A6.west);
    \draw (A6.east) -- (A7.west);
    \draw (A6.north) -- (A8.south);
\end{tikzpicture}

%% file: figures/Dh_os_flavor.tex
\begin{tikzpicture}
    \node[node, fill=blue]   (A1) [right=6mm of A0] {};
    \node[node, fill=red] (A2) [right=6mm of A1] {};
    \node[node, fill=blue]   (A3) [right=6mm of A2] {};
    \node (A4) [right=6mm of A3] {\dots};
    \node[node, fill=blue]   (A5) [right=6mm of A4] {};
    \node[node, fill=red] (A6) [right=6mm of A5] {};
    \node[node, fill=blue] (A7) [right=6mm of A6] {};
    \node[node, fill=blue] (A8) [above=6mm of A6] {};
    \node[node, fill=blue] (A0) [above=6mm of A2] {};
    \node[rectangle, fill=red] (F0) [above=4mm of A0] {};
    \node[rectangle, fill=red] (F1) [below=4mm of A1] {};
    \node[rectangle, fill=blue] (F2) [below=4mm of A2] {};
    \node[rectangle, fill=red] (F3) [below=4mm of A3] {};
    \node[rectangle, fill=red] (F5) [below=4mm of A5] {};
    \node[rectangle, fill=blue] (F6) [below=4mm of A6] {};
    \node[rectangle, fill=red] (F7) [below=4mm of A7] {};
    \node[rectangle, fill=red] (F8) [above=4mm of A8] {};
    \draw (A0.south) -- (A2.north);
    \draw (A1.east) -- (A2.west);
    \draw (A2.east) -- (A3.west);
    \draw (A3.east) -- (A4.west);
    \draw (A4.east) -- (A5.west);
    \draw (A5.east) -- (A6.west);
    \draw (A6.east) -- (A7.west);
    \draw (A6.north) -- (A8.south);
    \draw (A0.north) -- (F0.south);
    \draw (A1.south) -- (F1.north);
    \draw (A2.south) -- (F2.north);
    \draw (A3.south) -- (F3.north);
    \draw (A5.south) -- (F5.north);
    \draw (A6.south) -- (F6.north);
    \draw (A7.south) -- (F7.north);
    \draw (A8.north) -- (F8.south);
\end{tikzpicture}

%% file: figures/E6_os_flavor.tex
\begin{tikzpicture}
    \node[node, fill=blue] (A1)  {};
    \node[node, fill=red] (A2) [right=6mm of A1] {};
    \node[node, fill=blue] (A3) [right=6mm of A2] {};
    \node[node, fill=red] (A4) [right=6mm of A3] {};
    \node[node, fill=blue] (A5) [right=6mm of A4] {};
    \node[node, fill=red] (A6) [above=6mm of A3] {};
    \node[rectangle, fill=red] (F1) [below=4mm of A1] {};
    \node[rectangle, fill=blue] (F2) [below=4mm of A2] {};
	\node[rectangle, fill=red] (F3) [below=4mm of A3] {};
    \node[rectangle, fill=blue] (F4) [below=4mm of A4] {};
    \node[rectangle, fill=red] (F5) [below=4mm of A5] {};
    \node[rectangle, fill=blue] (F6) [above=4mm of A6] {};
    \draw (A1.east) -- (A2.west);
    \draw (A2.east) -- (A3.west);
    \draw (A3.east) -- (A4.west);
    \draw (A4.east) -- (A5.west);
    \draw (A3.north) -- (A6.south);
    \draw (A1.south) -- (F1.north);
    \draw (A2.south) -- (F2.north);
    \draw (A3.south) -- (F3.north);
    \draw (A4.south) -- (F4.north);
    \draw (A5.south) -- (F5.north);
    \draw (A6.north) -- (F6.south);
\end{tikzpicture}

%% file: figures/E6h_os.tex
\begin{tikzpicture}
    \node[node, fill=red] (A1)  {};
    \node[node, fill=blue] (A2) [right=6mm of A1] {};
    \node[node, fill=red] (A3) [right=6mm of A2] {};
    \node[node, fill=blue] (A4) [right=6mm of A3] {};
    \node[node, fill=red] (A5) [right=6mm of A4] {};
    \node[node, fill=blue] (A6) [above=6mm of A3] {};
	\node[node, fill=red] (A0) [above=6mm of A6] {};
    \draw (A1.east) -- (A2.west);
    \draw (A2.east) -- (A3.west);
    \draw (A3.east) -- (A4.west);
    \draw (A4.east) -- (A5.west);
    \draw (A3.north) -- (A6.south);
    \draw (A6.north) -- (A0.south);
\end{tikzpicture}

%% file: figures/E7_os_flavor.tex
\begin{tikzpicture}
	\node[node, fill=blue]   (A1) [right=6mm of A0] {};
	\node[node, fill=red] (A2) [right=6mm of A1] {};
	\node[node, fill=blue]   (A3) [right=6mm of A2] {};
	\node[node, fill=red] (A4) [right=6mm of A3] {};
	\node[node, fill=blue]   (A5) [right=6mm of A4] {};
	\node[node, fill=red] (A6) [right=6mm of A5] {};
	\node[node, fill=red]   (A7) [above=6mm of A3] {};
    \node[rectangle, fill=red] (F1) [below=4mm of A1] {};
    \node[rectangle, fill=blue] (F2) [below=4mm of A2] {};
	\node[rectangle, fill=red] (F3) [below=4mm of A3] {};
    \node[rectangle, fill=blue] (F4) [below=4mm of A4] {};
    \node[rectangle, fill=red] (F5) [below=4mm of A5] {};
    \node[rectangle, fill=blue] (F6) [below=4mm of A6] {};
    \node[rectangle, fill=blue] (F7) [above=4mm of A7] {};
    %
    \draw (A1.east) -- (A2.west);
    \draw (A2.east) -- (A3.west);
    \draw (A3.east) -- (A4.west);
    \draw (A4.east) -- (A5.west);
    \draw (A5.east) -- (A6.west);
    \draw (A3.north) -- (A7.south);
    \draw (A1.south) -- (F1.north);
    \draw (A2.south) -- (F2.north);
    \draw (A3.south) -- (F3.north);
    \draw (A4.south) -- (F4.north);
    \draw (A5.south) -- (F5.north);
    \draw (A6.south) -- (F6.north);
    \draw (A7.north) -- (F7.south);
\end{tikzpicture}

%% file: figures/E7h_os_flavor.tex
\begin{tikzpicture}
    \node[node, fill=blue] (A0)  {};
	\node[node, fill=red]   (A1) [right=6mm of A0] {};
	\node[node, fill=blue] (A2) [right=6mm of A1] {};
	\node[node, fill=red]   (A3) [right=6mm of A2] {};
	\node[node, fill=blue] (A4) [right=6mm of A3] {};
	\node[node, fill=red]   (A5) [right=6mm of A4] {};
	\node[node, fill=blue] (A6) [right=6mm of A5] {};
	\node[node, fill=blue]   (A7) [above=6mm of A3] {};
    \node[rectangle, fill=red] (F0) [below=4mm of A0] {};
    \node[rectangle, fill=blue] (F1) [below=4mm of A1] {};
    \node[rectangle, fill=red] (F2) [below=4mm of A2] {};
	\node[rectangle, fill=blue] (F3) [below=4mm of A3] {};
    \node[rectangle, fill=red] (F4) [below=4mm of A4] {};
    \node[rectangle, fill=blue] (F5) [below=4mm of A5] {};
    \node[rectangle, fill=red] (F6) [below=4mm of A6] {};
    \node[rectangle, fill=red] (F7) [above=4mm of A7] {};
    \draw (A0.east) -- (A1.west);
    \draw (A1.east) -- (A2.west);
    \draw (A2.east) -- (A3.west);
    \draw (A3.east) -- (A4.west);
    \draw (A4.east) -- (A5.west);
    \draw (A5.east) -- (A6.west);
    \draw (A3.north) -- (A7.south);
    \draw (A0.south) -- (F0.north);
    \draw (A1.south) -- (F1.north);
    \draw (A2.south) -- (F2.north);
    \draw (A3.south) -- (F3.north);
    \draw (A4.south) -- (F4.north);
    \draw (A5.south) -- (F5.north);
    \draw (A6.south) -- (F6.north);
    \draw (A7.north) -- (F7.south);
\end{tikzpicture}

%% file: figures/E8_os_flavor.tex
\begin{tikzpicture}
    \node[node, fill=blue]   (A1) {};
    \node[node, fill=red] (A2) [right=6mm of A1] {};
    \node[node, fill=blue]   (A3) [right=6mm of A2] {};
    \node[node, fill=red] (A4) [right=6mm of A3] {};
    \node[node, fill=blue]   (A5) [right=6mm of A4] {};
    \node[node, fill=red] (A6) [right=6mm of A5] {};
    \node[node, fill=blue] (A7) [right=6mm of A6] {};
    \node[node, fill=red]   (A8) [above=6mm of A5] {};
    \node[rectangle, fill=red] (F1) [below=4mm of A1] {};
    \node[rectangle, fill=blue] (F2) [below=4mm of A2] {};
	\node[rectangle, fill=red] (F3) [below=4mm of A3] {};
    \node[rectangle, fill=blue] (F4) [below=4mm of A4] {};
    \node[rectangle, fill=red] (F5) [below=4mm of A5] {};
    \node[rectangle, fill=blue] (F6) [below=4mm of A6] {};
    \node[rectangle, fill=blue] (F7) [below=4mm of A7] {};
    \node[rectangle, fill=blue] (F8) [above=4mm of A8] {};
    \draw (A1.east) -- (A2.west);
    \draw (A2.east) -- (A3.west);
    \draw (A3.east) -- (A4.west);
    \draw (A4.east) -- (A5.west);
    \draw (A5.east) -- (A6.west);
    \draw (A6.east) -- (A7.west);
    \draw (A5.north) -- (A8.south);
    \draw (A1.south) -- (F1.north);
    \draw (A2.south) -- (F2.north);
    \draw (A3.south) -- (F3.north);
    \draw (A4.south) -- (F4.north);
    \draw (A5.south) -- (F5.north);
    \draw (A6.south) -- (F6.north);
    \draw (A7.south) -- (F7.north);
    \draw (A8.north) -- (F8.south);
\end{tikzpicture}

%% file: figures/E8h_os_flavor.tex
\begin{tikzpicture}
    \node[node, fill=blue] (A0)  {};
    \node[node, fill=red]   (A1) [right=6mm of A0] {};
    \node[node, fill=blue] (A2) [right=6mm of A1] {};
    \node[node, fill=red]   (A3) [right=6mm of A2] {};
    \node[node, fill=blue] (A4) [right=6mm of A3] {};
    \node[node, fill=red]   (A5) [right=6mm of A4] {};
    \node[node, fill=blue] (A6) [right=6mm of A5] {};
    \node[node, fill=red] (A7) [right=6mm of A6] {};
    \node[node, fill=blue]   (A8) [above=6mm of A5] {};
    \node[rectangle, fill=red] (F0) [below=4mm of A0] {};
    \node[rectangle, fill=blue] (F1) [below=4mm of A1] {};
    \node[rectangle, fill=red] (F2) [below=4mm of A2] {};
	\node[rectangle, fill=blue] (F3) [below=4mm of A3] {};
    \node[rectangle, fill=red] (F4) [below=4mm of A4] {};
    \node[rectangle, fill=blue] (F5) [below=4mm of A5] {};
    \node[rectangle, fill=red] (F6) [below=4mm of A6] {};
    \node[rectangle, fill=blue] (F7) [below=4mm of A7] {};
	\node[rectangle, fill=red] (F8) [above=4mm of A8] {};
    \draw (A0.east) -- (A1.west);
    \draw (A1.east) -- (A2.west);
    \draw (A2.east) -- (A3.west);
    \draw (A3.east) -- (A4.west);
    \draw (A4.east) -- (A5.west);
    \draw (A5.east) -- (A6.west);
    \draw (A6.east) -- (A7.west);
    \draw (A5.north) -- (A8.south);
    \draw (A0.south) -- (F0.north);
    \draw (A1.south) -- (F1.north);
    \draw (A2.south) -- (F2.north);
    \draw (A3.south) -- (F3.north);
    \draw (A4.south) -- (F4.north);
    \draw (A5.south) -- (F5.north);
    \draw (A6.south) -- (F6.north);
    \draw (A7.south) -- (F7.north);
    \draw (A8.north) -- (F8.south);
\end{tikzpicture}

%% file: figures/Bk_os_flavor.tex
\begin{tikzpicture}
    \node[node, fill=red]   (A1) {};
    \node[node, fill=blue] (A2) [right=6mm of A1] {};
    \node[node, fill=red]   (A3) [right=6mm of A2] {};
    \node (A4) [right=6mm of A3] {\dots};
    \node[node, fill=red]   (A5) [right=6mm of A4] {};
    \node[node, fill=blue] (A6) [right=6mm of A5] {};
    \node[node, fill=black] (A7) [right=6mm of A6] {};
	\node[yscale=1.4] (C) [right=.2mm of A6] {$>$};
    \node[rectangle, fill=blue] (F1) [below=4mm of A1] {};
    \node[rectangle, fill=red] (F2) [below=4mm of A2] {};
    \node[rectangle, fill=blue] (F3) [below=4mm of A3] {};
    \node[rectangle, fill=blue] (F5) [below=4mm of A5] {};
    \node[rectangle, fill=red] (F6) [below=4mm of A6] {};
    \node[rectangle, fill=black] (F7) [below=4mm of A7] {};
    \draw (A1.east) -- (A2.west);
    \draw (A2.east) -- (A3.west);
    \draw (A3.east) -- (A4.west);
    \draw (A4.east) -- (A5.west);
    \draw (A5.east) -- (A6.west);
	\draw ([yshift=1.5pt]A6.east) -- ([yshift=1.5pt]A7.west);
    \draw ([yshift=-1.5pt]A6.east) -- ([yshift=-1.5pt]A7.west);
	\draw (A1.south) -- (F1.north);
	\draw (A2.south) -- (F2.north);
	\draw (A3.south) -- (F3.north);
	\draw (A5.south) -- (F5.north);
	\draw (A6.south) -- (F6.north);
	\draw (A7.south) -- (F7.north);

\end{tikzpicture}

%% file: figures/Bk_os_flavor_2.tex
\begin{tikzpicture}
    \node[node, fill=blue]   (A1) {};
    \node[node, fill=red] (A2) [right=6mm of A1] {};
    \node[node, fill=blue]   (A3) [right=6mm of A2] {};
    \node (A4) [right=6mm of A3] {\dots};
    \node[node, fill=red]   (A5) [right=6mm of A4] {};
    \node[node, fill=blue] (A6) [right=6mm of A5] {};
    \node[node, fill=black] (A7) [right=6mm of A6] {};
	\node[yscale=1.4] (C) [right=.2mm of A6] {$>$};
    \node[rectangle, fill=red] (F1) [below=4mm of A1] {};
    \node[rectangle, fill=blue] (F2) [below=4mm of A2] {};
    \node[rectangle, fill=red] (F3) [below=4mm of A3] {};
    \node[rectangle, fill=blue] (F5) [below=4mm of A5] {};
    \node[rectangle, fill=red] (F6) [below=4mm of A6] {};
    \node[rectangle, fill=black] (F7) [below=4mm of A7] {};
    \draw (A1.east) -- (A2.west);
    \draw (A2.east) -- (A3.west);
    \draw (A3.east) -- (A4.west);
    \draw (A4.east) -- (A5.west);
    \draw (A5.east) -- (A6.west);
	\draw ([yshift=1.5pt]A6.east) -- ([yshift=1.5pt]A7.west);
    \draw ([yshift=-1.5pt]A6.east) -- ([yshift=-1.5pt]A7.west);
	\draw (A1.south) -- (F1.north);
	\draw (A2.south) -- (F2.north);
	\draw (A3.south) -- (F3.north);
	\draw (A5.south) -- (F5.north);
	\draw (A6.south) -- (F6.north);
	\draw (A7.south) -- (F7.north);

\end{tikzpicture}

%% file: figures/Bh_os_flavor.tex
\begin{tikzpicture}
    \node[node, fill=red]   (A1) {};
    \node[node, fill=blue] (A2) [right=6mm of A1] {};
    \node[node, fill=red]   (A3) [right=6mm of A2] {};
    \node (A4) [right=6mm of A3] {\dots};
    \node[node, fill=red]   (A5) [right=6mm of A4] {};
    \node[node, fill=blue] (A6) [right=6mm of A5] {};
    \node[node, fill=black] (A7) [right=6mm of A6] {};
	\node[yscale=1.4] (C) [right=.2mm of A6] {$>$};
    \node[node, fill=red] (A0) [above=6mm of A2] {};
    \node[rectangle, fill=blue] (F1) [below=4mm of A1] {};
    \node[rectangle, fill=red] (F2) [below=4mm of A2] {};
    \node[rectangle, fill=blue] (F3) [below=4mm of A3] {};
    \node[rectangle, fill=blue] (F5) [below=4mm of A5] {};
    \node[rectangle, fill=red] (F6) [below=4mm of A6] {};
    \node[rectangle, fill=black] (F7) [below=4mm of A7] {};
    \node[rectangle, fill=blue] (F0) [above=4mm of A0] {};
    \draw (A0.south) -- (A2.north);
    \draw (A1.east) -- (A2.west);
    \draw (A2.east) -- (A3.west);
    \draw (A3.east) -- (A4.west);
    \draw (A4.east) -- (A5.west);
    \draw (A5.east) -- (A6.west);
	\draw ([yshift=1.5pt]A6.east) -- ([yshift=1.5pt]A7.west);
    \draw ([yshift=-1.5pt]A6.east) -- ([yshift=-1.5pt]A7.west);
    \draw (F0.south) -- (A0.north);
	\draw (A1.south) -- (F1.north);
	\draw (A2.south) -- (F2.north);
	\draw (A3.south) -- (F3.north);
	\draw (A5.south) -- (F5.north);
	\draw (A6.south) -- (F6.north);
	\draw (A7.south) -- (F7.north);

\end{tikzpicture}

%% file: figures/Bh_os_flavor_2.tex
\begin{tikzpicture}
    \node[node, fill=blue]   (A1) {};
    \node[node, fill=red] (A2) [right=6mm of A1] {};
    \node[node, fill=blue]   (A3) [right=6mm of A2] {};
    \node (A4) [right=6mm of A3] {\dots};
    \node[node, fill=red]   (A5) [right=6mm of A4] {};
    \node[node, fill=blue] (A6) [right=6mm of A5] {};
    \node[node, fill=black] (A7) [right=6mm of A6] {};
	\node[yscale=1.4] (C) [right=.2mm of A6] {$>$};
    \node[node, fill=blue] (A0) [above=6mm of A2] {};
    \node[rectangle, fill=red] (F1) [below=4mm of A1] {};
    \node[rectangle, fill=blue] (F2) [below=4mm of A2] {};
    \node[rectangle, fill=red] (F3) [below=4mm of A3] {};
    \node[rectangle, fill=blue] (F5) [below=4mm of A5] {};
    \node[rectangle, fill=red] (F6) [below=4mm of A6] {};
    \node[rectangle, fill=black] (F7) [below=4mm of A7] {};
    \node[rectangle, fill=red] (F0) [above=4mm of A0] {};
    \draw (A0.south) -- (A2.north);
    \draw (A1.east) -- (A2.west);
    \draw (A2.east) -- (A3.west);
    \draw (A3.east) -- (A4.west);
    \draw (A4.east) -- (A5.west);
    \draw (A5.east) -- (A6.west);
	\draw ([yshift=1.5pt]A6.east) -- ([yshift=1.5pt]A7.west);
    \draw ([yshift=-1.5pt]A6.east) -- ([yshift=-1.5pt]A7.west);
    \draw (F0.south) -- (A0.north);
	\draw (A1.south) -- (F1.north);
	\draw (A2.south) -- (F2.north);
	\draw (A3.south) -- (F3.north);
	\draw (A5.south) -- (F5.north);
	\draw (A6.south) -- (F6.north);
	\draw (A7.south) -- (F7.north);
\end{tikzpicture}

%% file: figures/Ck_os_flavor.tex
\begin{tikzpicture}
    \node[node, fill=black] (A2) [right=6mm of A1] {};
    \node[node, fill=black]   (A3) [right=6mm of A2] {};
    \node (A4) [right=6mm of A3] {\dots};
    \node[node, fill=black]   (A5) [right=6mm of A4] {};
    \node[node, fill=black] (A6) [right=6mm of A5] {};
    \node[node, fill=blue] (A7) [right=6mm of A6] {};
	\node[yscale=1.4] (D) [right=.2mm of A6] {$<$};
    \node[rectangle, fill=black] (F2) [below=4mm of A2] {};
    \node[rectangle, fill=black] (F3) [below=4mm of A3] {};
    \node[rectangle, fill=black] (F5) [below=4mm of A5] {};
    \node[rectangle, fill=black] (F6) [below=4mm of A6] {};
    \node[rectangle, fill=red] (F7) [below=4mm of A7] {};
    \draw (A2.east) -- (A3.west);
    \draw (A3.east) -- (A4.west);
    \draw (A4.east) -- (A5.west);
    \draw (A5.east) -- (A6.west);
	\draw ([yshift=1.5pt]A6.east) -- ([yshift=1.5pt]A7.west);
    \draw ([yshift=-1.5pt]A6.east) -- ([yshift=-1.5pt]A7.west);
	\draw (A2.south) -- (F2.north);
	\draw (A3.south) -- (F3.north);
	\draw (A5.south) -- (F5.north);
	\draw (A6.south) -- (F6.north);
	\draw (A7.south) -- (F7.north);
\end{tikzpicture}

%% file: figures/Ch_os_flavor.tex
\begin{tikzpicture}
    \node[node, fill=blue]   (A1) [right=6mm of A0] {};
    \node[node, fill=black] (A2) [right=6mm of A1] {};
    \node[node, fill=black]   (A3) [right=6mm of A2] {};
    \node (A4) [right=6mm of A3] {\dots};
    \node[node, fill=black]   (A5) [right=6mm of A4] {};
    \node[node, fill=black] (A6) [right=6mm of A5] {};
    \node[node, fill=blue] (A7) [right=6mm of A6] {};
	\node[yscale=1.4] (C) [right=.2mm of A1] {$>$};
	\node[yscale=1.4] (D) [right=.2mm of A6] {$<$};
	\draw ([yshift=1.5pt]A1.east) -- ([yshift=1.5pt]A2.west);
    \draw ([yshift=-1.5pt]A1.east) -- ([yshift=-1.5pt]A2.west);
    \node[rectangle, fill=red] (F1) [below=4mm of A1] {};
    \node[rectangle, fill=black] (F2) [below=4mm of A2] {};
    \node[rectangle, fill=black] (F3) [below=4mm of A3] {};
    \node[rectangle, fill=black] (F5) [below=4mm of A5] {};
    \node[rectangle, fill=black] (F6) [below=4mm of A6] {};
    \node[rectangle, fill=red] (F7) [below=4mm of A7] {};
    \draw (A2.east) -- (A3.west);
    \draw (A3.east) -- (A4.west);
    \draw (A4.east) -- (A5.west);
    \draw (A5.east) -- (A6.west);
	\draw ([yshift=1.5pt]A6.east) -- ([yshift=1.5pt]A7.west);
    \draw ([yshift=-1.5pt]A6.east) -- ([yshift=-1.5pt]A7.west);
	\draw (A1.south) -- (F1.north);
	\draw (A2.south) -- (F2.north);
	\draw (A3.south) -- (F3.north);
	\draw (A5.south) -- (F5.north);
	\draw (A6.south) -- (F6.north);
	\draw (A7.south) -- (F7.north);
\end{tikzpicture}

%% file: figures/Ch_os_flavor_2.tex
\begin{tikzpicture}
    \node[node, fill=blue]   (A1) [right=6mm of A0] {};
    \node[node, fill=black] (A2) [right=6mm of A1] {};
    \node[node, fill=black]   (A3) [right=6mm of A2] {};
    \node (A4) [right=6mm of A3] {\dots};
    \node[node, fill=black]   (A5) [right=6mm of A4] {};
    \node[node, fill=black] (A6) [right=6mm of A5] {};
    \node[node, fill=red] (A7) [right=6mm of A6] {};
	\node[yscale=1.4] (C) [right=.2mm of A1] {$>$};
	\node[yscale=1.4] (D) [right=.2mm of A6] {$<$};
	\draw ([yshift=1.5pt]A1.east) -- ([yshift=1.5pt]A2.west);
    \draw ([yshift=-1.5pt]A1.east) -- ([yshift=-1.5pt]A2.west);
    %
	%
    \draw (A2.east) -- (A3.west);
    \draw (A3.east) -- (A4.west);
    \draw (A4.east) -- (A5.west);
    \draw (A5.east) -- (A6.west);
	\draw ([yshift=1.5pt]A6.east) -- ([yshift=1.5pt]A7.west);
    \draw ([yshift=-1.5pt]A6.east) -- ([yshift=-1.5pt]A7.west);
	%
\end{tikzpicture}

%% file: figures/F4_flavor.tex
\begin{tikzpicture}
    \node[node, fill=red] (A1) {};
    \node[node, fill=blue] (A2) [right=6mm of A1] {};
    \node[node, fill=black] (A3) [right=6mm of A2] {};
    \node[node, fill=black] (A4) [right=6mm of A3] {};
	\node[yscale=1.4] (C) [right=.2mm of A2] {$>$};
    \node[rectangle, fill=blue] (F1) [below=4mm of A1] {};
    \node[rectangle, fill=red] (F2) [below=4mm of A2] {};
    \node[rectangle, fill=black] (F3) [below=4mm of A3] {};
    \node[rectangle, fill=black] (F4) [below=4mm of A4] {};
    \node (D) [right=6mm of C] {};
    \draw (A1.east) -- (A2.west);
	\draw ([yshift=1.5pt]A2.east) -- ([yshift=1.5pt]A3.west);
    \draw ([yshift=-1.5pt]A2.east) -- ([yshift=-1.5pt]A3.west);
    \draw (A3.east) -- (A4.west);
	\draw (A1.south) -- (F1.north);
	\draw (A2.south) -- (F2.north);
	\draw (A3.south) -- (F3.north);
	\draw (A4.south) -- (F4.north);
\end{tikzpicture}

%% file: figures/F4h_flavor.tex
\begin{tikzpicture}
    \node[node, fill=blue] (A0)  {};
    \node[node, fill=red] (A1) [right=6mm of A0] {};
    \node[node, fill=blue] (A2) [right=6mm of A1] {};
    \node[node, fill=black] (A3) [right=6mm of A2] {};
    \node[node, fill=black] (A4) [right=6mm of A3] {};
	\node[yscale=1.4] (C) [right=.2mm of A2] {$>$};
    \node (D) [right=6mm of C] {};
    \node[rectangle, fill=red] (F0) [below=4mm of A0] {};
    \node[rectangle, fill=blue] (F1) [below=4mm of A1] {};
    \node[rectangle, fill=red] (F2) [below=4mm of A2] {};
    \node[rectangle, fill=black] (F3) [below=4mm of A3] {};
    \node[rectangle, fill=black] (F4) [below=4mm of A4] {};
    \draw (A0.east) -- (A1.west);
    \draw (A1.east) -- (A2.west);
	\draw ([yshift=1.5pt]A2.east) -- ([yshift=1.5pt]A3.west);
    \draw ([yshift=-1.5pt]A2.east) -- ([yshift=-1.5pt]A3.west);
    \draw (A3.east) -- (A4.west);
	\draw (A0.south) -- (F0.north);
	\draw (A1.south) -- (F1.north);
	\draw (A2.south) -- (F2.north);
	\draw (A3.south) -- (F3.north);
	\draw (A4.south) -- (F4.north);
\end{tikzpicture}

%% file: figures/D3_2_flavor.tex
\begin{tikzpicture}
    \node[node, fill=black] (A0) {};
    \node[node, fill=blue] (A1) [right=6mm of A0] {};
    \node[node, fill=black] (A2) [right=6mm of A1] {};
	\node[yscale=1.4] (C) [right=.2mm of A0] {$<$};
	\node[yscale=1.4] (D) [right=.2mm of A1] {$>$};
    \node[rectangle, fill=black] (F0) [below=4mm of A0] {};
    \node[rectangle, fill=red] (F1) [below=4mm of A1] {};
    \node[rectangle, fill=black] (F2) [below=4mm of A2] {};
	\draw ([yshift=1.5pt]A0.east) -- ([yshift=1.5pt]A1.west);
    \draw ([yshift=-1.5pt]A0.east) -- ([yshift=-1.5pt]A1.west);
	\draw ([yshift=1.5pt]A1.east) -- ([yshift=1.5pt]A2.west);
    \draw ([yshift=-1.5pt]A1.east) -- ([yshift=-1.5pt]A2.west);
	\draw (A0.south) -- (F0.north);
	\draw (A1.south) -- (F1.north);
	\draw (A2.south) -- (F2.north);
\end{tikzpicture}

%% file: figures/E6_2_flavor.tex
\begin{tikzpicture}
    \node[node, fill=black] (A0)  {};
    \node[node, fill=black] (A1) [right=6mm of A0] {};
    \node[node, fill=black] (A2) [right=6mm of A1] {};
    \node[node, fill=blue] (A3) [right=6mm of A2] {};
    \node[node, fill=red] (A4) [right=6mm of A3] {};
	\node[yscale=1.4] (C) [right=.2mm of A2] {$<$};
    \node[rectangle, fill=black] (F0) [below=4mm of A0] {};
    \node[rectangle, fill=black] (F1) [below=4mm of A1] {};
    \node[rectangle, fill=black] (F2) [below=4mm of A2] {};
    \node[rectangle, fill=red] (F3) [below=4mm of A3] {};
    \node[rectangle, fill=blue] (F4) [below=4mm of A4] {};
    \draw (A0.east) -- (A1.west);
    \draw (A1.east) -- (A2.west);
	\draw ([yshift=1.5pt]A2.east) -- ([yshift=1.5pt]A3.west);
    \draw ([yshift=-1.5pt]A2.east) -- ([yshift=-1.5pt]A3.west);
    \draw (A3.east) -- (A4.west);
	\draw (A0.south) -- (F0.north);
	\draw (A1.south) -- (F1.north);
	\draw (A2.south) -- (F2.north);
	\draw (A3.south) -- (F3.north);
	\draw (A4.south) -- (F4.north);
\end{tikzpicture}

%% file: figures/Dk_2_flavor.tex
\begin{tikzpicture}
    \node[node, fill=black]   (A1) [right=6mm of A0] {};
    \node[node, fill=red] (A2) [right=6mm of A1] {};
    \node[node, fill=blue]   (A3) [right=6mm of A2] {};
    \node (A4) [right=6mm of A3] {\dots};
    \node[node, fill=blue]   (A5) [right=6mm of A4] {};
    \node[node, fill=red] (A6) [right=6mm of A5] {};
    \node[node, fill=black] (A7) [right=6mm of A6] {};
	\node[yscale=1.4] (C) [right=.2mm of A1] {$<$};
	\node[yscale=1.4] (D) [right=.2mm of A6] {$>$};
	\draw ([yshift=1.5pt]A1.east) -- ([yshift=1.5pt]A2.west);
    \draw ([yshift=-1.5pt]A1.east) -- ([yshift=-1.5pt]A2.west);
    \node[rectangle, fill=black] (F1) [below=4mm of A1] {};
    \node[rectangle, fill=red] (F2) [below=4mm of A2] {};
    \node[rectangle, fill=blue] (F3) [below=4mm of A3] {};
    \node[rectangle, fill=red] (F5) [below=4mm of A5] {};
    \node[rectangle, fill=blue] (F6) [below=4mm of A6] {};
    \node[rectangle, fill=black] (F7) [below=4mm of A7] {};
    \draw (A2.east) -- (A3.west);
    \draw (A3.east) -- (A4.west);
    \draw (A4.east) -- (A5.west);
    \draw (A5.east) -- (A6.west);
	\draw ([yshift=1.5pt]A6.east) -- ([yshift=1.5pt]A7.west);
    \draw ([yshift=-1.5pt]A6.east) -- ([yshift=-1.5pt]A7.west);
	\draw (A1.south) -- (F1.north);
	\draw (A2.south) -- (F2.north);
	\draw (A3.south) -- (F3.north);
	\draw (A5.south) -- (F5.north);
	\draw (A6.south) -- (F6.north);
	\draw (A7.south) -- (F7.north);
\end{tikzpicture}

%% file: figures/Dk_2_flavor_2.tex
\begin{tikzpicture}
    \node[node, fill=black]   (A1) [right=6mm of A0] {};
    \node[node, fill=red] (A2) [right=6mm of A1] {};
    \node[node, fill=blue]   (A3) [right=6mm of A2] {};
    \node (A4) [right=6mm of A3] {\dots};
    \node[node, fill=red]   (A5) [right=6mm of A4] {};
    \node[node, fill=blue] (A6) [right=6mm of A5] {};
    \node[node, fill=black] (A7) [right=6mm of A6] {};
	\node[yscale=1.4] (C) [right=.2mm of A1] {$<$};
	\node[yscale=1.4] (D) [right=.2mm of A6] {$>$};
	\draw ([yshift=1.5pt]A1.east) -- ([yshift=1.5pt]A2.west);
    \draw ([yshift=-1.5pt]A1.east) -- ([yshift=-1.5pt]A2.west);
    \node[rectangle, fill=black] (F1) [below=4mm of A1] {};
    \node[rectangle, fill=red] (F2) [below=4mm of A2] {};
    \node[rectangle, fill=blue] (F3) [below=4mm of A3] {};
    \node[rectangle, fill=blue] (F5) [below=4mm of A5] {};
    \node[rectangle, fill=red] (F6) [below=4mm of A6] {};
    \node[rectangle, fill=black] (F7) [below=4mm of A7] {};
    \draw (A2.east) -- (A3.west);
    \draw (A3.east) -- (A4.west);
    \draw (A4.east) -- (A5.west);
    \draw (A5.east) -- (A6.west);
	\draw ([yshift=1.5pt]A6.east) -- ([yshift=1.5pt]A7.west);
    \draw ([yshift=-1.5pt]A6.east) -- ([yshift=-1.5pt]A7.west);
	\draw (A1.south) -- (F1.north);
	\draw (A2.south) -- (F2.north);
	\draw (A3.south) -- (F3.north);
	\draw (A5.south) -- (F5.north);
	\draw (A6.south) -- (F6.north);
	\draw (A7.south) -- (F7.north);
\end{tikzpicture}

%% file: figures/Ak_2_flavor.tex
\begin{tikzpicture}
    \node[node, fill=black]   (A1) [right=6mm of A0] {};
    \node[node, fill=black] (A2) [right=6mm of A1] {};
    \node[node, fill=black]   (A3) [right=6mm of A2] {};
    \node (A4) [right=6mm of A3] {\dots};
    \node[node, fill=black]   (A5) [right=6mm of A4] {};
    \node[node, fill=blue] (A6) [right=6mm of A5] {};
    \node[node, fill=black] (A0) [above=6mm of A2] {};
    \node[rectangle, fill=black] (F0) [above=4mm of A0] {};
    \node[rectangle, fill=black] (F1) [below=4mm of A1] {};
    \node[rectangle, fill=black] (F2) [below=4mm of A2] {};
    \node[rectangle, fill=black] (F3) [below=4mm of A3] {};
    \node[rectangle, fill=black] (F5) [below=4mm of A5] {};
    \node[rectangle, fill=red] (F6) [below=4mm of A6] {};
	\node[yscale=1.4] (C) [right=.2mm of A5] {$<$};
	\draw ([yshift=1.5pt]A5.east) -- ([yshift=1.5pt]A6.west);
    \draw ([yshift=-1.5pt]A5.east) -- ([yshift=-1.5pt]A6.west);
    \draw (A0.south) -- (A2.north);
    \draw (A1.east) -- (A2.west);
    \draw (A2.east) -- (A3.west);
    \draw (A3.east) -- (A4.west);
    \draw (A4.east) -- (A5.west);
    \draw (A0.north) -- (F0.south);
    \draw (A1.south) -- (F1.north);
    \draw (A2.south) -- (F2.north);
    \draw (A3.south) -- (F3.north);
    \draw (A5.south) -- (F5.north);
    \draw (A6.south) -- (F6.north);
\end{tikzpicture}